\documentclass[aps,pra,twocolumn,amsfonts, amssymb, amsmath, amsthm,groupedaddress]{revtex4-2}
\usepackage{url}
\usepackage{color}
\usepackage{dcolumn}
\usepackage{bm}
\usepackage{cancel}

\usepackage{graphicx}
\DeclareMathOperator{\Tr}{\operatorname{Tr}}
\usepackage{subcaption}
\usepackage{relsize}

\begin{document}

\title{Continuous quantum error detection and suppression with pairwise local interactions}

\author{Yi-Hsiang Chen}
\email{yihsianc@usc.edu}

\affiliation{Department of Physics, University of Southern California, Los Angeles, California 90089, USA}

\author{Todd A. Brun}
\email{tbrun@usc.edu}
\affiliation{Communication Sciences Institute, University of Southern California, Los Angeles, California 90089, USA}

\begin{abstract}

Performing measurements for high-weight operators has been a practical problem in quantum computation, especially for quantum codes in the stabilizer formalism. The conventional procedure of measuring a high-weight operator requires multiple pairwise unitary operations, which can be slow and prone to errors.  We provide an alternative method to passively detect the value of a high-weight operator using only two-local interactions and single-qubit continuous measurements. This approach involves joint interactions between the system and continuously-monitored ancillary qubits. The measurement outcomes from the monitor qubits reveal information about the value of the operator. This information can be retrieved by using a numerical estimator or by evaluating the time average of the signals.  The interaction Hamiltonian can be effectively built using only two-local operators, based on techniques from perturbation theory. We apply this indirect detection scheme to the four-qubit Bacon-Shor code, where the two stabilizers are indirectly monitored using four ancillary qubits. Due to the fact that the four-qubit Bacon-Shor code is an error-detecting code and that the Quantum Zeno Effect can suppress errors, we also study the error suppression under the indirect measurement process. In this example, we show that various types of non-Markovian errors can be suppressed.  

\end{abstract}


\maketitle

\section{\label{intro}Introduction}
In many conventional quantum algorithms, circuits are presented in discrete time---unitary operations and measurements are treated as if they happened instantly. However, each quantum gate requires an operational duration, during which, errors can also happen. These effects can be captured in a continuous-time description of the evolution of a quantum system. Continuous measurement can be naturally incorporated into this framework. In fact, continuous weak measurement has been extensively studied and used for both practical applications and fundamental understanding \cite{Wiseman2009,Kurt2014,Oreshkov2005,Korotkov1999,Vool2016,Murch2013,Zoller2020}. In particular, continuous measurement for single-qubit observables has been well studied both theoretically and experimentally \cite{Korotkov1999,Korotkov2001,Weber2014,Murch2013}. There have been earlier studies of simultaneous continuous measurements of the non-commuting operators in the Bacon-Shor code \cite{Atalaya2017,Atalaya2019}, under the assumption that continuous measurements of the two-local operators exist. The same two-qubit continuous measurement is used in a more recent work \cite{atalaya2020continuous} for error correction and suppression under time-dependent Hamiltonians. However, methods to perform practical continuous measurement of multi-qubit observables have not been fully developed.

In the context of continuous quantum error correction, early work by Paz and Zurek \cite{Paz1998} introduces a jump-like error correction process, where the recovery operation is applied with probability $\gamma dt$ at each time step $dt$ with rate $\gamma$. This continuous-time jump-like error correction process can be realized as applying a sequence of weak measurements \cite{OreshkovBook}, and the minimum number of the required ancillary qubits is found to be $n-k+1$ for an $[[n,k,d]]$ code \cite{KC2016}. Another framework proposed by Ahn, Doherty, and Landahl (ADL) \cite{Ahn2002} uses continuous measurements with feedback control to maintain the fidelity of an unknown quantum state. Some feedback-based error correcting protocols related to the ADL scheme are studied in \cite{Ahn2003,Ahn2004,Ramon2005,Chase2008}. 

A major practical difficulty of almost all continuous quantum error correction schemes is that they assume that it is possible to continuously measure multi-qubit operators. Measuring high-weight operators is crucial for many quantum codes in the stabilizer formalism \cite{gottesman1997stabilizer}. The surface code \cite{bravyi1998quantum,Fowler2012}, for example, has stabilizer generators of weight four, and other codes can have even higher-weight stabilizers. To continuously measure these high-weight operators is challenging, since it requires Hamiltonians with many-body terms. 

In this paper, we introduce a method to indirectly detect the value of a high-weight operator using local two-body interactions and single-qubit continuous measurements. The approach involves applying an interaction Hamiltonian for the system and additional qubits that are being continuously monitored. The information of the system's value for the observable translates into different signatures of the monitored qubits, which we can identify. The setup can be applied to quantum codes where we detect errors by measuring the stabilizers. As an example, we apply this detection scheme to the four-qubit Bacon-Shor, which is an error-detecting code that can detect errors by measuring two-local operators \cite{Bacon2006}. In this paper, we focus on the measurements of the weight-four stabilizers in the error-detecting Bacon-Shor code.

It is well-known that the Quantum Zeno Effect can freeze a state in an eigenstate of an observable that is frequently measured. Since the four-qubit Bacon-Shor code is an error-detecting code, we examine whether errors can be suppressed when we apply the continuous indirect measurement of the stabilizers. In \cite{Oreshkov2007}, it is shown that non-Markovian errors can be suppressed by the Zeno effect when the system is being directly measured. Our results for indirect detection also agree with this observation.

The paper is organized as follows. In Sec.~\ref{setup}, we introduce in detail the theory behind continuous indirect detection. We also show how to retrieve the information in the monitored qubits. In Sec.~\ref{6-qubit}, we demonstrate the application to the four-qubit Bacon-Shor code including the process of error detection and error suppression. In the last section~\ref{construction}, we provide a construction of the target Hamiltonian using only two-local interactions.

\section{\label{setup}Indirect detections}
Suppose we want to detect the value of a Pauli operator $\mathcal{O}$ with eigenvalues $+1$ and $-1$ for a system. We design a Hamiltonian $H=(k/2)(I-\mathcal{O})\otimes X_m$ coupling the system to an additional monitor qubit $m$, where $X_m$ is the Pauli $X$ operator for the ancillary qubit $m$. It is convenient to rewrite the Hamiltonian in terms of projectors, i.e., $H=k\Pi_- \otimes X_m$, where $\Pi_-$ is the projector onto the $-1$ eigenspace of $\mathcal{O}$. The intuition behind this construction is that $m$ will be static when the system is in the $\mathcal{O}=+1$ eigenspace, while $Z_m$ will rotate when the system is in the $\mathcal{O}=-1$ eigenspace. By measuring $Z_m$, we gain information about which eigenspace the system is in. Therefore, we continuously measure $Z_m$ with measurement rate $\lambda$ to indirectly measure the value of $\mathcal{O}$. The measurement outcomes are given by a continuous output current $I(t)$ \cite{Wiseman2009,Kurt2014}, with
\begin{equation}
dI=\langle Z_m\rangle dt + \frac{dW}{2\sqrt{\lambda}}, \label{signal}
\end{equation}
where $dW$, representing the measurement noise, is a Wiener process with zero mean and variance $dt$. The expectation value is $\langle \ \cdot \ \rangle\equiv \Tr\left[\ \cdot\  \rho\right]$ on the total system $\rho$, including the system and the ancillary qubit. The whole system evolves according to 
\begin{equation}
\rho(t+dt)=\frac{\mathcal{A}(dI)\rho(t)\mathcal{A}^{\dagger}(dI)}{\Tr\left[ \mathcal{A}(dI)\rho(t)\mathcal{A}^{\dagger}(dI)\right]}, \label{stateevolution}
\end{equation}
where 
\begin{equation}
\mathcal{A}(dI)=e^{-iH dt -\lambda \left(\frac{dI}{dt}-Z_m\right)^2dt}. \label{Aoperator}
\end{equation}
This process drives $\rho$ towards one of the eigenspace of $\mathcal{O}$. 

To observe this, we expand Eq.~(\ref{stateevolution}) using Ito's rule \cite{ito1944,Kurt2010}:
\begin{align}
d\rho&=-i[H,\rho]dt+ \lambda (Z_m \rho Z_m -\rho)dt \nonumber \\
&\ \ \ +\sqrt{\lambda}(Z_m \rho+\rho Z_m-2 \rho\langle Z_m\rangle)dW. \label{eqn4}
\end{align}
(A detailed derivation can be found in Appendix A.) The expectation value of $\mathcal{O}$ evolves as
\begin{equation}
d\langle\mathcal{O}\rangle=2\sqrt{\lambda}(\langle Z_m\mathcal{O}\rangle-\langle Z_m\rangle \langle\mathcal{O} \rangle)dW. \label{eqn5}
\end{equation} 
Here, $\langle\mathcal{O}\rangle$ is a time-dependent stochastic variable. Since $d\langle\mathcal{O}\rangle$ is proportional to the Weiner increment, the evolution of $\langle\mathcal{O}\rangle$ is a random walk with a time-varying step size. This implies the following two properties: (1) the ensemble average of $\langle \mathcal{O} \rangle$ remains constant at its initial value. (2) the variance of $\langle \mathcal{O} \rangle$ tends to increase with time. The first property can be observed by the fact that 
\begin{equation}\label{ensembleO}
\text{E}\left[ d\langle\mathcal{O}\rangle\right]=d \text{E}\left[\langle\mathcal{O}\rangle\right]=0 \implies \text{E}\left[\langle\mathcal{O}\rangle\right]=\langle\mathcal{O}\rangle_{t=0}. 
\end{equation}
The change of the variance of $\langle \mathcal{O}\rangle$ is 
\begin{align}\label{variance}
&d\left(\text{E}\left[\langle\mathcal{O}\rangle^2\right]-\left(\text{E}\left[\langle \mathcal{O}\rangle \right]\right)^2\right) \nonumber \\ 
&= \text{E}\left[d (\langle \mathcal{O}\rangle^2) \right] =4\lambda\text{E}\left[(\langle Z_m\mathcal{O}\rangle-\langle Z_m\rangle \langle\mathcal{O} \rangle)^2dt\right]\geq 0.
\end{align}
 
Eq.~(\ref{variance}) implies that $\langle\mathcal{O}\rangle$ tends to deviate from its average which remains at its initial value due to Eq.~(\ref{ensembleO}). However, $\langle\mathcal{O}\rangle$ is bounded between $-1$ and $1$. The increase of the variance implies that $\langle\mathcal{O}\rangle$ approaches either $+1$ or $-1$ at later times. As $\langle\mathcal{O}\rangle$ becomes close to $\pm1$, we have $\langle Z_m\mathcal{O}\rangle\approx\langle Z_m\rangle \langle\mathcal{O} \rangle$, and the step size of the random walk becomes small. Hence, $\langle\mathcal{O}\rangle$ tends to stabilize at $\pm1$. When $\langle\mathcal{O}\rangle$ is either $+1$ or $-1$, we have $\langle Z_m\mathcal{O}\rangle=\langle Z_m\rangle \langle\mathcal{O} \rangle$ and $d\langle\mathcal{O}\rangle=0$ for all later times. Therefore $\langle\mathcal{O}\rangle=\pm1$ is stable. This shows that when $\rho$ is constantly monitored by $Z_m$, the process drives it towards an eigenspace of $\mathcal{O}$. We call this property A.

The probabilities of $\rho$ approaching the $\mathcal{O}=\pm1$ eigenspaces, i.e., $P(\langle \mathcal{O} \rangle\to \pm1)$, also match the probabilities of getting the outcomes $\pm1$ when an $\mathcal{O}$ measurement is directly applied to $\rho$. We call this property B. This is a direct consequence of Eq.~(\ref{ensembleO}) and the fact that $\langle \mathcal{O} \rangle \to \pm1$ at later times. In fact, after a period when $\langle \mathcal{O} \rangle \to \pm1$,
 \begin{align}
 &\text{E}\left[\langle\mathcal{O}\rangle\right]=P(\langle \mathcal{O} \rangle\to +1)-P(\langle \mathcal{O} \rangle\to -1) \nonumber \\
 &=\langle \mathcal{O} \rangle_{t=0}=\Tr\left[ \frac{I+\mathcal{O}}{2} \rho(0)\right]- \Tr\left[ \frac{I-\mathcal{O}}{2} \rho(0)\right], \label{probabilities}
 \end{align}
 where $\Tr\left[ \frac{I\pm\mathcal{O}}{2} \rho(0)\right]$ are the probabilities of getting the results $\pm1$ when an $\mathcal{O}$ measurement is performed on the system. Because the probabilities add to unity, Eq.~(\ref{probabilities}) implies $\Tr\left[ \frac{I\pm\mathcal{O}}{2} \rho(0)\right]=P(\langle \mathcal{O} \rangle\to \pm1)$, which is property B.
 
 Property A and B validate the whole process as a proper $\mathcal{O}$ measurement on the quantum system.

 \subsection{Detection methods}
The value of $\langle \mathcal{O}\rangle$, however, is not directly accessible because only $Z_m$ is being continuously measured. In order to learn the value of $\mathcal{O}$, we can use a numerical estimator $\hat{\rho}$, initially proportional to the identity, to evolve according to Eq.~(\ref{stateevolution}) with the outcomes $dI$ from the $Z_m$ measurements of $\rho$. The information contained in $dI$ steers $\hat{\rho}$ to the correct eigenspace $\rho$ is in. The following explains this behavior.

Since $H$ commutes with $\mathcal{O}$, the evolution from Eq.~(\ref{stateevolution}) does not cause transitions between the eigenspaces of $\mathcal{O}$. We have Property I: if a state starts in a block diagonal form, i.e.,
\begin{equation}
\rho(0)=p_+(0) \rho_+(0) + p_-(0)\rho_-(0),
\end{equation}
such that $\Tr\left[\mathcal{O}\rho_{\pm}(0)\right]=\pm1$, $p_{\pm}(0)\geq 0$ and $p_+(0) +p_-(0)=1$, then the state maintains the same block diagonal structure at all later times:
\begin{equation}
\rho(t)=p_+(t) \rho_+(t) + p_-(t)\rho_-(t),
\end{equation}
where $\Tr\left[\mathcal{O}\rho_{\pm}(t)\right]=\pm1$, $p_{\pm}(t)\geq 0$ and $p_+(t) +p_-(t)=1$ for all $t\geq0$.

To evaluate how $p_{\pm}$ evolves with time, we look at the expectation values of the eigenspace projectors, 
\begin{align}
p_{\pm}(t+dt)&=\Tr\left[\rho(t+dt)\Pi_{\pm} \right] \nonumber\\
&=\frac{1}{\mathcal{N}}p_{\pm}(t)\Tr\left[\mathcal{A}(dI)\rho_{\pm}(t)\mathcal{A}^{\dagger}(dI)\right],
\end{align}
where $\mathcal{N}=\Tr\left[ \mathcal{A}(dI)\rho(t)\mathcal{A}^{\dagger}(dI)\right]$. For infinitesimal $dt$, one can deduce that
\begin{align} 
p_{\pm}(t+dt)&\approx\frac{1}{\mathcal{N}} p_{\pm}(t)\Tr \left[\rho_{\pm}(t)e^{-2\lambda\left(\frac{dI}{dt}-Z_m\right)^2dt}\right]  \nonumber\\
&\approx \frac{1}{\mathcal{N}} p_{\pm}(t) e^{-2\lambda\left(\frac{dI}{dt}-\langle Z_m\rangle_{\rho_{\pm}}\right)^2dt}, \label{prob1}
\end{align}
where $\langle Z_m\rangle_{\rho_{\pm}}=\Tr \left[Z_m \rho_{\pm}(t)\right]$. (The derivation is in Appendix B.) 
This form is essentially the same as Bayes's theorem---our knowledge of the probability of $\pm1$ given the outcome $dI$ is 
\begin{align}
P(\pm1|dI)&=\frac{P(dI| \pm1) P(\pm1)}{P(dI| +1) P(+1)+P(dI| -1) P(-1)} \nonumber\\
&=\frac{1}{\mathcal{N}}e^{-2\lambda \left(\frac{dI-(\pm1)dt}{\sqrt{dt}}\right)^2}P(\pm1),
\end{align}
where the exponential represents the Gaussian distribution of the stochastic variable $dI$, given the value $+1$ or $-1$. 

The evolution for $\rho_{\pm}(t)$ is 
\begin{align}
&\rho_{\pm}(t+dt)=\frac{\Pi_{\pm}\rho(t+dt)\Pi_{\pm}}{\Tr\left[\Pi_{\pm}\rho(t+dt)\Pi_{\pm}\right]} \nonumber\\
&=\frac{\mathcal{A}(dI)\rho_{\pm}(t)\mathcal{A}^{\dagger}(dI)}{\Tr\left[ \mathcal{A}(dI)\rho(t)\mathcal{A}^{\dagger}(dI)\right]}\frac{\Tr\left[ \mathcal{A}(dI)\rho(t)\mathcal{A}^{\dagger}(dI)\right]}{\Tr\left[\mathcal{A}(dI)\rho_{\pm}(t)\mathcal{A}^{\dagger}(dI)\right]} \nonumber\\
&=\frac{\mathcal{A}(dI)\rho_{\pm}(t)\mathcal{A}^{\dagger}(dI)}{\Tr\left[\mathcal{A}(dI)\rho_{\pm}(t)\mathcal{A}^{\dagger}(dI)\right]},  \label{pmevolution}
\end{align}
which has the same form as Eq.~(\ref{stateevolution}). This shows that the states $\rho_{\pm}$ evolve independently. We have Property II: if two initial states, $\rho_{1,2}(0)=\sigma_{1,2}\otimes |0\rangle_m\langle0|$, are both in the $\mathcal{O}=+1$ or both in the $\mathcal{O}=-1$ eigenspace, and they both evolve according to Eq.~(\ref{stateevolution}) with the same $\mathcal{A}(dI)$, then we have $\langle Z_m\rangle_{\rho_1}=\langle Z_m\rangle_{\rho_2}$ for any time. 

Property II is true because for any state strictly in either $\mathcal{O}=\pm 1$ eigenspace, the monitor qubit $m$ is the only part of the system with nontrivial evolution. Since both $m$'s of $\rho_{1,2}$ are initially prepared in $|0\rangle_m\langle0|$, it is true that $\langle Z_m\rangle_{\rho_1}=\langle Z_m \rangle_{\rho_2}$ for all times. 

Properties I and II and Eq.~(\ref{prob1}) are sufficient to show the steering effect of the estimator.

Let us use a numerical estimator $\hat{\rho}$ to represent our knowledge of a real system $\rho_{\rm real}$ that is constantly monitored through the measurements of $Z_m$. The initial state of the estimator is
\begin{align}
\hat{\rho}(0)&=\frac{I_d}{d}\otimes |0\rangle_m\langle0| \nonumber \\
&=\frac{1}{2}\left(\frac{2}{d}\Pi_+\right) \otimes |0\rangle_m\langle0|+\frac{1}{2}\left(\frac{2}{d}\Pi_-\right) \otimes |0\rangle_m\langle0| \nonumber \\
&= p_+(0) \rho_+(0) + p_-(0)\rho_-(0),
\end{align}
where $d$ is the system dimension excluding qubit $m$. The estimator initially has $p_{\pm}(0)=1/2$ and $\rho_{\pm}(0)=(2/d)\Pi_{\pm}\otimes  |0\rangle_m\langle0|$, and it satisfies Property I.  
Suppose the real system $\rho_{\rm real}(0)$ is in the $\mathcal{O}=-1$ eigenspace and the monitor qubit $m$ is prepared in state $|0\rangle_m\langle0|$. Continuously measuring $Z_m$ gives outcomes 
\begin{equation}
dI=\langle Z_m\rangle_{\rho_{\rm real}}dt +\frac{dW}{2\sqrt{\lambda}}. \label{realrecord}
\end{equation}
We use the signal $dI$ from $\rho_{\rm real}$ to evolve $\hat{\rho}$ according to Eq.~(\ref{stateevolution}). Since both $\rho_{\rm real}(t)$ and $\rho_-(t)$ are in the $\mathcal{O}=-1$ eigenspace and have the same initial state of $m$, we have 
\begin{equation}
\langle Z_m\rangle_{\rho_{\rm real}}=\langle Z_m\rangle_{\rho_-}
\end{equation}
for any time $t\geq0$, due to property II. 

From Eq.~(\ref{prob1}), the ratio of the $p_{\pm}$ in the estimator becomes
\begin{align}\label{minusratio}
&\frac{p_+(t+dt)}{p_-(t+dt)}=\frac{p_{+}(t) e^{-2\lambda\left(\langle Z_m\rangle_{\rho_{\rm real}}-\langle Z_m\rangle_{\rho_+}+\frac{dW}{2\sqrt{\lambda}} \right)^2dt}}{p_{-}(t) e^{-2\lambda\left(\frac{dW}{2\sqrt{\lambda}}\right)^2dt}} \nonumber\\
&\xrightarrow{\text{on\ average}}\frac{p_{+}(t) }{p_{-}(t) }e^{-2\lambda\left(\langle Z_m\rangle_{\rho_{\rm real}}-\langle Z_m\rangle_{\rho_+} \right)^2dt}.
\end{align} 
It shows that the ratio of  $p_+/p_-$ decreases on average due to the difference between $\langle Z_m\rangle_{\rho_{\rm real}}$ and $\langle Z_m\rangle_{\rho_+}$. Since $H=k\Pi_- \otimes X_m$, only the negative eigenspace causes transitions. Therefore, it is evident that $\langle Z_m\rangle_{\rho_{\rm real}}\ne \langle Z_m\rangle_{\rho_+}=1$ for most times. It is expected that $p_{-}\to 1$ and $p_{+}\to 0$ at later times. This means that the estimator $\hat{\rho}$ is driven to the $\mathcal{O}=-1$ eigenspace at later times.

If $\rho_{\rm real}$ is in the $\mathcal{O}=+1$ eigenspace, the ratio becomes
\begin{equation}\label{plusratio}
\frac{p_+(t+dt)}{p_-(t+dt)}=\frac{p_{+}(t) }{p_{-}(t) }e^{2\lambda\left(\langle Z_m\rangle_{\rho_{\rm real}}-\langle Z_m\rangle_{\rho_-} \right)^2dt}, 
\end{equation}
where $\langle Z_m\rangle_{\rho_{\rm real}}=\langle Z_m \rangle_{\rho_+}=1$. We will have $p_+\to 1$ and $p_-\to 0$ instead. The estimator $\hat{\rho}$ evolves to the $\mathcal{O}=+1$ eigenspace in this case.

The above shows that when a physical system $\rho_{\rm real}$ is in an eigenspace of $\mathcal{O}$, its measurement records $dI$ drive $\hat{\rho}$ to that eigenspace. If $\langle \mathcal{O} \rangle_{\hat{\rho}}$ approaches +1 (equivalently $p_+\to1$), we learn that $\rho_{\rm real}$ is in the eigenspace  of $\mathcal{O}=+1$. If $\langle \mathcal{O} \rangle_{\hat{\rho}}$ approaches $-1$ (equivalently $p_-\to1$) then $\rho_{\rm real}$ is in the eigenspace of $\mathcal{O}=-1$. These results are sufficient for error detections on general stabilizer codes, where we prepare the encoded state in the joint +1 eigenspace of a set of commuting operators. For each stabilizer $\mathcal{O}_i$, we attach an extra qubit $m_i$ to the system with Hamiltonian $(1/2)(I-\mathcal{O}_i)\otimes X_{m_i}$ and continuously measure $Z_{m_i}$. From the signals of measuring $Z_{m_i}$, we are able to detect if errors have taken the state out of the stabilized space. However, simulating the evolution of the estimator requires computational overhead. As the system size grows, the exponential increase of the matrix dimension makes the method of simulating the estimator impractical. We provide in the following an alternative method to retrieve the information contained in the outcomes $dI$ without simulating the whole quantum state.

Note that in this particular setup where $H=k\Pi_-\otimes X_{m}$, it is clear that if the state $\rho$ is in the $+1$ eigenspace then $\langle Z_m\rangle=1$ at all times. The signal becomes a Wiener process with a constant drift, i.e., $dI=1 dt+(dW/2\sqrt{\lambda})$. We can evaluate an average function of $dI$ defined by
\begin{equation} \label{avgdI}
\overline{I}(t)\equiv \begin{cases}
\frac{1}{t}\int^t_0 dI        & \quad \text{if} \ 0\leq t \leq w \\
\frac{1}{w}\int^t_{t-w} dI & \quad \text{if} \ w < t 
\end{cases},
\end{equation}
where $w$ is the window width which is short compared to the average time between errors (1/the rate of errors) but long compared to the inverse of the measurement rate on the monitor qubits ($1/\lambda$), i.e., $(1/\lambda)\ll w \ll$ (1/the rate of errors). In the case where $\rho$ is in the $+1$ eigenspace, the average function reads
\begin{equation}
\overline{I}(t)=1+ \begin{cases}
\frac{1}{t}\int^t_0 \frac{dW}{2\sqrt{\lambda}}        & \quad \text{if} \ 0\leq t \leq w \\
\frac{1}{w}\int^t_{t-w} \frac{dW}{2\sqrt{\lambda}} & \quad \text{if} \ w < t 
\end{cases}.
\end{equation}
The variance of $\overline{I}(t)$ is 
\begin{equation}
\text{Var}\left[ \overline{I}(t)\right]= \begin{cases}
\text{E}\left[\left(\frac{1}{t}\int^t_0 \frac{dW}{2\sqrt{\lambda}}\right)^2\right]=\frac{1}{4\lambda t}  &  \text{if} \ 0\leq t \leq w \\
\text{E}\left[\left(\frac{1}{w}\int^t_{t-w} \frac{dW}{2\sqrt{\lambda}}\right)^2\right]=\frac{1}{4\lambda w} &  \text{if} \ w<t 
\end{cases}.
\end{equation}
Because $\text{E}\left[\overline{I}(t)\right]=1$ and $\text{Var}\left[ \overline{I}(t)\right]$ is inversely proportional to time, we should expect that $\overline{I}(t)$ converges to 1 after $t\geq w$. If $\rho$ is in the $-1$ eigenspace, there will be oscillations of $\langle Z_m\rangle$. The dynamics of $\langle Z_m\rangle$ involve
\begin{align}
d\langle Z_m\rangle&= 2k \langle Y_m\rangle dt+ 2\sqrt{\lambda} (1-\langle Z_m\rangle^2)dW, \\
d\langle Y_m\rangle&= -2k \langle Z_m\rangle dt -2 \lambda \langle Y_m\rangle dt -2 \sqrt{\lambda} \langle Z_m\rangle \langle Y_m \rangle dW.
\end{align}
The $X_m$ in $H$ causes a rotation of the $y$-$z$ plane in the Bloch sphere for the monitor qubit $m$. The first terms in above equations indicate such a rotation. The exponential suppression in the second term for $\langle Y_m\rangle$ is due to the measurements on $Z_m$. The average function of $dI$ becomes
\begin{equation}
\overline{I}(t)=\overline{\langle Z_m\rangle}+ \begin{cases}
\frac{1}{t}\int^t_0 \frac{dW}{2\sqrt{\lambda}}        & \quad \text{if} \ 0\leq t \leq w \\
\frac{1}{w}\int^t_{t-w} \frac{dW}{2\sqrt{\lambda}} & \quad \text{if} \ w < t 
\end{cases},
\end{equation}
where $\overline{\langle Z_m\rangle}$ denotes the average value of $\langle Z_m\rangle$ over an integration period, i.e.,
\begin{equation}
\overline{\langle Z_m\rangle}=\begin{cases}
\frac{1}{t}\int^t_0 \langle Z_m\rangle dt         & \quad \text{if} \ 0\leq t \leq w \\
\frac{1}{w}\int^t_{t-w}  \langle Z_m\rangle dt & \quad \text{if} \ w < t 
\end{cases}.
\end{equation}

 Since there are oscillations of $\langle Z_m\rangle$ between $-1$ to 1, $\overline{\langle Z_m \rangle}$ should be noticeably smaller than $1$. The later simulation shows that $\overline{I}(t)$ approaches zero after a period of time, when $\rho$ is in the $-1$ eigenspace. By directly evaluating $\overline{I}(t)$ from the measurement outcomes, one can determine whether the state is in the $+1$ eigenspace. Although this method is noisier than the method of calculating $\langle \mathcal{O}\rangle_{\hat{\rho}}$ from the estimator, it significantly speeds up the process of detecting errors. 

The relative size between the strength of the Hamiltonian $k$ and the measurement rate $\lambda$ plays a role in determining the effectiveness of this indirect detection scheme. If $\lambda$ is too large, then the frequent measurements on $Z_m$ freeze $m$ in the state $|0\rangle_m\langle0|$ due to the quantum Zeno effect. In this case, $\langle Z_m\rangle$ stays close to 1 for a much longer time, and the information gain is greatly reduced. If $\lambda$ is too small, the ratio, in Eqs.~(\ref{minusratio}) and (\ref{plusratio}), between $p_+$ and $p_-$ changes slowly. The rate at which the estimator approaches either $\pm1$ eigenspace becomes small. This is also not an ideal limit for learning the value of $\mathcal{O}$ for $\rho$. From our testing, the most efficient regime is around $\lambda=0.5k\sim1.5k$.

In most cases, the stabilizers are high weight operators, e.g., weight four stabilizers in the surface code. Directly measuring these high weight operators requires multiple gate operations, which can be more inaccurate. This passive indirect detection scheme can provide an alternative way to measure these stabilizers. In Sec.~\ref{construction}, we show how the desired Hamiltonians can be effectively constructed by 2-local operators. In the following subsection, we provide a minimal example demonstrating the process and the behavior of the indirect detection method. We set $\lambda=0.6k$ and the time unit to be $1/k$ throughout the rest of the paper. We also omit the tensor product notation ``$\otimes$'' for the rest of the paper.

\begin{figure*}
 \begin{subfigure}{0.23\textwidth}
 \centering
  \includegraphics[width=\textwidth]{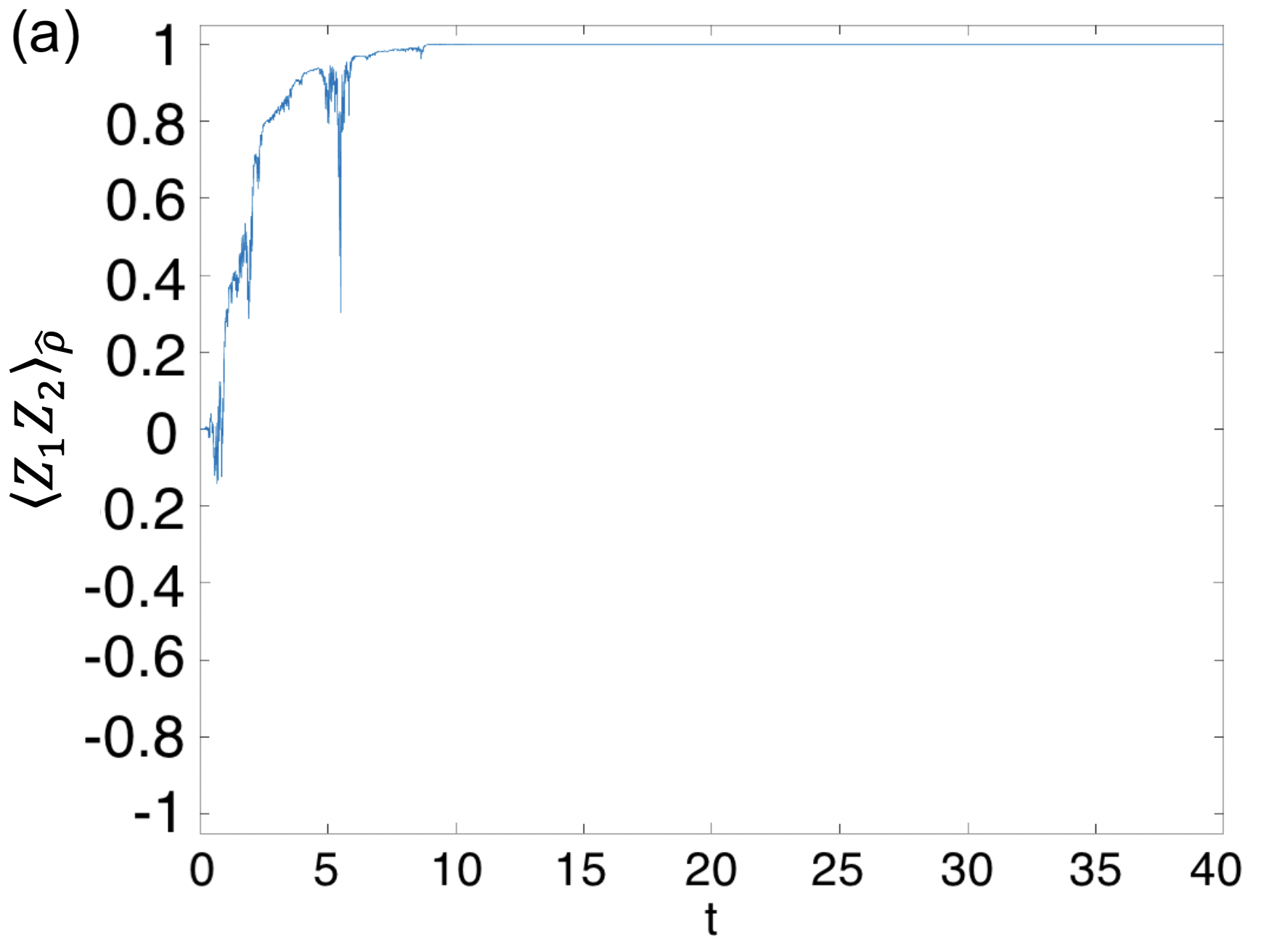} 
     \label{ZZplus_sample}
  \end{subfigure}
  ~
  \begin{subfigure}{0.23\textwidth}
  \centering
  \includegraphics[width=\textwidth]{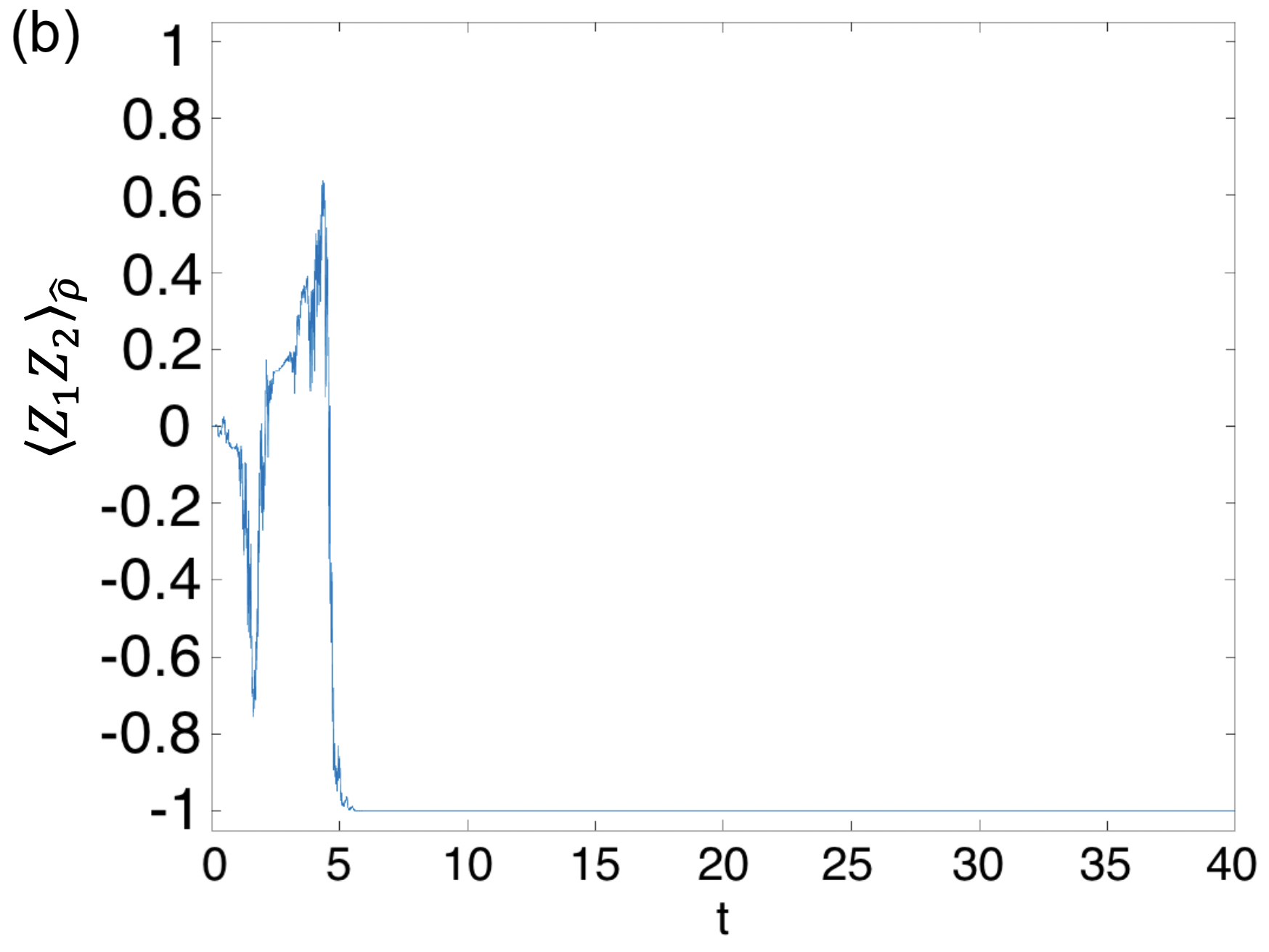}
 \label{ZZminus_sample}
\end{subfigure}
~
  \begin{subfigure}{0.23\textwidth}
  \includegraphics[width=\textwidth]{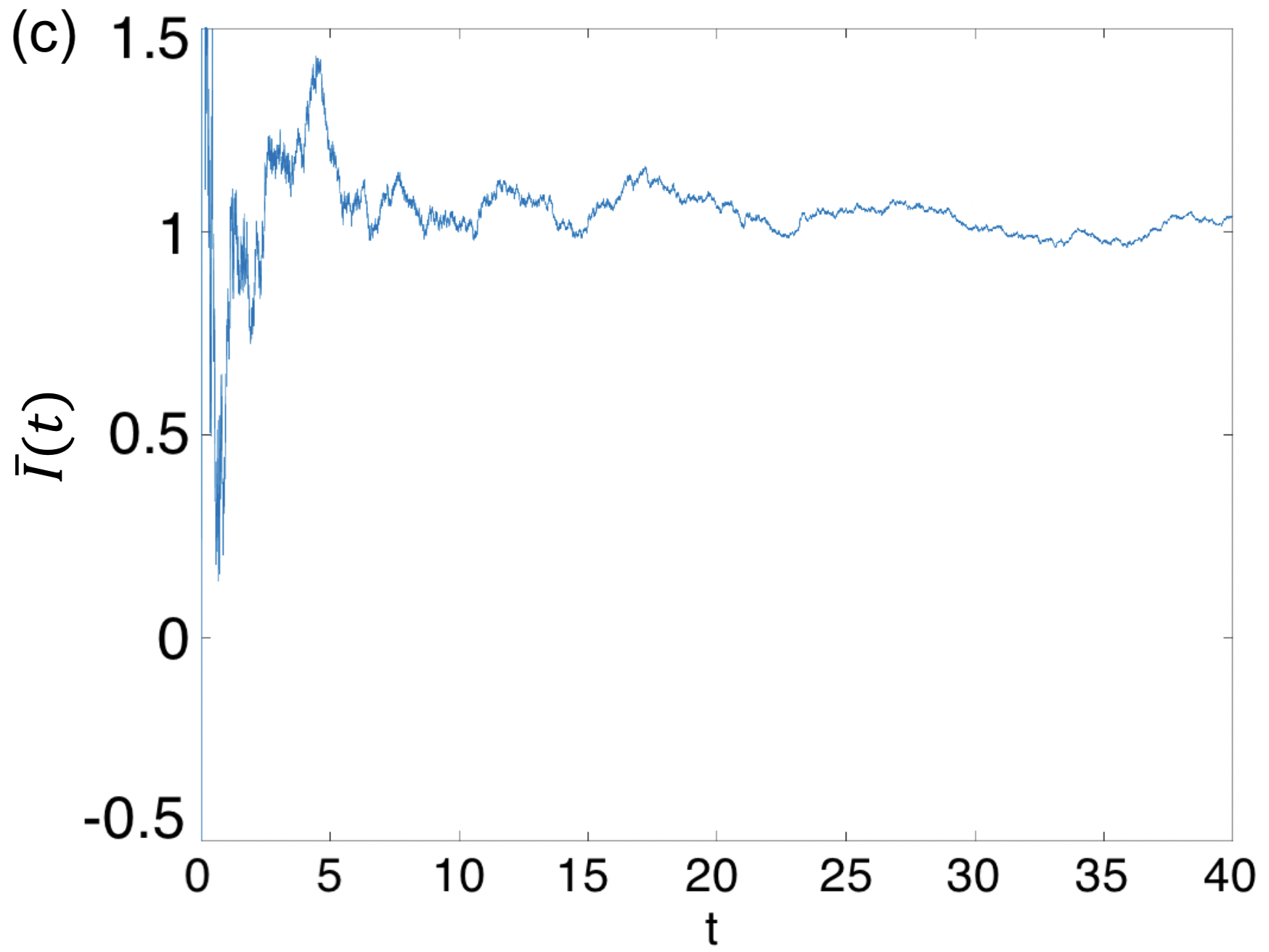} 
     \label{ZZsignalplus_sample}
  \end{subfigure}
  ~
  \begin{subfigure}{0.23\textwidth}
  \includegraphics[width=\textwidth]{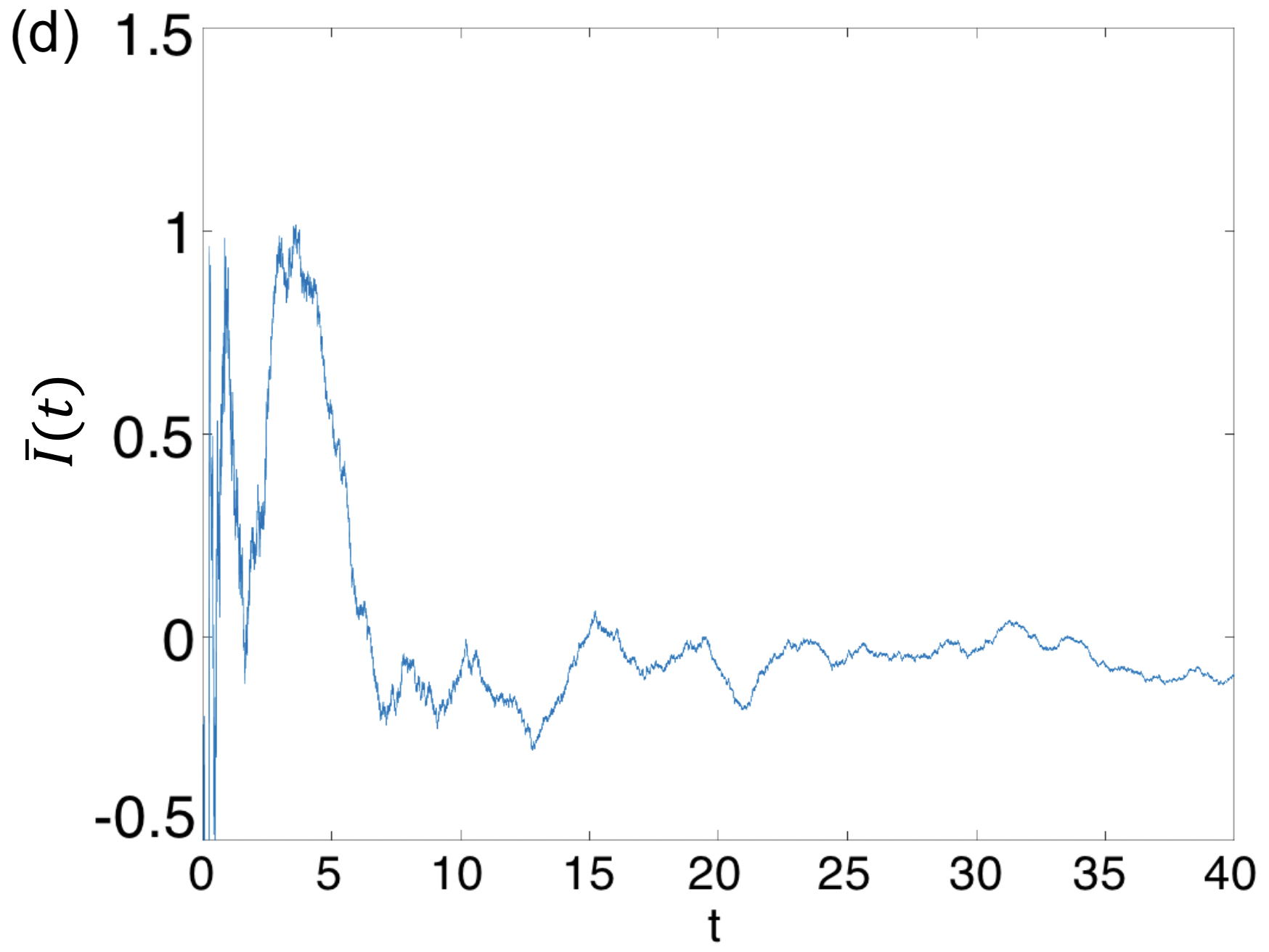}
 \label{ZZsignalminus_sample}
\end{subfigure}
~
\begin{subfigure}{0.23\textwidth}
 \centering
  \includegraphics[width=\textwidth]{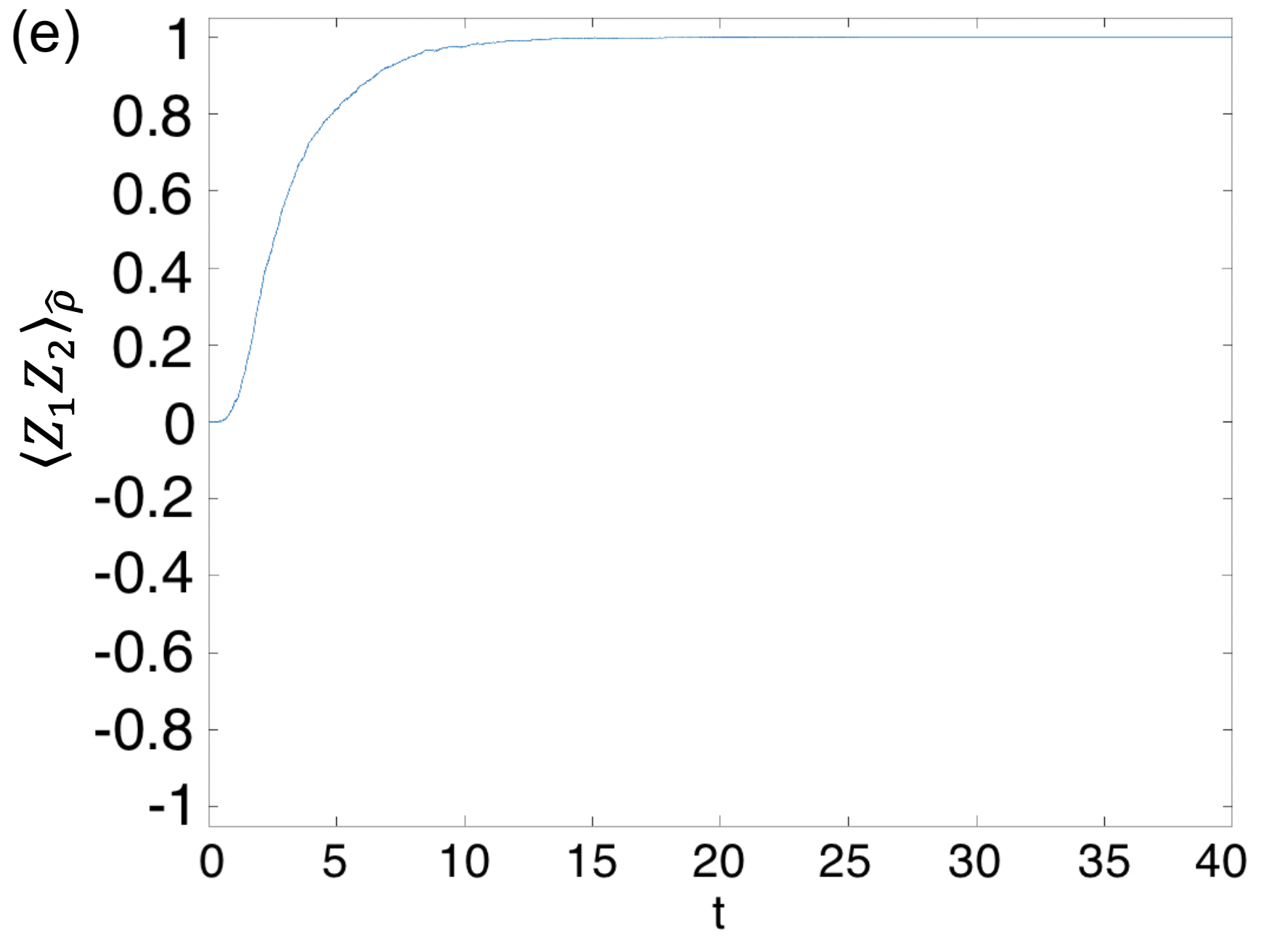} 
     \label{ZZplus_avg}
  \end{subfigure}
  ~
  \begin{subfigure}{0.23\textwidth}
  \centering
  \includegraphics[width=\textwidth]{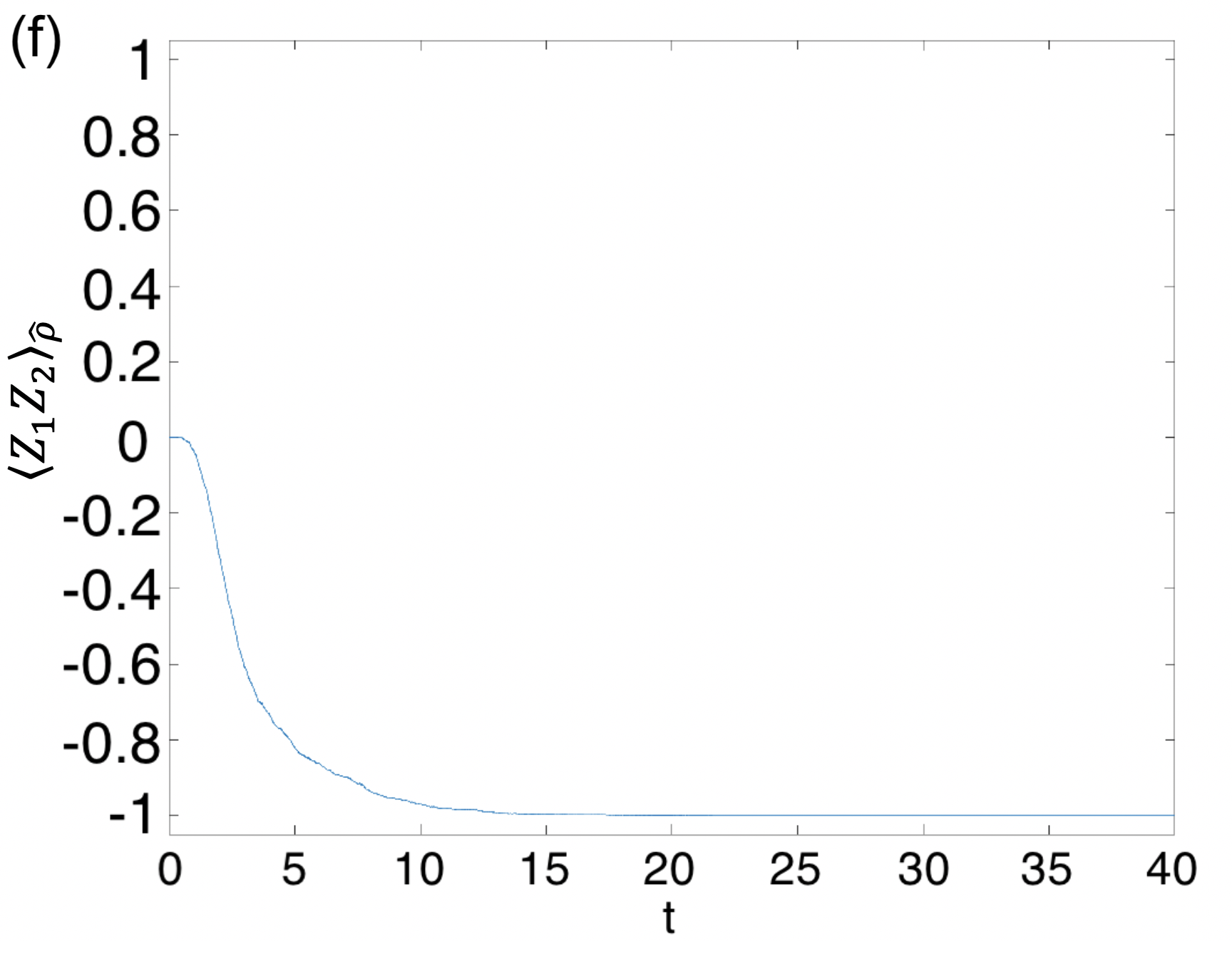}
 \label{ZZminus_avg}
\end{subfigure}
~
  \begin{subfigure}{0.23\textwidth}
  \includegraphics[width=\textwidth]{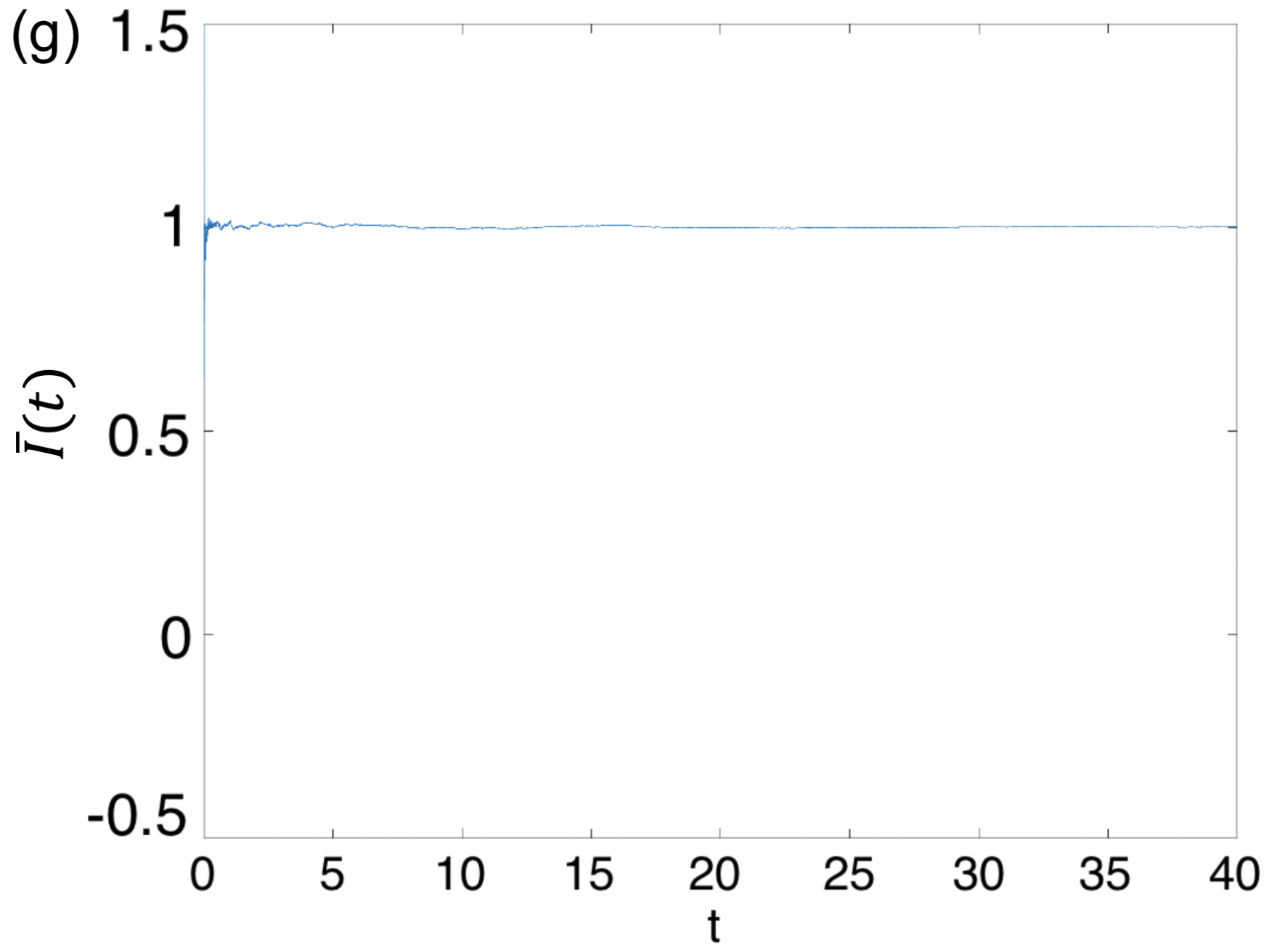} 
     \label{ZZsignalplus_avg}
  \end{subfigure}
  ~
  \begin{subfigure}{0.23\textwidth}
  \includegraphics[width=\textwidth]{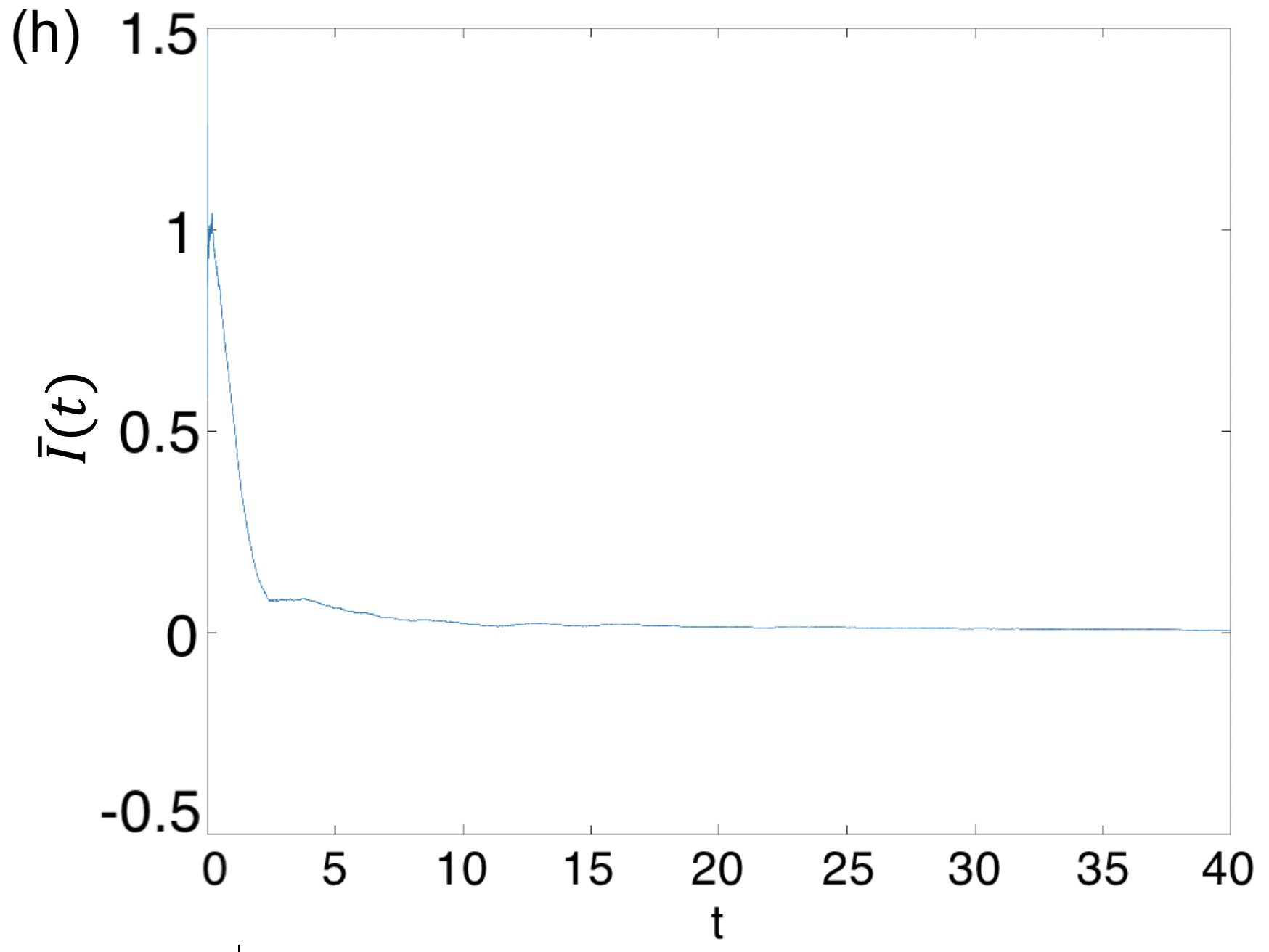}
 \label{ZZsignalminus_avg}
\end{subfigure}

 \caption{\label{ZZplot}(a), (b), (e) and (f) are the estimator approach. Each of (a) and (b) is a sample trajectory, and each of (e) and (f) is an ensemble average over 500 trajectories. The signals from measuring the physical state drive the estimator to the $Z_1Z_2=\pm1$ eigenspace the physical state is in.  (a) and (e) are the evolutions of $\langle Z_1Z_2\rangle_{\hat{\rho}}$ when $\rho$ is in the $+1$ eigenspace of $Z_1Z_2$. (b) and (f) are the cases when $\rho$ is in the $-1$ eigenspace of $Z_1Z_2$. (c), (d), (g) and (h) represent the average function $\bar{I}(t)$. Each of (c) and (d) is a sample trajectory, and each of (g) and (h) is an ensemble average over 500 trajectories. It converges to $1$ when the physical state is in the $+1$ eigenspace, and it converges to 0 when the state is in the $-1$ eigenspace.}
\end{figure*}

\subsection{$ZZ$ example}

We provide a simple 3-qubit example to demonstrate the indirect detection scheme. Suppose we want to know the value of the operator $Z_1Z_2$ for qubits $1$ and $2$. We bring in an additional monitor qubit $m$ and turn on the joint Hamiltonian $H=(k/2)(I-Z_1Z_2)X_m$. Under continuous measurements of $Z_m$ with outcomes $dI$, the whole state $\rho$ evolves according to Eq.~(\ref{stateevolution}). In experiment, $dI$ are obtained from the measurement apparatus. For simulation, the outcomes are generated using $dI=\Tr\left[ Z_m\rho\right] dt +dW/(2\sqrt{\lambda})$, where $dW$ is a Wiener process. Figs.~\ref{ZZplot}(a) and \ref{ZZplot}(b) show the dragging effect of the estimator---if the state $\rho$ is in the $Z_1Z_2=\pm1$ eigenspace then the estimator $\hat{\rho}$ approaches the eigenspace $\rho$ is in. From the value that $\langle Z_1Z_2\rangle_{\hat{\rho}}$ approaches, one can know which eigenspace $\rho$ is in. The plots Figs.~\ref{ZZplot}(e) and \ref{ZZplot}(f) are the ensemble average over $500$ trajectories of $\langle Z_1Z_2\rangle_{\hat{\rho}}$. The other method to translate the information contained in $dI$ is to evaluate the average function $\overline{I}(t)$ defined by Eq.~(\ref{avgdI}). For convenience, we choose $w=40/k$ and evaluate $\overline{I}(t)$ from time $0$ to $w$. Figs.~\ref{ZZplot}(c) and \ref{ZZplot}(d) illustrates the difference between the state $\rho$ being in the $\pm 1$ eigenspaces. $\overline{I}(t)$ converges to 1 if $\rho$ is in the $+1$ eigenspace, and it converges to 0 if $\rho$ is in the $-1$ eigenspace. In these example, the initial state of $\rho$ is $\rho(0)=(1/2)(|00\rangle+|11\rangle)(\langle 00|+\langle 11|)\otimes |0\rangle_m\langle0|$ for the case of $Z_1Z_2=+1$ and is $\rho(0)=(1/2)(|01\rangle+|10\rangle)(\langle 01|+\langle 10|)\otimes |0\rangle_m\langle0|$ for the case of $Z_1Z_2=-1$. 

 Note that there is a trade-off between accuracy and efficiency for the two methods---the estimator approach gives a more stable readout comparing to $\bar{I}(t)$ but requires computational overhead. The estimator approach can be more accurate for theoretical analysis while the average function is more experimentally feasible.

\begin{table*}
\begin{ruledtabular}
\begin{tabular}{cccc}
$ |\bar{0}\bar{0}\bar{0}\bar{0}\rangle=\frac{1}{\sqrt{2}}\left( |0000\rangle+|1111\rangle \right)$, & $|\bar{0}\bar{1}\bar{0}\bar{0}\rangle=\frac{1}{\sqrt{2}}\left( |0101\rangle+|1010\rangle \right)$, & $|\bar{0}\bar{0}\bar{1}\bar{0}\rangle=\frac{1}{\sqrt{2}}\left( |0000\rangle-|1111\rangle \right)$, &   $|\bar{0}\bar{1}\bar{1}\bar{0}\rangle=\frac{1}{\sqrt{2}}\left( |1010\rangle-|0101\rangle \right)$,  \\
$|\bar{1}\bar{0}\bar{0}\bar{0}\rangle=\frac{1}{\sqrt{2}}\left(|0011\rangle+|1100\rangle \right)$, & $|\bar{1}\bar{1}\bar{0}\bar{0}\rangle=\frac{1}{\sqrt{2}}\left( |1001\rangle+|0110\rangle \right)$, & $|\bar{1}\bar{0}\bar{1}\bar{0}\rangle=\frac{1}{\sqrt{2}}\left( |1100\rangle-|0011\rangle \right)$, &  $|\bar{1}\bar{1}\bar{1}\bar{0}\rangle=\frac{1}{\sqrt{2}}\left( |0110\rangle-|1001\rangle \right)$, \\
 $|\bar{0}\bar{0}\bar{0}\bar{1}\rangle=\frac{1}{\sqrt{2}}\left( |0001\rangle+|1110\rangle \right)$, & $|\bar{0}\bar{1}\bar{0}\bar{1}\rangle=\frac{1}{\sqrt{2}}\left( |0100\rangle+|1011\rangle \right)$, & $|\bar{0}\bar{0}\bar{1}\bar{1}\rangle=\frac{1}{\sqrt{2}}\left( |1110\rangle-|0001\rangle \right)$, & $|\bar{0}\bar{1}\bar{1}\bar{1}\rangle=\frac{1}{\sqrt{2}}\left( |0100\rangle-|1011\rangle \right)$, \\
$|\bar{1}\bar{0}\bar{0}\bar{1}\rangle=\frac{1}{\sqrt{2}}\left( |1101\rangle+|0010\rangle \right)$, & $|\bar{1}\bar{1}\bar{0}\bar{1}\rangle=\frac{1}{\sqrt{2}}\left( |1000\rangle+|0111\rangle \right)$, &  $|\bar{1}\bar{0}\bar{1}\bar{1}\rangle=\frac{1}{\sqrt{2}}\left( |0010\rangle -|1101\rangle\right)$, & $|\bar{1}\bar{1}\bar{1}\bar{1}\rangle=\frac{1}{\sqrt{2}}\left( |1000\rangle-|0111\rangle \right)$.
\end{tabular}
\end{ruledtabular}
\caption{\label{codebasis}Code basis: bar/un-bar represents encoded/physical basis.}
\end{table*}

\section{\label{6-qubit}An application to the four-qubit Bacon-Shor code}

\begin{figure*}
 \begin{subfigure}{0.23\textwidth}
 \includegraphics[width=\textwidth]{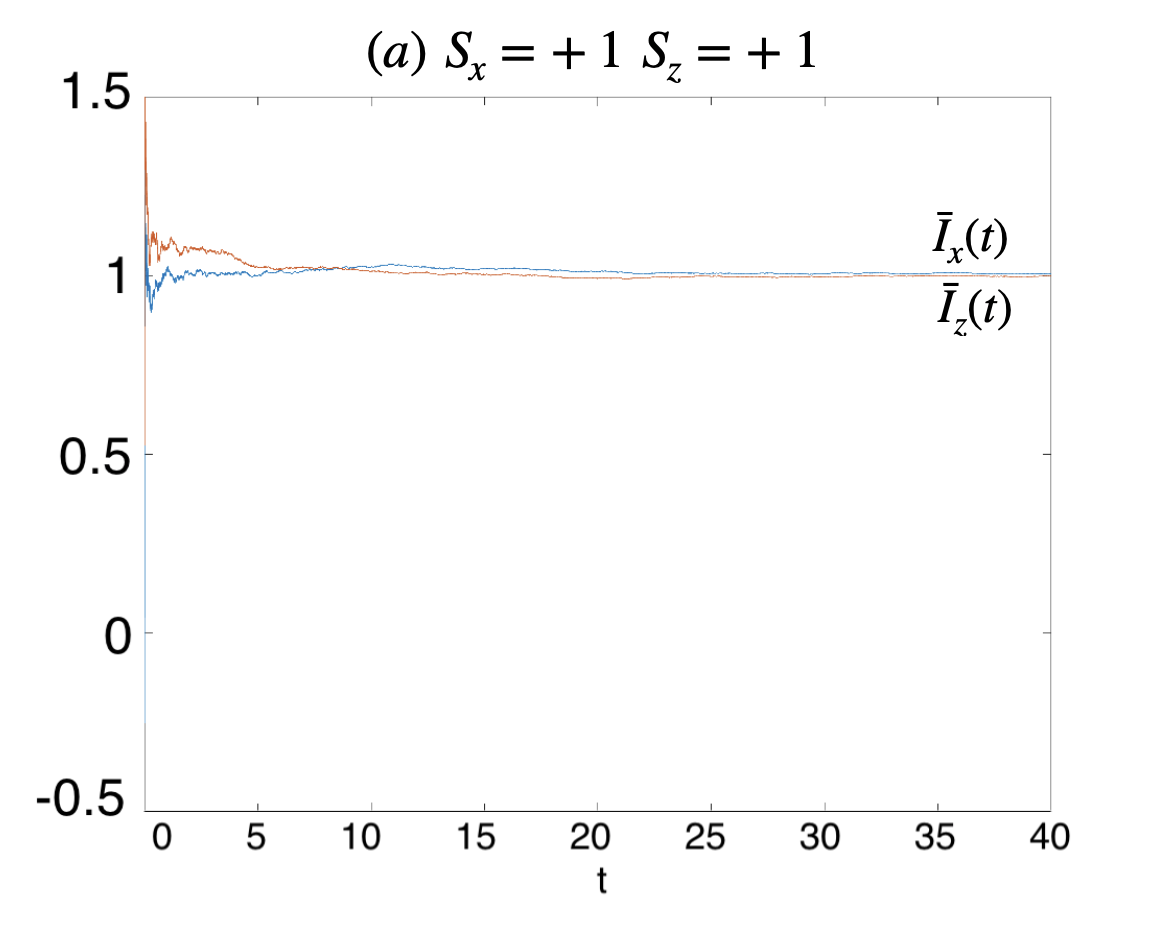}
  \label{sig++avg} 
   \end{subfigure}
  ~
  \begin{subfigure}{0.23\textwidth}
 \includegraphics[width=\textwidth]{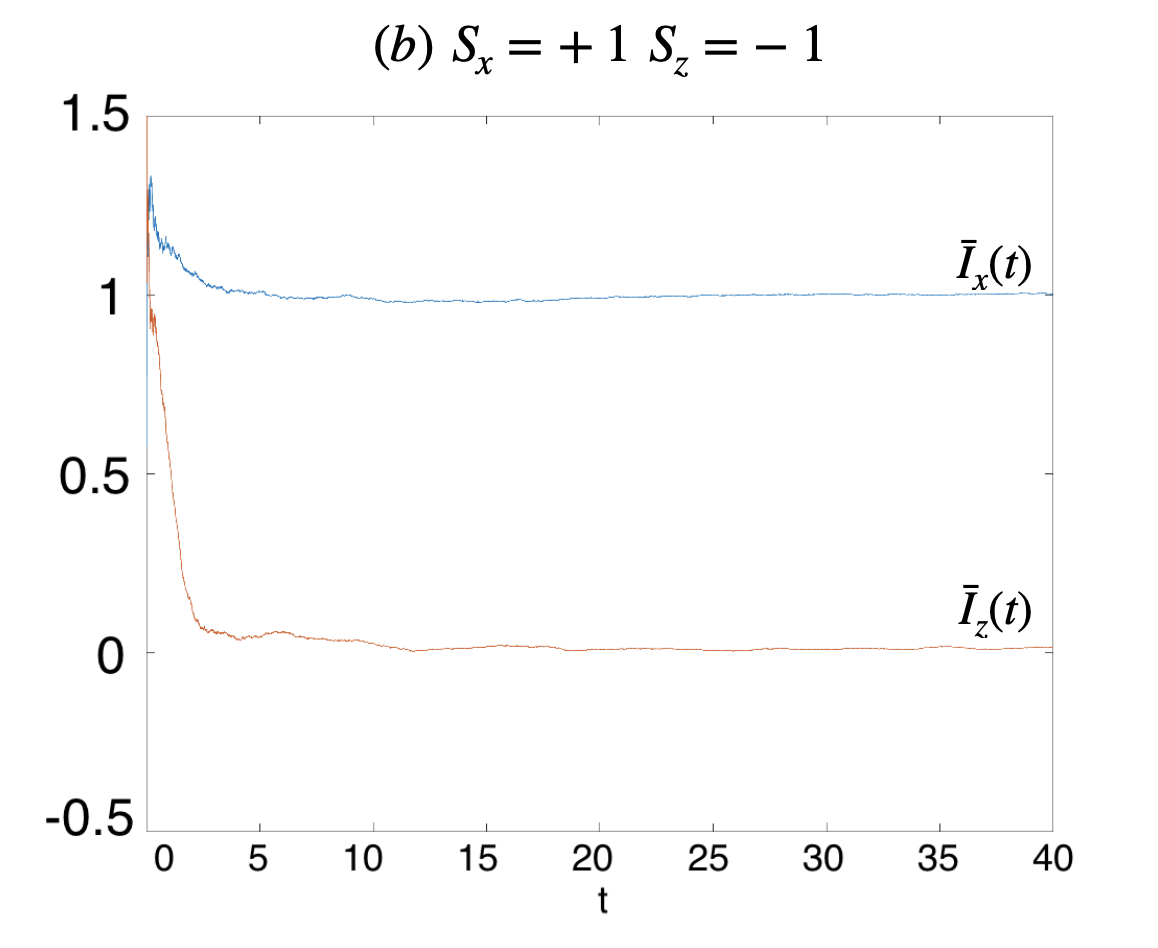}
  \label{sig+-avg} 
  \end{subfigure}
  ~
   \begin{subfigure}{0.23\textwidth}
 \includegraphics[width=\textwidth]{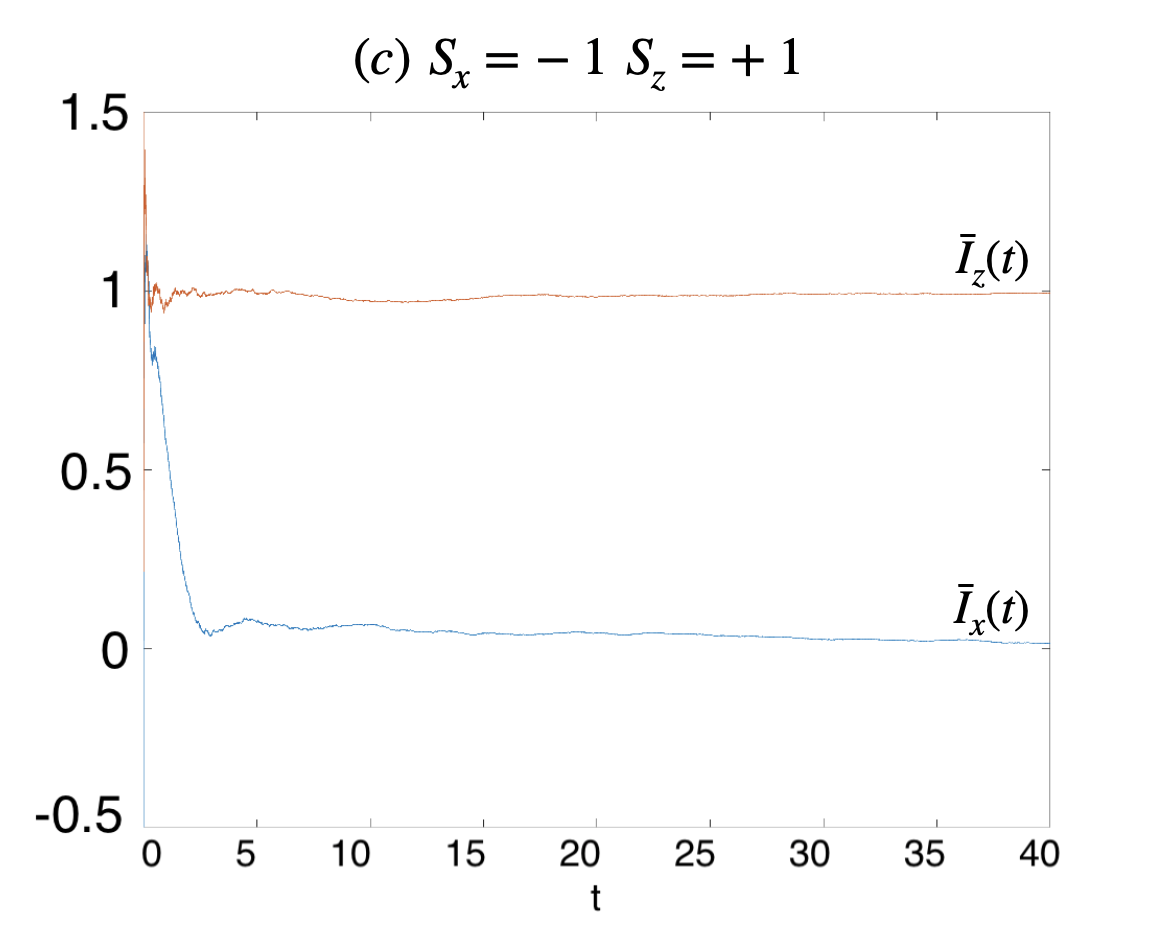}
  \label{sig-+avg} 
  \end{subfigure}
  ~
   \begin{subfigure}{0.23\textwidth}
 \includegraphics[width=\textwidth]{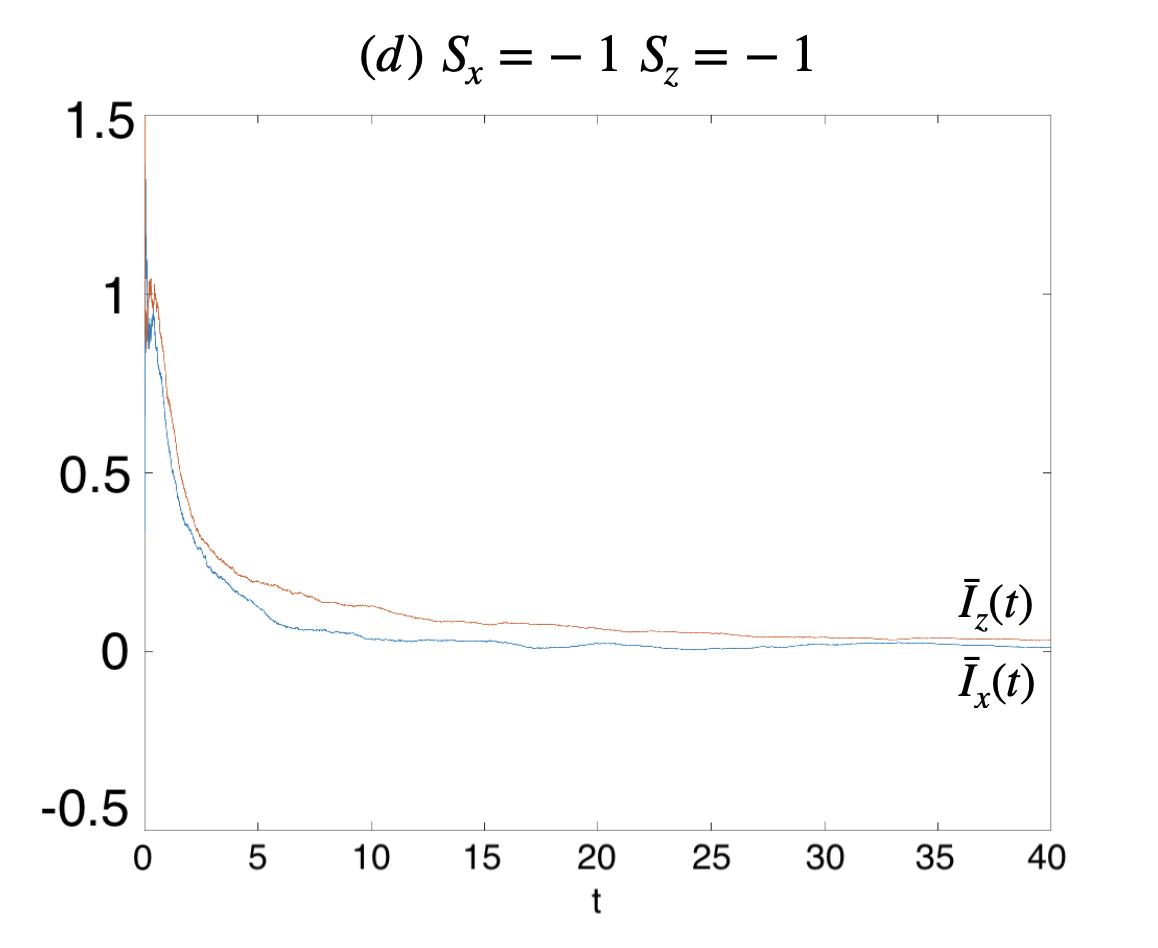}
  \label{sig--avg} 
  \end{subfigure}
  \caption{\label{sigplot}$\bar{I}_{x,z}$ verse time for the four eigenspaces. The red is $\bar{I}_z(t)$ and the blue is $\bar{I}_x(t)$. The average function converges to 1 (or 0) if the corresponding stabilizer is in the $+1$ (or $-1$) eigenspace. }
 
\end{figure*}

\begin{figure*}
 \begin{subfigure}{0.23\textwidth}
 \includegraphics[width=\textwidth]{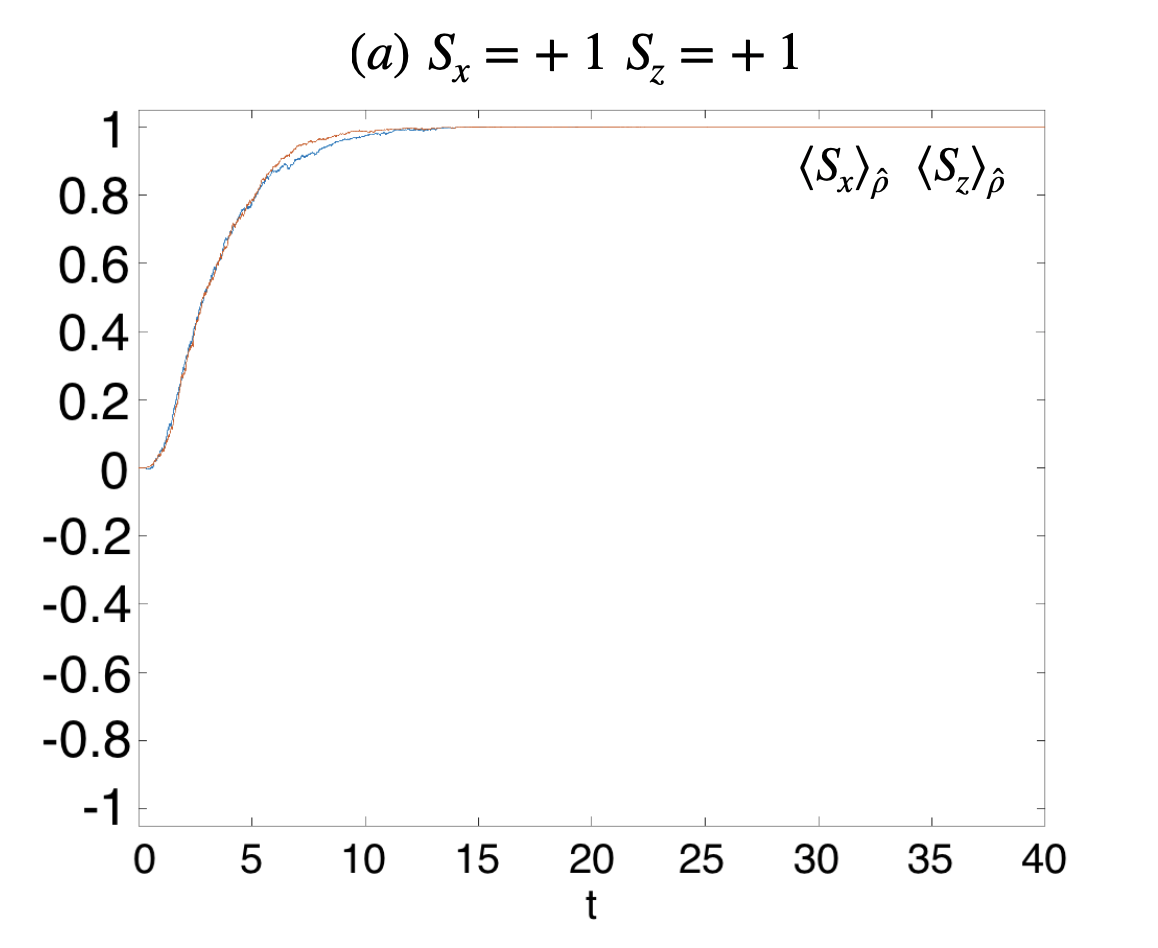}
  \label{est++avg} 
   \end{subfigure}
  ~
  \begin{subfigure}{0.23\textwidth}
 \includegraphics[width=\textwidth]{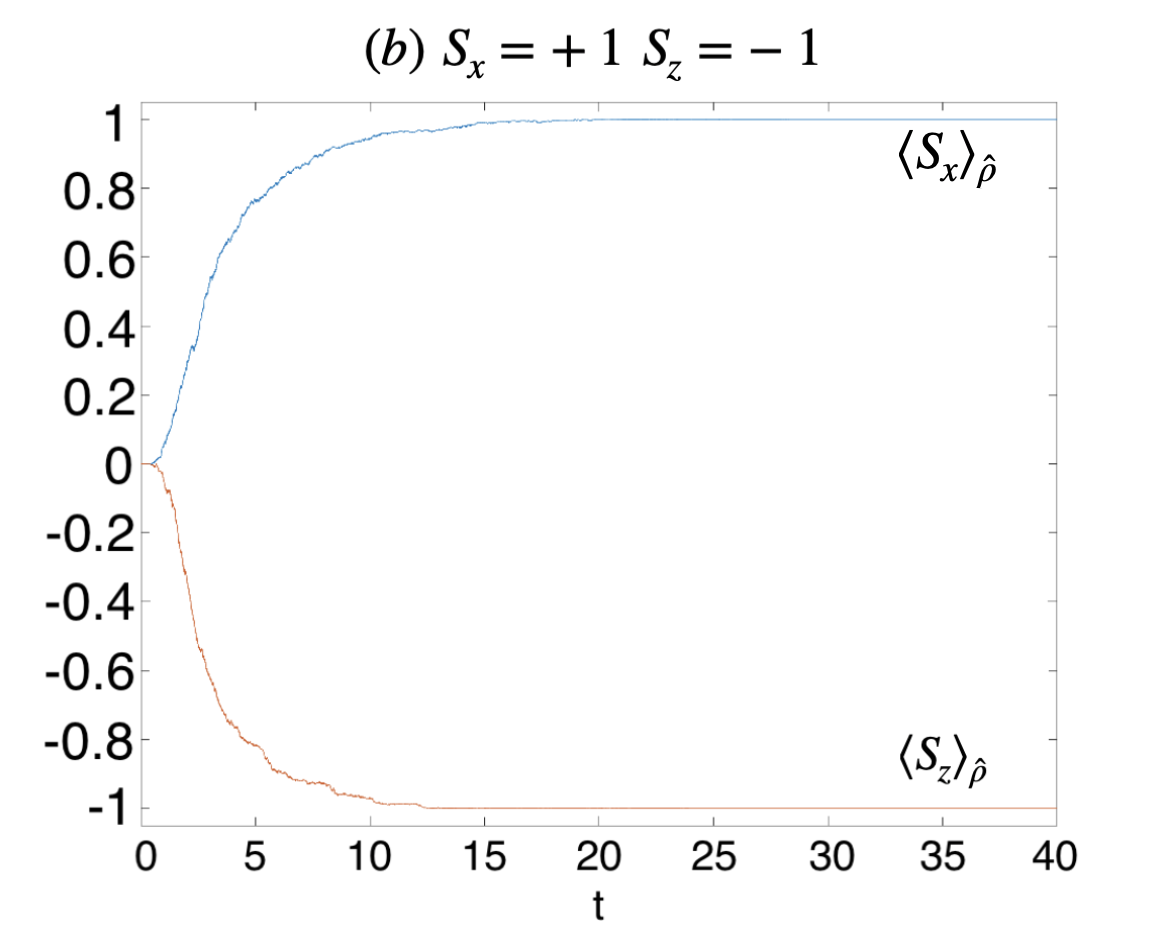}
  \label{est+-avg} 
  \end{subfigure}
  ~
   \begin{subfigure}{0.23\textwidth}
 \includegraphics[width=\textwidth]{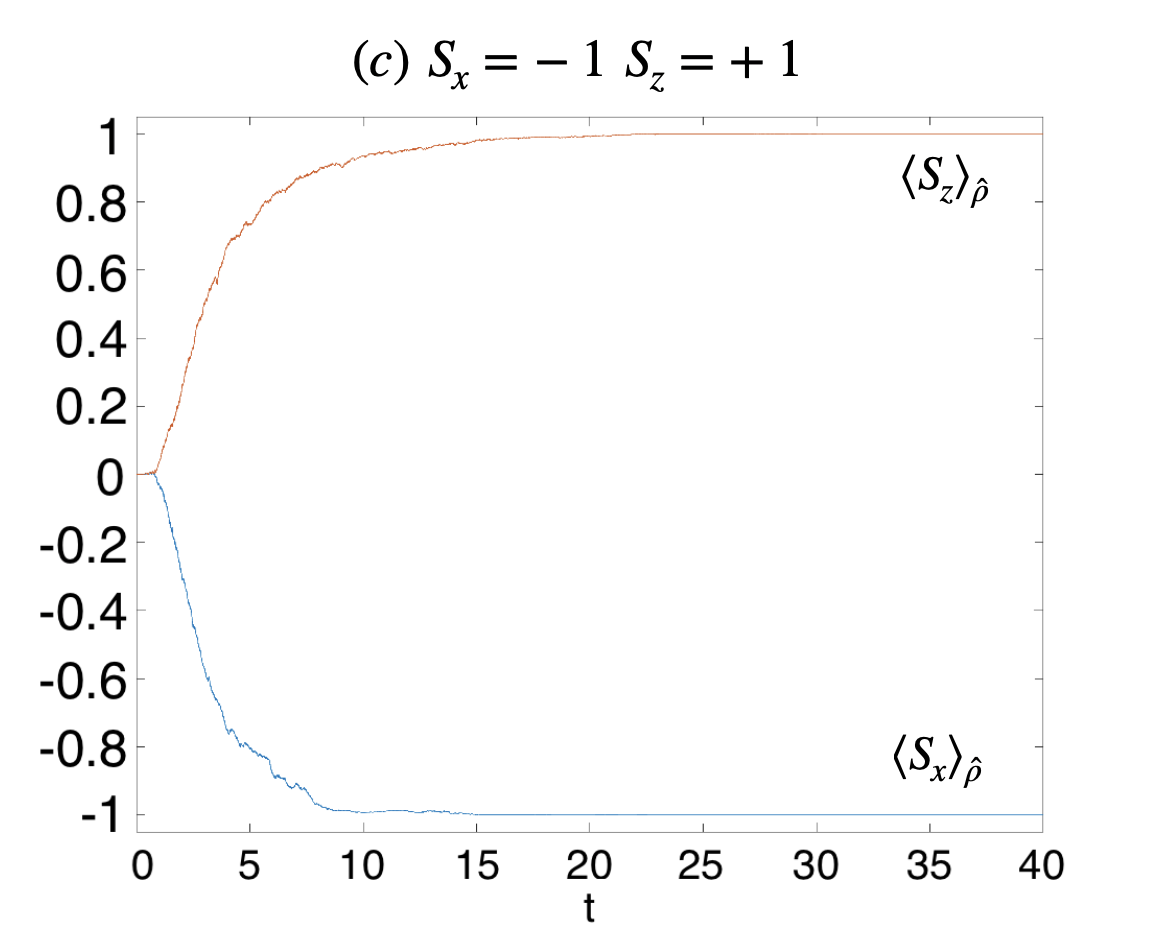}
  \label{est-+avg} 
  \end{subfigure}
  ~
   \begin{subfigure}{0.23\textwidth}
 \includegraphics[width=\textwidth]{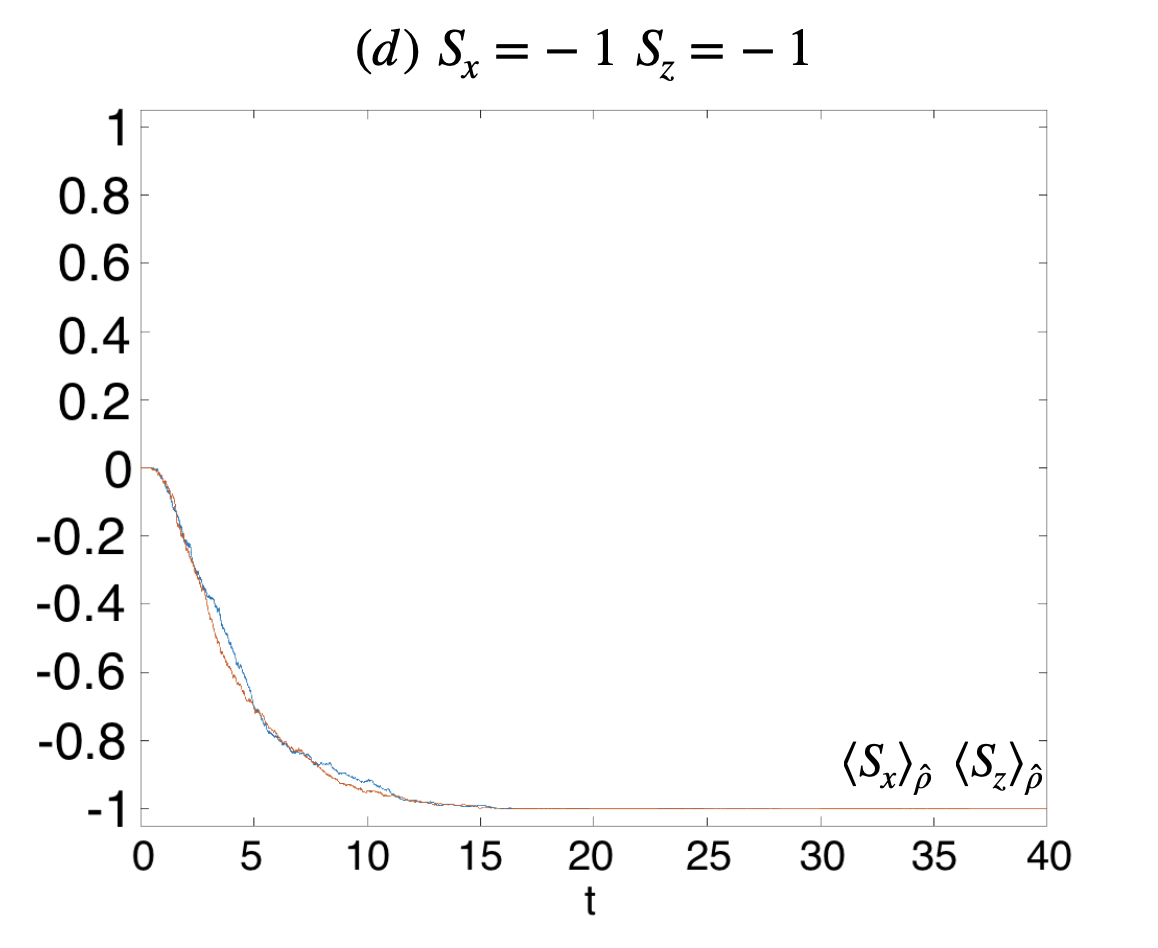}
  \label{est--avg} 
  \end{subfigure}
  \caption{\label{estplot}The estimator approach. The blue is the evolution of $\langle S_x\rangle_{\hat{\rho}}$. The red is the evolution of $\langle S_z\rangle_{\hat{\rho}}$. It is shown that the estimator approaches the eigenspace the system belongs to. }
  
\end{figure*}

The 4-qubit Bacon-Shor code is an error-detecting code that can detect errors by measuring only weight-two operators. In the stabilizer formalism, it has two weight-four stabilizers, $S_z=ZZZZ$ and $S_x=XXXX$. Checking if the system stays in the joint $S_z=+1$ and $S_x=+1$ eigenspace allows us to detect single-qubit errors. To measure the stabilizers, we could in principle bring in two extra qubits $m_z$ and $m_x$ and apply the Hamiltonian 
\begin{equation*}
H=\frac{k}{2}(I-Z_1Z_2Z_3Z_4)X_{m_z}+\frac{k}{2}(I-X_1X_2X_3X_4)X_{m_x},
\end{equation*}
with continuous measurements on $Z_{m_z}$ and $Z_{m_x}$. However, applying the weight-five Hamiltonian requires many-body interactions and is experimentally hard. As we show in Sec.~\ref{construction}, the above Hamiltonian would appear in the fifth-order expansion of the perturbation construction. It means that the base Hamiltonian should be five orders of magnitude stronger than the Hamiltonian needed for indirect detection. This poses a practical challenge for experiments. To reduce the energy scale, we can instead use 
\begin{equation}
H=\frac{k}{2}(Z_1Z_2-Z_3Z_4)X_{m_z}+\frac{k}{2}(X_1X_3-X_2X_4)X_{m_x}, \label{Hphysical}
\end{equation}
which involves only 3-local interactions. As shown in Sec.~\ref{construction}, this Hamiltonian appears in the second order expansion of the perturbative construction, where $m_z$ and $m_x$ are effective two-level systems. To gain insight into this setup, let us first recall the stabilizer formalism for the 4-qubit Bacon-Shor code. The code uses four physical qubits to encode one logical qubit and can detect any single-qubit error. The Hilbert space decomposes into tensor products of four subsystems with the following set of commuting operators and their complements,
\begin{align}
\text{Logical qubit}:& \ Z_L=Z_1Z_3& &X_L=X_1X_2 \nonumber\\
\text{Gauge qubit}:&\  Z_G=Z_1Z_2& &X_G=X_1X_3 \nonumber\\
\text{Stabilizers}:&\  S_x=X_1X_2X_3X_4& &\overline{S_x}=Z_4 \nonumber\\ 
&\ S_z=Z_1Z_2Z_3Z_4& &\overline{S_z}=X_1X_2X_3. \nonumber
\end{align}
The encoded basis is $|Z_LZ_GS_xS_z\rangle$. We add a bar on each bit for the encoded basis to distinguish it from the physical basis. For example, $|\bar{0}\bar{1}\bar{0}\bar{1}\rangle$ represents the basis vector corresponding to $Z_L=+1$, $Z_G=-1$, $S_x=+1$ and $S_z=-1$. The relationship between the two bases can be found in Table~\ref{codebasis}. In this encoded basis, it is convenient to rewrite Eq.~(\ref{Hphysical}) as
\begin{equation}\label{Hcoded}
\begin{split}
H&=\frac{k}{2}(I-S_z)Z_GX_{m_z}+\frac{k}{2}(I-S_x)X_GX_{m_x}  \\
&=k\Pi^{S_z}_{-}Z_GX_{m_z}+k\Pi^{S_x}_{-}X_GX_{m_x}, 
\end{split}
\end{equation}
where $\Pi^{S_z}_{-}$ and $\Pi^{S_x}_{-}$ are projectors onto their $-1$ eigenspaces. The signature for the state being in either combination of $S_z=\pm1$ and $S_x=\pm1$ is clear: when the state is in the $S_z=+1$ and $S_x=+1$ eigenspace, the monitor qubits are static; when either $S_z$ or $S_x$ is $-1$, there are oscillations for $Z_{m_z}$ or $Z_{m_x}$. Note that since there is no term involving $Z_L$ or $X_L$, the logical qubit is perfectly preserved during the process of indirect detection. The gauge qubit can be treated as an external degree of freedom for the system, where its dynamics are irrelevant. The non-commutativity between the two terms in $H$ is on the gauge system, and does not affect the detection process. We prepare the state in the simultaneous $+1$ eigenspace of $S_z$ and $S_x$ and continuously monitor $Z_{m_z}$ and $Z_{m_x}$. If there is no error, we should observe static values of $\langle Z_{m_{z,x}}\rangle$, which are both one in our setting. If an error takes the state out of the $+1$ eigenspace of a stabilizer then we can detect it by the non-static evolution of $\langle Z_{m_{z,x}}\rangle$. However, these expectation values are not directly obtained from experiments. The outcomes of the continuous measurements are $dI_{z,x}=\langle Z_{m_{z,x}}\rangle dt +dW_{z,x}/2\sqrt{\lambda}$. To retrieve information contained in $\langle Z_{m_{z,x}}\rangle$, we can evaluate the time average of the signals defined in Eq.~(\ref{avgdI}). We can also use an estimator $\hat{\rho}$ to learn the stabilizer values as described in Sec.~\ref{setup}. From the outcomes $dI_{z,x}=\langle Z_{m_{z,x}}\rangle_{\rho} dt +dW_{z,x}/2\sqrt{\lambda}$, we evolve the estimator according to
\begin{equation}
\hat{\rho}(t+dt)=\frac{\mathcal{A}\hat{\rho}(t)\mathcal{A}^{\dagger}}{\Tr\left[ \mathcal{A}\hat{\rho}(t)\mathcal{A}^{\dagger}\right]}, 
\end{equation}
where 
\begin{equation}
\mathcal{A}=e^{-iH dt -\lambda \left(\frac{dI_z}{dt}-Z_{m_z}\right)^2dt-\lambda \left(\frac{dI_x}{dt}-Z_{m_x}\right)^2dt}.
\end{equation}
The estimator is initially maximally mixed and can be decomposed into four blocks, i.e.,
\begin{equation}
\hat{\rho}(t)=\sum_{\alpha=\pm1,\beta=\pm1}p_{\alpha\beta}(t) \rho_{\alpha\beta}(t),
\end{equation} 
where $p_{\alpha\beta}(t)=\Tr\left[\Pi^{S_x}_{\alpha}\Pi^{S_z}_{\beta}\hat{\rho}(t)\right]$ and $\rho_{\alpha \beta}(t)=\Pi^{S_x}_{\alpha}\Pi^{S_z}_{\beta}\hat{\rho}(t)\Pi^{S_z}_{\beta}\Pi^{S_x}_{\alpha}$. The evolutions for the probabilities become
\begin{align}
&p_{\alpha \beta}(t+dt) \\
&\approx\frac{p_{\alpha\beta}(t)}{\mathcal{N}}e^{-2\lambda\left[\left(\frac{dI_z}{dt}-\langle Z_{m_z}\rangle_{\alpha\beta}\right)^2-\left(\frac{dI_x}{dt}-\langle Z_{m_x}\rangle_{\alpha\beta}\right)^2\right]dt}. \nonumber
\end{align}
When $\rho$ is in the eigenspace of $S_x=\alpha$ and $S_z=\beta$, the $p_{\alpha\beta}$ of the estimator has the largest increase on average. Hence, the estimator approaches the eigenspace of $S_x=\alpha$ and $S_z=\beta$. The argument mostly follows the discussion in Sec.~\ref{setup} for each $S_x$ and $S_z$. 

Simulations of the time-averaged signal and of the estimator approach, over 500 trajectories, are shown in Fig.~\ref{sigplot} and Fig.~\ref{estplot}. We use the following initial states as examples for the system being in the four eigenspaces of $S_x=\pm1$ and $S_z=\pm1$:
\begin{equation}
\begin{split}
\sigma_{++}&\equiv \frac{1}{2}(|\bar{0}\rangle+|\bar{1}\rangle)(\langle \bar{0}|+\langle\bar{1}|)\otimes |\bar{0}\bar{0}\bar{0}\rangle\langle\bar{0} \bar{0}\bar{0}|, \\
\sigma_{+-}&\equiv \frac{1}{2}(|\bar{0}\rangle+|\bar{1}\rangle)(\langle \bar{0}|+\langle\bar{1}|)\otimes  |\bar{0}\bar{0}\bar{1}\rangle\langle\bar{0} \bar{0}\bar{1}|, \\
\sigma_{-+}&\equiv \frac{1}{2}(|\bar{0}\rangle+|\bar{1}\rangle)(\langle \bar{0}|+\langle\bar{1}|)\otimes  |\bar{0}\bar{1}\bar{0}\rangle\langle\bar{0} \bar{1}\bar{0}|, \\
\sigma_{--}&\equiv \frac{1}{2}(|\bar{0}\rangle+|\bar{1}\rangle)(\langle \bar{0}|+\langle\bar{1}|)\otimes   |\bar{0}\bar{1}\bar{1}\rangle\langle\bar{0} \bar{1}\bar{1}|,
\end{split}
\end{equation}
where they are expressed in the encoded basis $|Z_LZ_GS_xS_z\rangle$.  $\rho(0)$ is one of the states above with monitor qubits initialized in state $|0\rangle\langle 0|$.

\subsection{Error analysis}

\subsubsection{Error detection}

\begin{figure}

\begin{subfigure}{0.45\linewidth}
 \includegraphics[width=\linewidth]{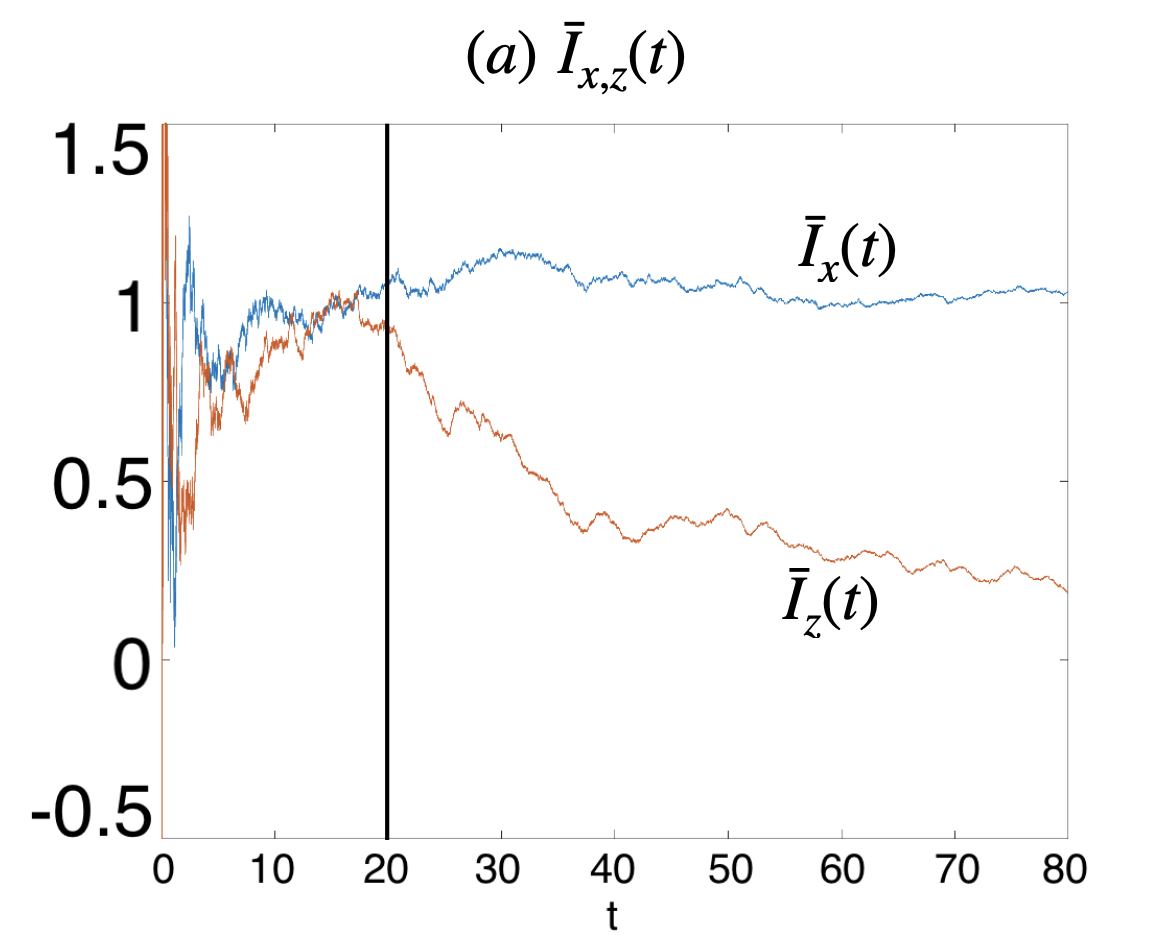}
   \end{subfigure}
  ~
  \begin{subfigure}{0.45\linewidth}
 \includegraphics[width=\linewidth]{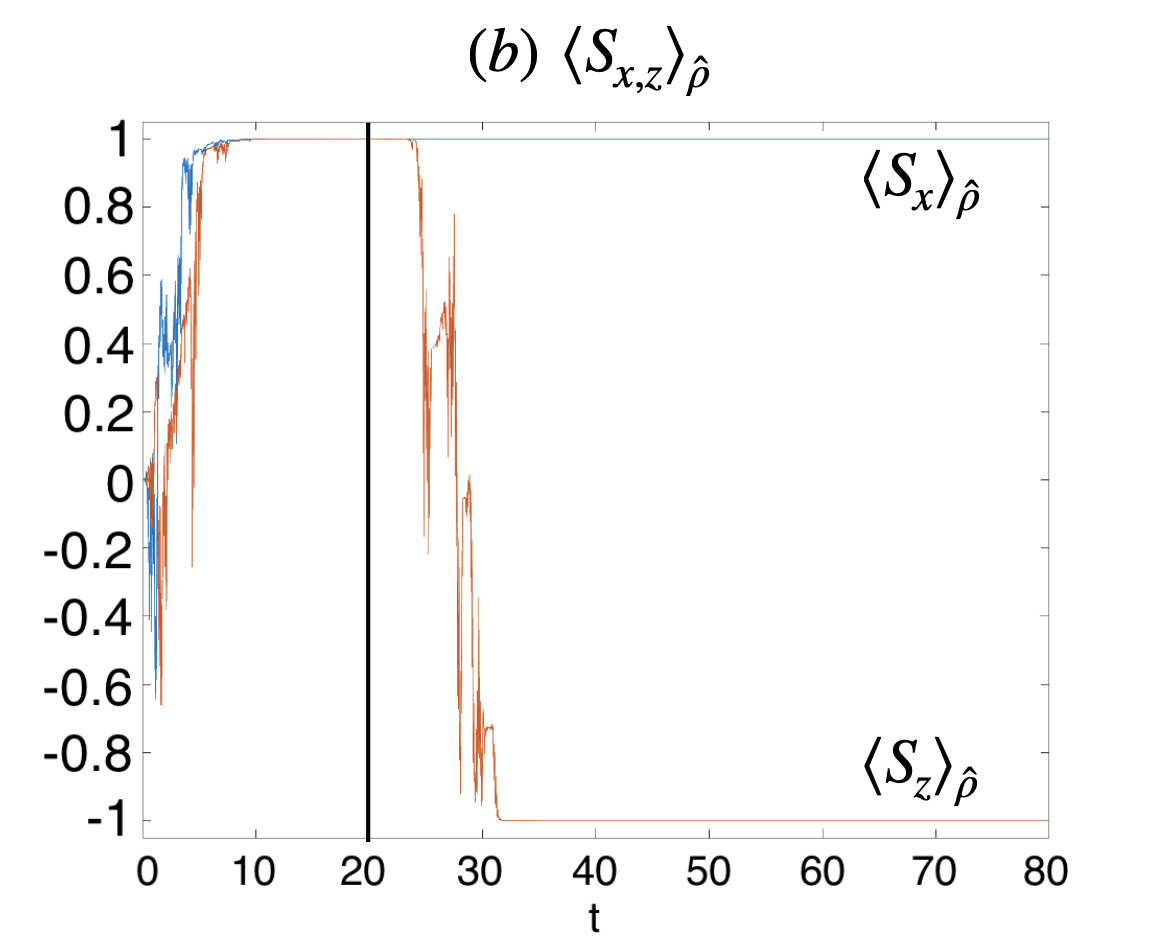}
  \end{subfigure}
\caption{An $X_1$ error happened at $t=20$ as indicated by the black line. After the error, $\bar{I}_z$ starts to approach 0 and $\langle S_z\rangle_{\hat{\rho}}$ flips to $-1$.}
\label{X1Err}

\end{figure}

When we apply the four-qubit Bacon-Shor code, we prepare the state in the $S_x=+1$ and $S_z=+1$ eigenspace and store information in the logical qubit of the state. To detect errors, we attach monitor qubits to the system and continuously measure them. If there are no errors, the monitor qubits are static and both $\bar{I}_{z,x}(t)$ converge to 1. Or we can simulate the estimator, which will approach the joint $+1$ eigenspace.  Let us first consider single-qubit errors. Suppose an $X_1$ error happened on the first system qubit.  The error anticommutes with $S_z$ and the state is taken to the $S_z=-1$ eigenspace.  A sample trajectory is shown in Fig.~\ref{X1Err}, where the error is detected by observing that $\bar{I}_z(t)$ drifts to 0 and $\langle S_z\rangle_{\hat{\rho}}$ flips to $-1$. 

\begin{figure*}

\begin{subfigure}{0.3\linewidth}
\includegraphics[width=\linewidth]{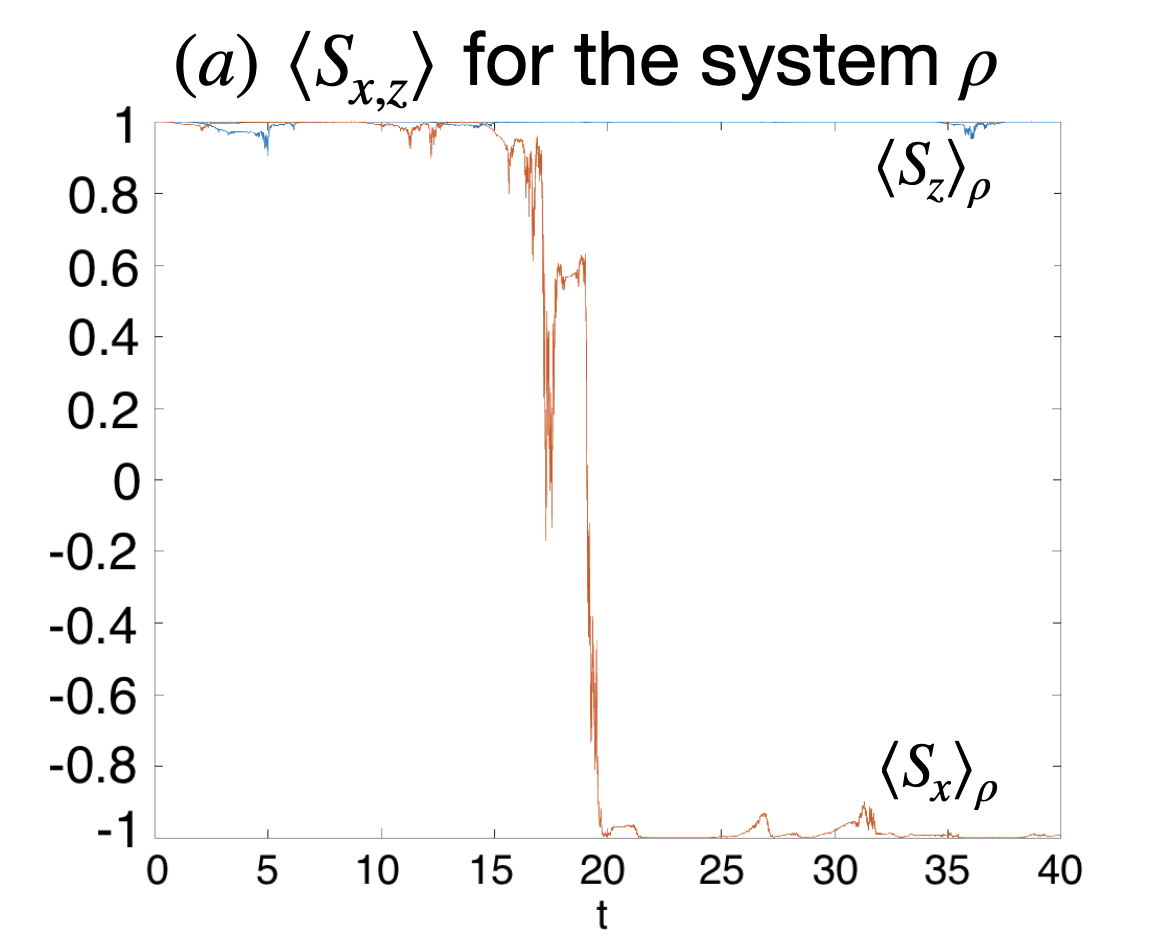}
\end{subfigure}
~
\begin{subfigure}{0.3\linewidth}
\includegraphics[width=\linewidth]{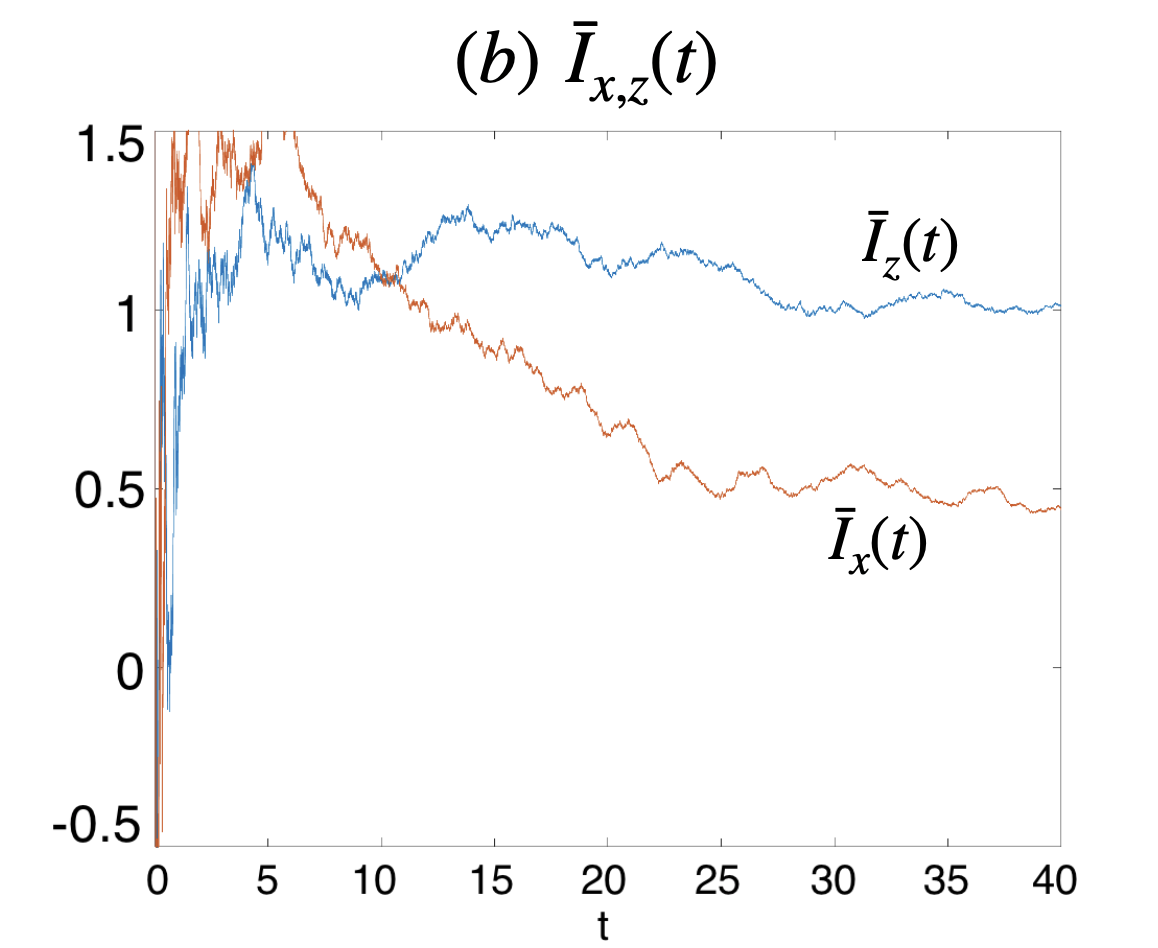}
\end{subfigure}
~
\begin{subfigure}{0.3\linewidth}
\includegraphics[width=\linewidth]{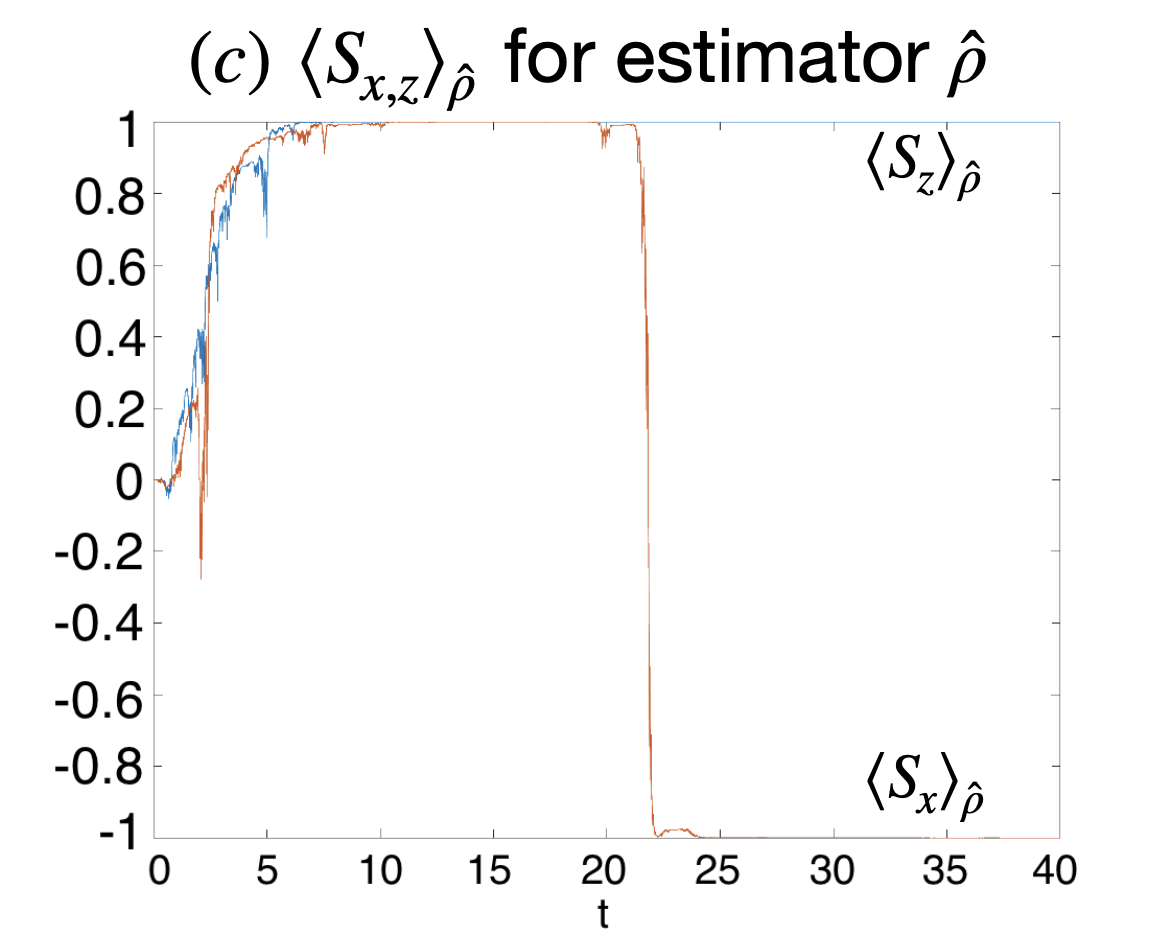}
\end{subfigure}

\caption{(a) shows $\langle S_{z,x}\rangle$ for the system. After a period, $S_x$ is flipped to $-1$ while $S_z$ remains at $+1$. (b) shows $\bar{I}_{z,x}(t)$. After a period, $\bar{I}_x(t)$ starts to decay while $\bar{I}_z(t)$ remains at $+1$. (c) shows $\langle S_{z,x}\rangle_{\hat{\rho}}$ for the estimator. $\langle S_x\rangle_{\hat{\rho}}$ approaches $+1$ initially, but it flips to $-1$ after $\langle S_x\rangle\to -1$. }\label{errsample}

\end{figure*}

We present another example where the errors are continuous-in-time 1/f Hamiltonian errors, i.e.,
\begin{equation}
H_{err}(t)=\displaystyle\sum_i \epsilon_i(t)\sigma_i, \label{1/fnoise}
\end{equation}
where each $\sigma_i$ is a single-qubit Pauli matrix acting on the $i$th qubit and $\epsilon_i(t)$ is a time-dependent scalar function. Each $\epsilon_i(t)$ consists of exponentially decaying random pulses with magnitude $\epsilon$, i.e., 
\begin{equation}
\epsilon_i(t)=\epsilon\sum_{\alpha_i}\theta(t-t_{\alpha_i}) \exp{\left(-\frac{t-t_{\alpha_i}}{\tau}\right)}, \label{1/fepsilon}
\end{equation}
where $\theta(t)$ is the Heaviside step function \cite{milotti20021f}. In Sec.~\ref{Errsuppress}, it is shown that this type of error can be suppressed by implementing continuous indirect measurements. For most trajectories, the system stays close to the $S_{z,x}=+1$ eigenspace. $\bar{I}_{z,x}(t)$ converges to $1$ while $\langle S_{z,x}\rangle_{\hat{\rho}}$ approaches 1 and stays at 1. Occasionally, the error can cause the system to jump to the $-1$ eigenspace of $S_{z,x}$. A sample trajectory of this case is shown in Fig.~\ref{errsample}, where we detect the system's $\langle S_x\rangle$ jumping to $-1$ by observing that $\bar{I}_x(t)$ starts to decay to 0 and $\langle S_x\rangle_{\hat{\rho}}$ flips to $-1$.
 In general, since any single-qubit Pauli error anticommutes with at least one of the $S_{z,x}$, any single-qubit error can be detected. Multi-qubit errors can be detected when they consist of operators anticommuting with one of the $S_{x,z}$. The undetectable errors are those commuting with both $S_{x,z}$. However, they must be at least weight 2. They happen at lower rates than single-qubit errors.

\begin{figure}
 \includegraphics[width=6cm,height=6cm,keepaspectratio]{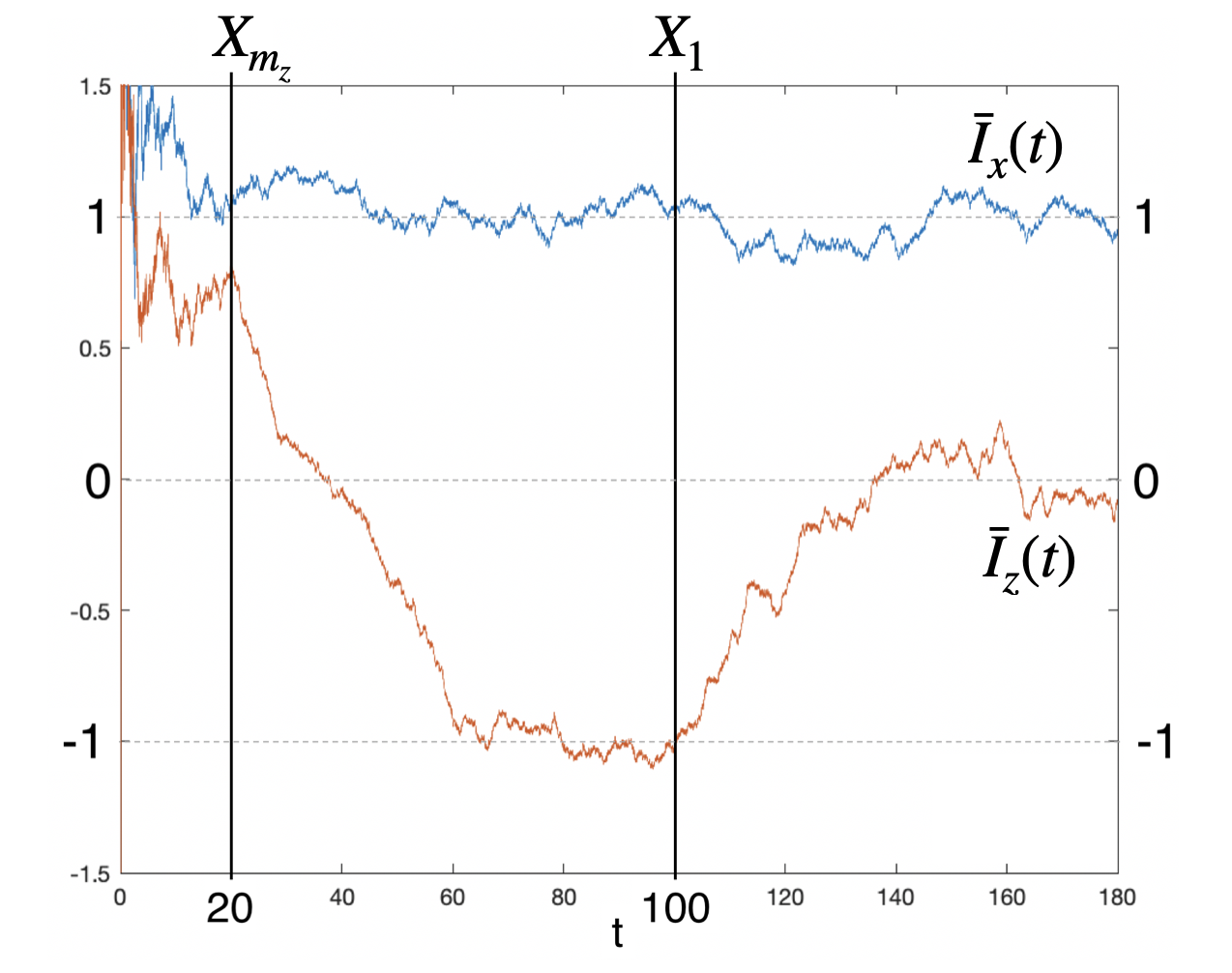}
\caption{A sample of $\bar{I}_{x,z}(t)$ with an $X_{m_z}$ error at $t=20$ and an $X_1$ error at $t=100$. The red curve represents $\bar{I}_z(t)$, and the blue curve represents $\bar{I}_x(t)$. $\bar{I}_x(t)$ remains at 1 because the errors commute with $S_x$. $\bar{I}_z(t)$ flips to $-1$ due to the $X_{m_z}$ error and then converges to $0$ after the $X_1$ error happened. The window width $w$ is set to $40/k$ in this example.}
\label{instantErrXm1X1}
\end{figure}

We now consider the cases when errors happen on the monitor qubits. Note that the essential indicator that allows us to distinguish the four eigenspaces of $S_{x,z}=\pm1$ is whether $\langle Z_{m_{z,x}}\rangle$ is static or oscillatory in motion. When $S_{x,z}$ is $+1$, $\langle Z_{m_{x,z}}\rangle$ is static. When $S_{x,z}$ is $-1$, $\langle Z_{m_{x,z}}\rangle$ is oscillatory. It turns out that the process of stabilizer detection can be preserved with low rate errors on the monitor qubits. Let us begin with the case of instantaneous errors on the monitor qubits. Suppose an $X_{m_z}$ error happened on the monitor $m_z$. If $\langle Z_{m_z}\rangle$ was static (because the system is in the $S_z=+1$ eigenspace), $\langle Z_{m_z}\rangle$ flips from $+1$ to $-1$ but remains static. As shown in Fig.~\ref{instantErrXm1X1}, $\bar{I}_z(t)$ converges to $-1$ after an $X_{m_z}$ error happened at $t=20$, and then a subsequent $X_1$ error happened at $t=100$ is detected by observing $\bar{I}_z(t)$ evolving to $0$. 
If $\langle Z_{m_z}\rangle$ is oscillatory, an $X_{m_z}$ flips the value of $\langle Z_{m_z}\rangle$ but does not change the oscillatory motion. In general, $\bar{I}_{x,z}(t)$ converging to $\pm1$ indicates that there was no error, and either $\bar{I}_{x,z}(t)$ converging to $0$ indicates there was an error. When such instantaneous errors happen with rates much smaller than $~0.1\lambda$, which is approximately the inverse of the measurement time for the indirect detection, the error detection process is preserved. For continuous Hamiltonian errors, since the monitors are being continuously measured, errors with strength much smaller than the measurement rate $\lambda$ are partially suppressed by the quantum Zeno effect. Hence, the process of detecting errors on the encoded four-qubit system can be preserved even with low-rate errors on the monitors. However, if the errors happen at high rates, it can cause rapid flipping that mimics the oscillatory effect of $Z_{m_z}$, which normally would occur only when $S_z=-1$. In that limit indirect detection becomes ineffective. Of course, these conclusions are for the particular error model we have been considering.  Experimental measurements of the error process might suggest alternative versions of this scheme, e.g., measuring $X_{m_z}$ instead of $Z_{m_z}$ if there are mainly $X_{m_z}$ errors. 

In the full construction described in Sec.~\ref{construction}, the qubits $m_z$ and $m_x$ are two pairs of physical qubits $(a,b)$ and $(c,d)$. Each pair $(a,b)$ and $(c,d)$ is confined to the ground space of the strong base Hamiltonian, $(K/2)(I-Z_aZ_b)+(K/2)(I-Z_cZ_d)$. $X_aX_b$ and $X_cX_d$ in the effective Hamiltonian cause transitions only within the ground space, i.e., $|00\rangle \leftrightarrow|11\rangle$. Hence they act as $X_{m_{z,x}}$ for the effective two qubits $m_z$ and $m_x$. The process of detecting errors for the encoded system is similar: when $S_z=+1$, both $a$ and $b$ are static; when $S_z=-1$, both $a$ and $b$ are oscillatory. The same applies to $c$ and $d$ for $S_x$. We only need to continuously measure one qubit for each pair, e.g., measuring $a$ and $c$. If errors can happen on the monitor qubits, it follows similarly from the above argument that the error detection process for the system qubit can be preserved under low-rate errors.

\subsubsection{\label{Errsuppress}Error suppression}

\begin{figure*}
 \begin{subfigure}{0.32\linewidth}
 \includegraphics[width=\linewidth]{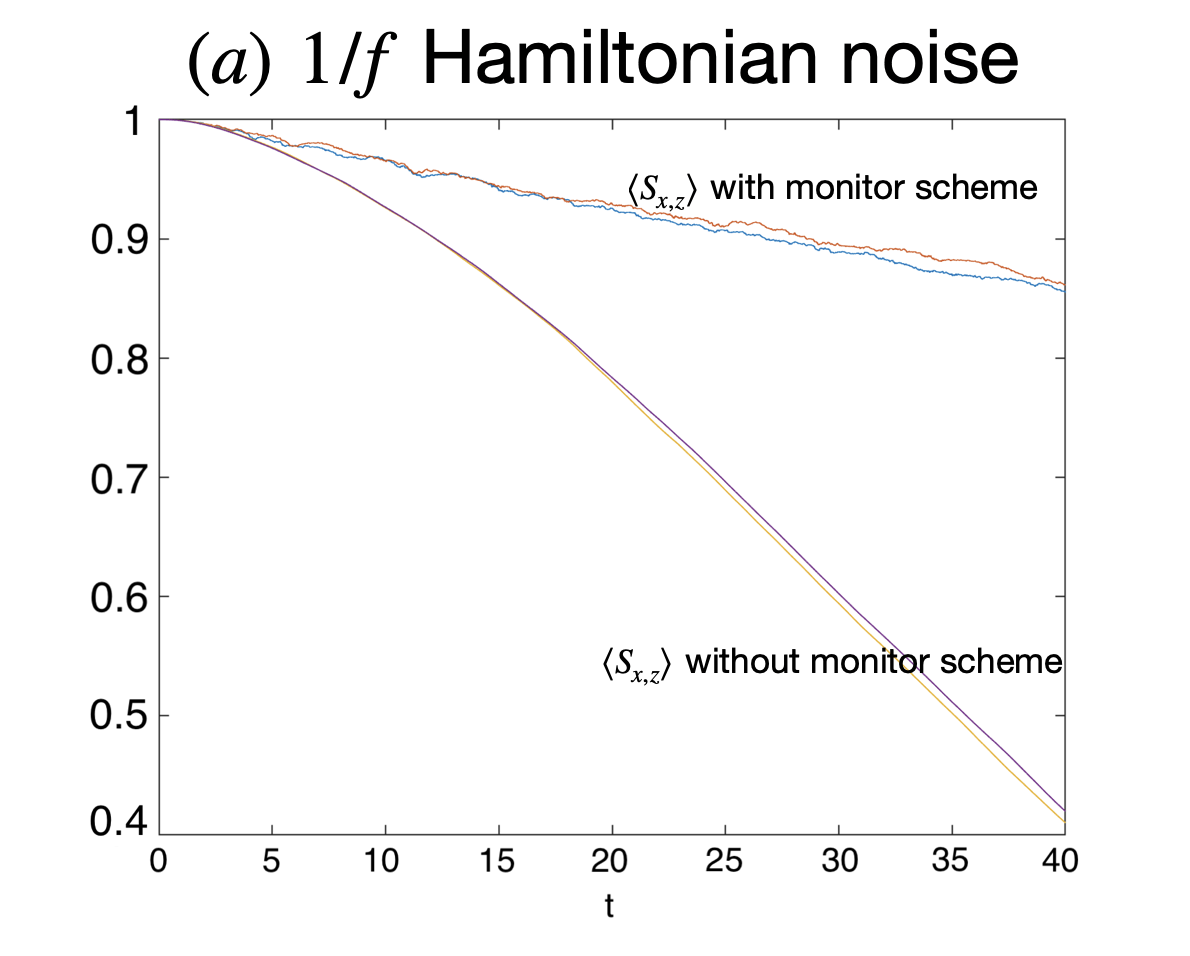}
  \label{1/f} 
  \end{subfigure}
  ~
   \begin{subfigure}{0.32\linewidth}
 \includegraphics[width=\linewidth]{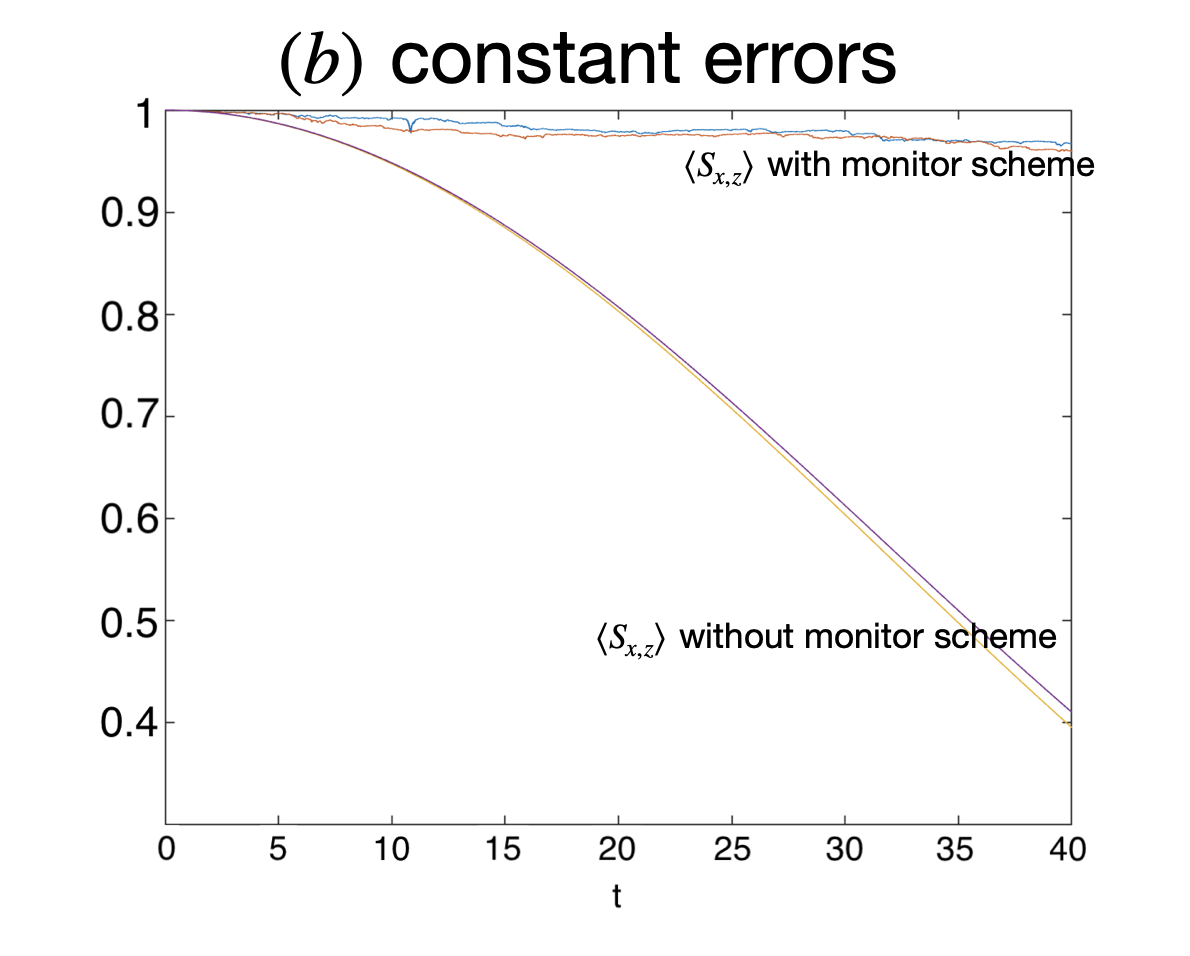} \label{consErr} 
  \end{subfigure}
  ~
   \begin{subfigure}{0.32\linewidth}
 \includegraphics[width=\linewidth]{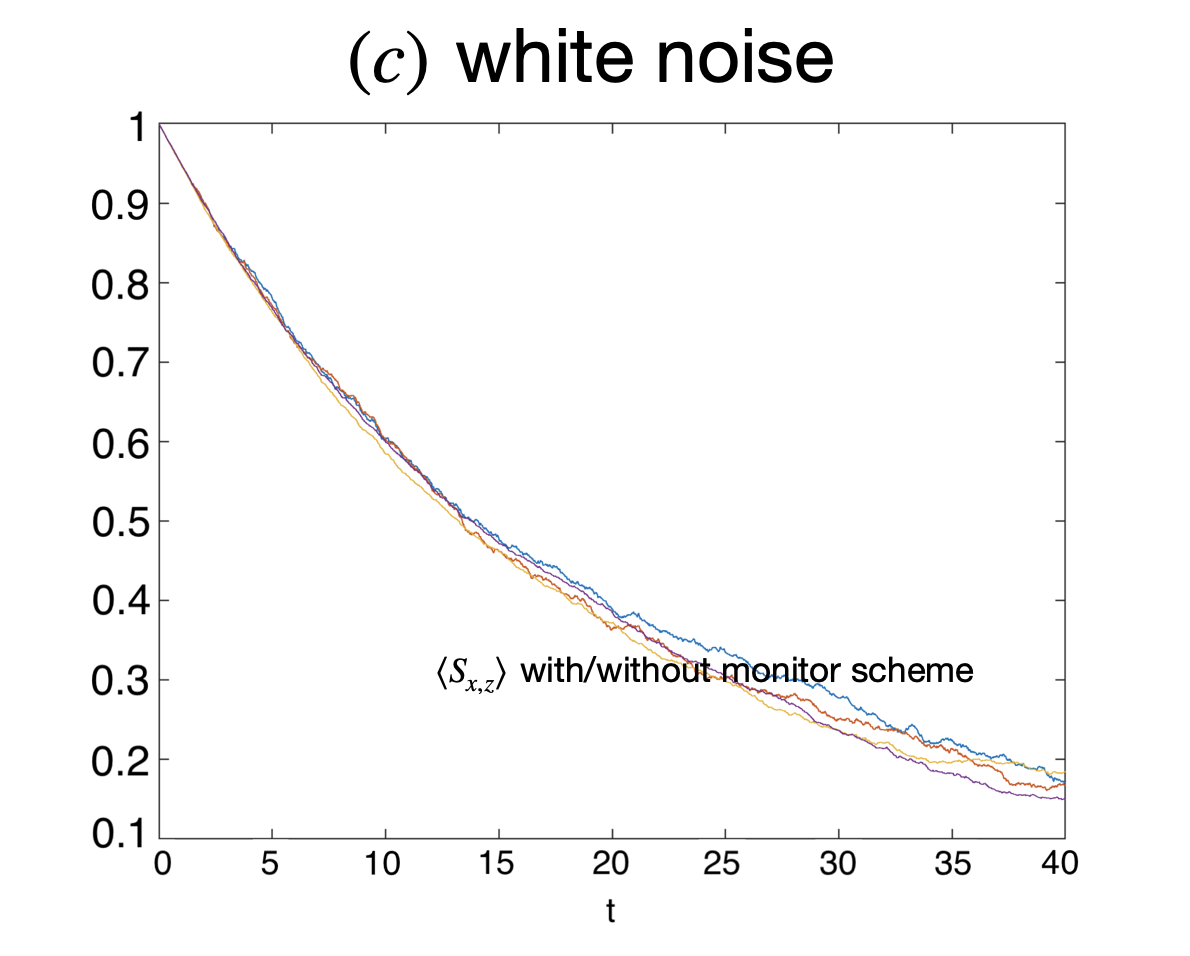}
  \label{whitenoise} 
  \end{subfigure}
  
  \caption{The evolutions of the stabilizers under various error models. (a) compares evolutions of $\langle S_{z,x}\rangle$ with/without indirect detection, under 1/f Hamiltonian noise. (b) compares evolutions of $\langle S_{z,x}\rangle$ with/without indirect detection, under constant Hamiltonian errors. (c) is the case for white noise, where there shows no suppression of error. They are ensemble averages over 1000 trajectories.}
  \label{errorplot}
\end{figure*}

It is well-known that frequent measurements can freeze a system in an eigenspace of the measurement observable due to the quantum Zeno effect. There have been many efforts to harness the Zeno effect for error suppression \cite{Paz-Silva2012,Wuster2017,Kondo_2016}. In \cite{Oreshkov2007}, it is shown that non-Markovian errors can be suppressed by the quantum Zeno effect while Markovian errors can not.  In this subsection, we investigate error suppression for various models under continuous indirect measurements. We first consider the 1/f Hamiltonian errors defined in Eq.~(\ref{1/fnoise}), where the sum is over all physical qubits. In Fig.~\ref{errorplot}, we plot the ensemble average of the system's stabilizer values under this 1/f Hamiltonian error. As shown in Fig.~\ref{errorplot}(a), the red and blue curves represent the case with indirect detection while the purple and yellow curves represent the case without the measurement setup. The pulse rate and $\epsilon$ are $0.1k$ and $\tau\sim 1/k$, where $k$ is the strength of the Hamiltonian. The measurement rate $\lambda$ is set to $0.6k$. The red and blue curves decay noticeably more slowly than the purple and yellow, which shows that the system state tends to remain in the $S_{z,x}=+1$ eigenspace in the presence of indirect stabilizer detection. Note that 1/f Hamiltonian noise is a type of non-Markovian error process. The exhibited suppression aligns with the result in \cite{Oreshkov2007} that non-Markovian errors can be suppressed by the quantum Zeno effect. We present another example of non-Markovian errors, where the errors are constant Hamiltonian terms, i.e., $H_{err}=\epsilon\sum \sigma_i$. To keep the same error magnitude as in the 1/f noise case, the $\epsilon$ is set to $0.01k$, which is the average strength of $\epsilon_i(t)$ in the 1/f noise. The result is shown in Fig.~\ref{errorplot}(b). Convergence to the joint $+1$ eigenspace of $S_{z,x}$ is apparent when indirect detection is applied.  

Another method to benchmark the state protection is to evaluate the trace distance between the state at time $t$ and the initial state \cite{Paz-Silva2012}. 
\begin{figure}
 \begin{subfigure}{0.45\linewidth}
 \includegraphics[width=\linewidth]{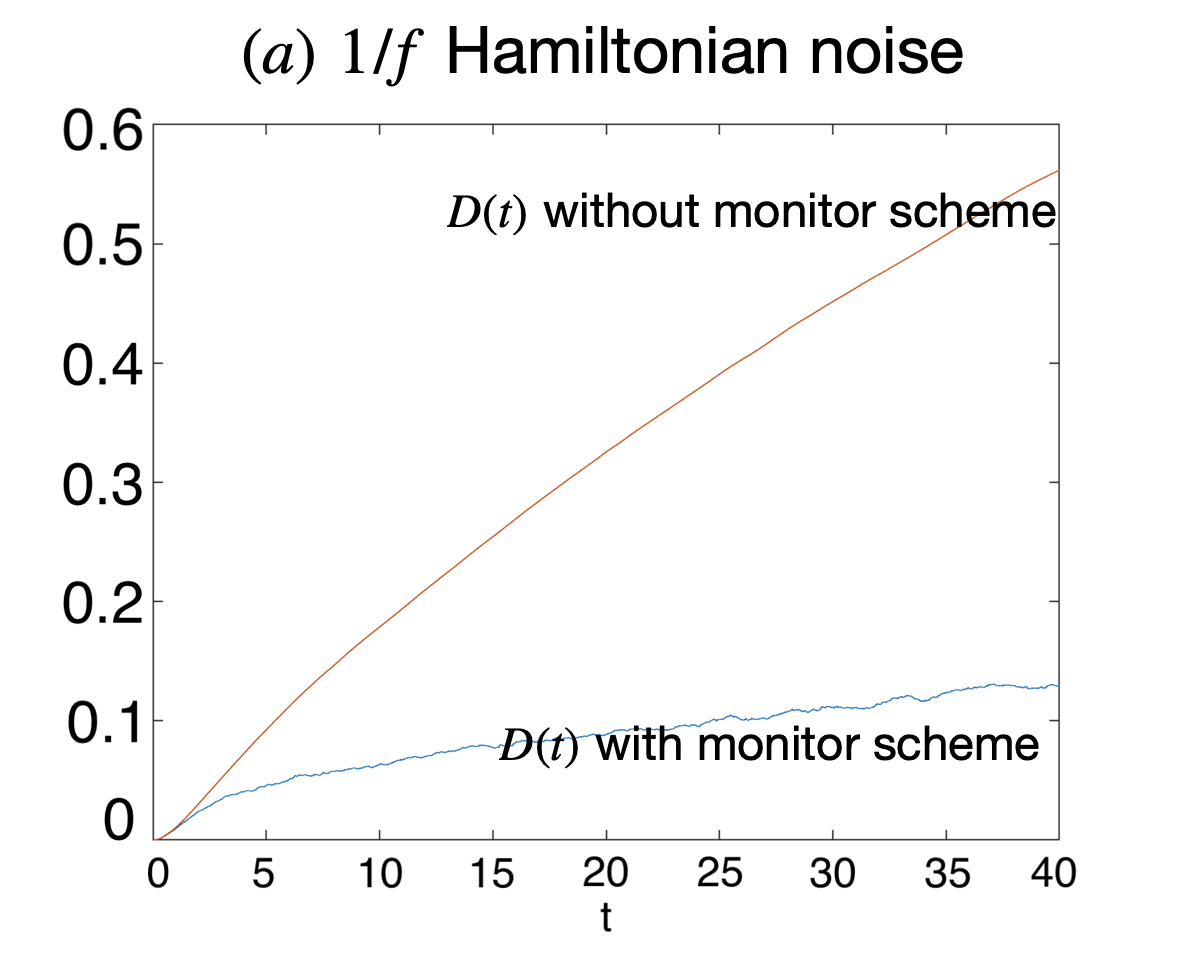}
  \label{Trd1/f} 
  \end{subfigure}
  ~
   \begin{subfigure}{0.45\linewidth}
 \includegraphics[width=\linewidth]{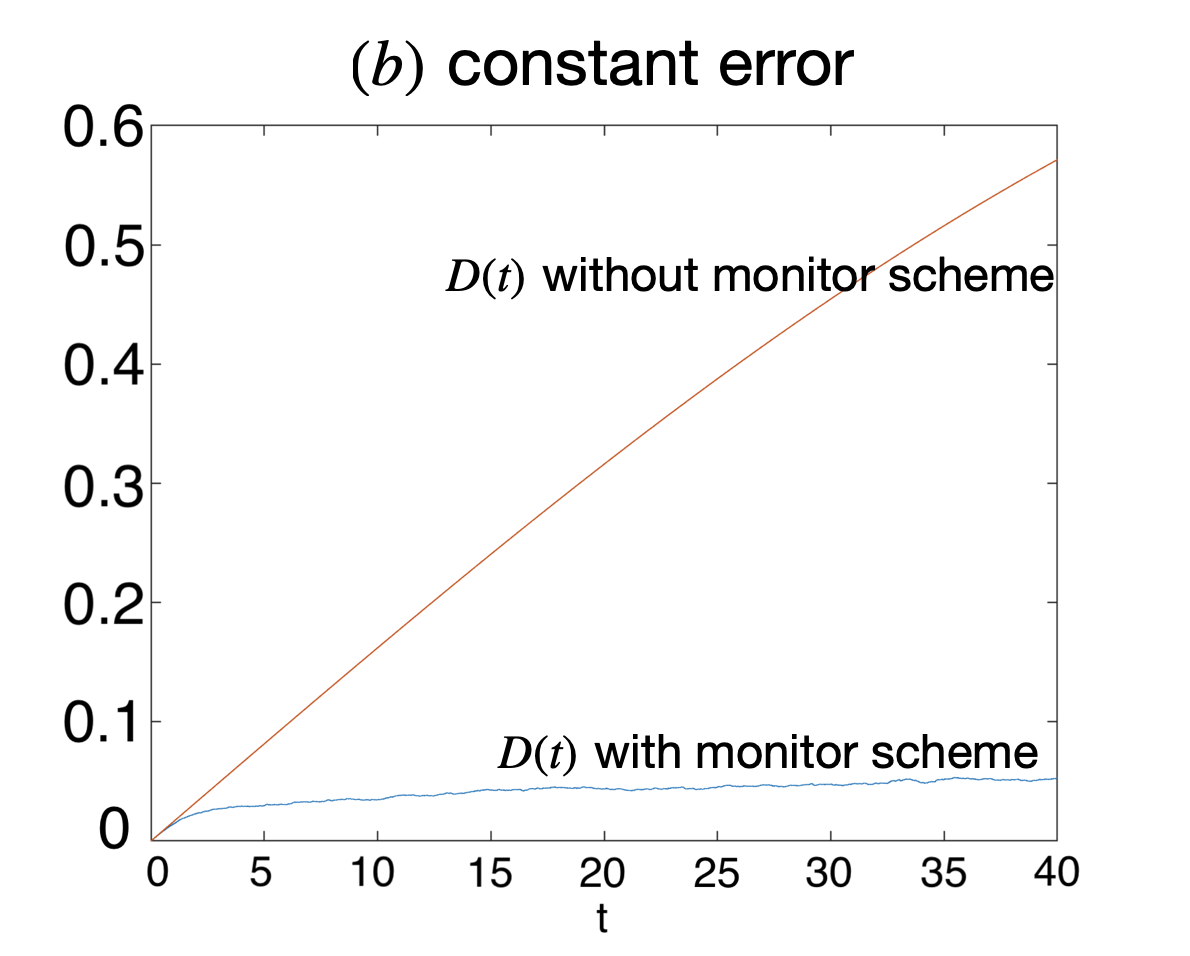}
  \label{TrdconsErr} 
  \end{subfigure}
\caption{ $\mathcal{D}(t)=\frac{1}{2}||\rho(t)-\rho(0) ||_1$. The red curve is the case without measurements while the blue curve is the case with continuous indirect detection. They are ensemble averages over 1000 trajectories.}
\label{tracedistance}
\end{figure}
The smaller the trace distance the closer the state remains to its initial state. Fig.~\ref{tracedistance} shows a clear protection of the state when the system is being measured. 

However, when the errors are Markovian (white noise) the measurements do not appear to fix the stabilizer values, as shown in Fig.~\ref{errorplot}(c). For Markovian noise, the probability of a state transition is of order $dt$ for a time step. Errors of this type cannot be suppressed by frequent measurements and full error correction is required to protect the states. For non-Markovian noise, by contrast, the probability of a state transition is of order $dt^2$ in a time step. This is why these transitions can be suppressed by the quantum Zeno effect \cite{Oreshkov2007}. The above simulations for continuous indirect measurements agree with these results.

\begin{figure}
 \includegraphics[width=6cm,height=6cm,keepaspectratio]{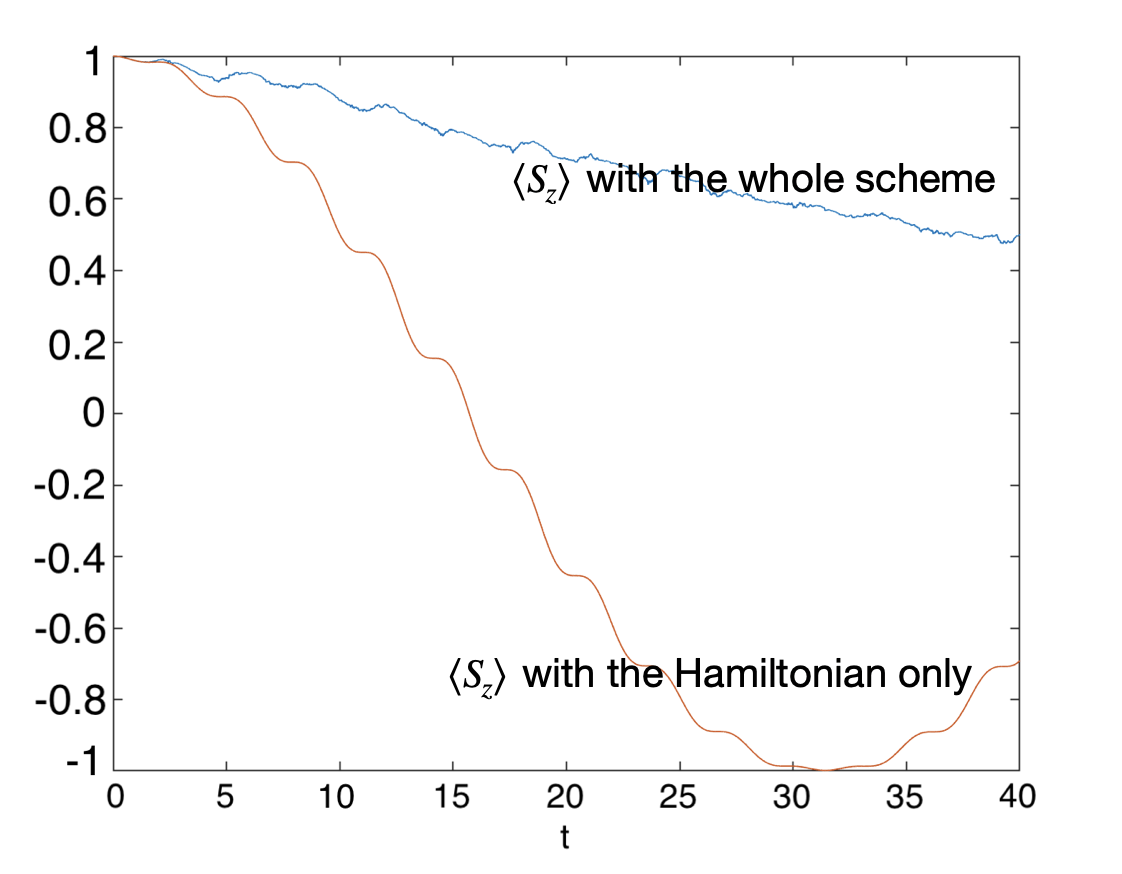}
 \caption{The evolution of $\langle S_z\rangle$ under an on-resonance $X_1$ error. The blue curve is the case when the full indirect detection scheme is applied. The red curve is the case when the Hamiltonian for indirect detection is applied but no measurements are made on the monitor qubits.}
\label{onresonant}
\end{figure}

It is worth noting that for the purpose of error prevention, it is possible in principle to suppress Hamiltonian errors by applying a strong Hamiltonian alone. For example, suppose we have a qubit prepared in the $|0\rangle$ state in the $Z$ basis with the presence of an $X$ Hamiltonian error. The error can cause the state to rotate on the y-z plane in the Bloch sphere. However, if we apply a $Z$ Hamiltonian, which is strong comparing to the error term $X$, the rotating axis becomes closely aligned with the z-axis. The evolution for the state will be confined in a small region near the north pole. The region can be made smaller as we increase the strength for the $Z$ term. Therefore, the state is maintained close to its initial state $|0\rangle$. Recall the setup in the indirect measurements. We require interaction Hamiltonians between the system and the monitor qubits. These Hamiltonians also contribute to the suppression of errors because of the above axis-pinning behavior. However, if an error term has a time-dependent coefficient with a frequency component on resonance with energy differences in the system Hamiltonian, transitions are not suppressed. In this case, applying the Hamiltonian alone is not effective against the error. However, applying both the Hamiltonian and the continuous measurements on the monitor qubits performs better in these cases as shown in Fig.~\ref{onresonant}. Overall, continuous indirect measurements can protect encoded states against errors.

\section{\label{construction}Constructing the Hamiltonian for indirect detection}

In this section, we show how to build an effective Hamiltonian for indirect detection. The method is based on the idea of perturbation gadgets \cite{Jordan2008}. It uses 2-local Hamiltonians to produce an effective $k$-local Hamiltonian that appears in the first non-vanishing order for the low-lying energy eigenstates. We begin by briefly recapping the theory presented in \cite{Jordan2008}. 

Suppose we have a strong base Hamiltonian $H^{(0)}$ and a weak potential $\epsilon V$. $H^{(0)}$ has zero ground state energy with a degenerate ground space $\mathcal{G}^{(0)}$ spanned by eigenvectors $|e^{0}_1\rangle\dots|e^{0}_d\rangle$, and $\epsilon V$ weakly perturbs it. The total Hamiltonian $H=H^{(0)}+\epsilon V$ will have a $d$-dimensional vector space $\mathcal{G}$ spanned by the $d$ lowest energy eigenstates $|e_1\rangle\dots |e_d\rangle$. For small enough $\epsilon$, $\mathcal{G}$ largely overlaps with $\mathcal{G}^{0}$. The space spanned by the lowest $d$ energy eigenstates has an effective Hamiltonian 
\begin{equation}
H_{\text{eff}}\equiv\sum_{i=1}^{d}E_i |e_i\rangle\langle e_i|,
\end{equation}
which can be expanded in powers of $\epsilon$, i.e.,
\begin{equation}
H_{\text{eff}}= \mathcal{U} \left(\sum^{\infty}_{m=1}  \epsilon^{m}\sum_{(m-1)} P_0 VS^{l_1}V\cdots S^{l_{m-1}}VP_0\right)\mathcal{U}^{\dagger}. \label{Heff}
\end{equation}
The operator $P_0$ projects any vector onto the unperturbed ground space $\mathcal{G}^{(0)}$, and the linear operator $\mathcal{U}$ satisfies 
\begin{equation}
\mathcal{U}P_0 |e_i\rangle=|e_i\rangle \ \ \text{and}\ \ \mathcal{U}\mathcal{G}^{(0)\perp}=0.
\end{equation}
The operator $S^l$ is 
\begin{equation}
S^l=
\begin{cases}
   \displaystyle\sum_{i>0}\frac{P_i}{(-E^{(0)}_i)^l}    & \quad \text{if } \ l>0,\\
     -P_0 & \quad \text{if } \ l=0,
  \end{cases}
\end{equation}
where $P_i$ is the projector corresponding to the energy level $E^{(0)}_i$ of the base Hamiltonian $H^{(0)}$. The summation $\sum_{(m-1)}$ is over nonnegative integers $l_1,l_2,\dots ,l_{m-1}$ such that $l_1+\cdots +l_{m-1}=m-1$ and $l_1+ \cdots +l_x\geq x $ for any $x$ from 1 to $m-1$. $\mathcal{U}$ and $\mathcal{U}^{\dagger}$ can also be expanded in powers of $\epsilon$ but only their zeroth order terms, which are both $P_0$, will contribute in the later discussion. A more detailed derivation of these results can be found in \cite{Jordan2008}. Note that the expansion converges only if $||\epsilon V||<\Delta E^{(0)}/4$, where $\Delta E^{(0)}$ is the energy gap between the ground energy (assumed zero) and the second lowest energy. To have a good approximation from the perturbation, we would expect $||\epsilon V||$ to be much smaller than $\Delta E^{(0)}$. In this limit, the effect of adding $\epsilon V$ to $H^{(0)}$ becomes a small splitting of the degenerate ground space with a small deviation from the ground space $\mathcal{G}^{(0)}$ to $\mathcal{G}$. When an initial state is prepared in $\mathcal{G}^{(0)}$, its evolution stays mainly in $\mathcal{G}$ and the effective Hamiltonian $H_{\text{eff}}$ will be a good approximation for $H$. 

The following construction for the indirect measurement requires us to design 2-local Hamiltonian terms $H^{(0)}$ and $V$ such that the first non-vanishing order of the expansion gives the desired Hamiltonian.

\subsection{First example: $ZZ$ detection}
Suppose we want to measure $Z_1Z_2$ for qubits 1 and 2, and the desired Hamiltonian is 
\begin{equation}
H_{\text{target}}=\frac{k}{2}(I-Z_1Z_2)X_m. \nonumber
\end{equation}
We bring in two ancillary qubits  $m_1$ and $m_2$, and turn on a Hamiltonian $H=H^{(0)}+\epsilon V$, where
\begin{equation}
H^{(0)}=\frac{K}{2}(I-Z_{m_1}Z_{m_2})
\end{equation}
and the perturbing term is
\begin{equation}
V= K\left( Z_1X_{m_1}+ Z_2 X_{m_2} + 2\epsilon X_{m_1}X_{m_2}\right).
\end{equation}
$K$ is a constant and $\epsilon\ll 1$. (Note that the identity term in the base Hamiltonian is unnecessary but we keep it for convenience.) The expansion of $H_{\text{eff}}$ in Eq.~(\ref{Heff}) up to second order in $\epsilon$ gives
\begin{align}
&H_{\text{eff}}=\mathcal{U}\left[ \epsilon P_0 V P_0 +\epsilon^2 P_0 V S^1 V P_0 +\mathcal{O}(\epsilon^3)\right]  \mathcal{U}^{\dagger} \nonumber \\
&=K \epsilon^2  P_0\left[2\left(I-Z_1Z_2\right)X_{m_1}X_{m_2}-2\right] P_0+\mathcal{O}(\epsilon^3).
\end{align}
The ancillary qubits are prepared in the ground space, $\mathcal{G}^{(0)}=\mathcal{H}_{12}\otimes\text{span}\{|00\rangle, |11\rangle \}_{m_1m_2}$, and the effective Hamiltonian for the 4-qubit system can be approximated by 
\begin{equation}
\tilde{H}_{\text{eff}}=2K\epsilon^2\left(I-Z_1Z_2\right)X_{m_1}X_{m_2},
\end{equation}
in the limit of $\epsilon\ll 1$. The shifted term proportional to $P_0$ is neglected since it acts as the identity in the subspace. Note that since the ancillary qubits are restricted to $\mathcal{G}^{(0)}$, which is a 2-dimensional subspace, we can treat $m_1$ and $m_2$ as an effective qubit $m$ and the operator $X_{m_1}X_{m_2}$ behaves as $X_m$ that flips $m$. Hence, it can be simplified as a 3-body system with Hamiltonian
\begin{equation}
\tilde{H}_{\text{eff}}=2K\epsilon^2\left(I-Z_1Z_2\right)X_m,
\end{equation}
 which is in the desired form for the indirect measurement (with $k=4K\epsilon^2$). Since $m_1$ and $m_2$ are confined to the ground space $\mathcal{G}^{(0)}$ and are simultaneously rotated by $X_{m_1}X_{m_2}$, we can detect the value of $Z_1Z_2$ by continuously measure only one of $Z_{m_1}$ or $Z_{m_2}$. When the state is in the eigenspace of $Z_1Z_2=+1$, both $m_1$ and $m_2$ are static. When the state is in the $Z_1Z_2=-1$ eigenspace, $\langle Z_{m_1}\rangle$ and $\langle Z_{m_2}\rangle$ are oscillatory. The system's $Z_1Z_2$ value can be obtained by calculating the estimator or evaluating the time average of the signal as described above.

\subsection{Construction for the four-qubit Bacon-Shor code}
To indirectly measure the stabilizers, $S_z=Z_1Z_2Z_3Z_4$ and $S_x=X_1X_2X_3X_4$, for the 4-qubit Bacon-Shor code, we apply the Hamiltonian,
\begin{align}\label{target}
H=\frac{k}{2}(Z_1Z_2-Z_3Z_4)X_{m_z}+\frac{k}{2}(X_1X_3-X_2X_4)X_{m_x},
\end{align}
and continuously measure $Z_{m_z}$ and $Z_{m_x}$. However, to obtain this Hamiltonian using only 2-local operators requires four ancillary qubits, which we call $a, b, c$ and $d$. The full physical system becomes an 8-qubit state, where $1,2,3,4$ are the system qubits and $a,b,c,d$ are the monitor qubits for the indirect measurements. The full perturbative construction has a Hamiltonian $H=H^{(0)}+\epsilon V$, where the base Hamiltonian is 
\begin{equation}\label{base}
H^{(0)}=\frac{K}{2}(I-Z_{a}Z_b)+\frac{K}{2}(I-Z_{c}Z_d),
\end{equation}
and the perturbing term is 
\begin{align}\label{perturb}
 V&=\frac{K}{2\sqrt{2}}(Z_3+Z_4-Z_1-Z_2)X_a \nonumber\\
 &+\frac{K}{2\sqrt{2}}(Z_3+Z_4+Z_1+Z_2)X_b \nonumber\\
&+\frac{K}{2\sqrt{2}}(X_2+X_4-X_1-X_3)X_c  \nonumber\\
&+\frac{K}{2\sqrt{2}}(X_2+X_4+X_1+X_3)X_d  \nonumber\\
&+\frac{K\epsilon}{2}(Z_1Z_2+Z_3Z_4+X_1X_3+X_2X_4).
\end{align}
The monitor qubits are prepared in the ground space of $H^{(0)}$, which consists of two two-level subspaces for the monitors. The unperturbed ground space is $\mathcal{G}^{(0)}=\mathcal{H}_{sys}\otimes \text{span}\{|00\rangle,|11\rangle\}_{ab}\otimes \text{span}\{|00\rangle,|11\rangle\}_{cd}$. After adding $\epsilon V$, the perturbed ground space has an effective Hamiltonian that reads
\begin{align}
&H_{\text{eff}} = \epsilon P_0 V P_0 +\epsilon^2 P_0 V S^1 V P_0 +\mathcal{O}(\epsilon^3) \nonumber \\
&=  K\epsilon^2\frac{1}{2}(Z_1Z_2-Z_3Z_4)X_aX_b P_0 \\
&+K\epsilon^2\frac{1}{2} (X_1X_3-X_2X_4)X_cX_d P_0 - K\epsilon^2 2 P_0 +\mathcal{O}(\epsilon^3). \nonumber
\end{align}
Since the ancillary qubits are prepared in the ground space of $H^{(0)}$, the full system effectively has the Hamiltonian
\begin{align}
\tilde{H}_{\text{eff}}=&\frac{K\epsilon^2}{2}[\left(Z_1Z_2-Z_3Z_4\right)X_aX_b \nonumber\\
& \ \ \ \ \ \ +\left(X_1X_3-X_2X_4\right)X_cX_d ]. 
\end{align}
The $X_aX_b$ and $X_cX_d$ only cause transitions within the ground space $\mathcal{G}^{(0)}$, and they act as a single-qubit $X$ for an effective qubit confined in the space spanned by $\{|00\rangle,|11\rangle\}$. We obtain the target Hamiltonian (\ref{target}) by identifying $ X_{a}X_b\to X_{m_z}$ and $X_c X_d\to X_{m_x}$. The monitors are initially prepared in $|0000\rangle_{abcd}$.  $X_aX_b$ $(X_cX_d)$ simultaneously rotates $Z_a$ and $Z_b$ ($Z_c$ and $Z_d$) when the state is in the $S_z=-1$ $(S_x=-1)$ eigenspace. To measure the values of $S_z$ and $S_x$, we continuously measure $Z_a$ (or $Z_b$) and $Z_c$ (or $Z_d$). The information of the system being in either eigenspace of $S_{z,x}=\pm1$ can be obtained by evaluating $\bar{I}_a(t)$ and $\bar{I}_c(t)$ or by calculating $\langle S_{z,x}\rangle_{\hat{\rho}}$ using an estimator $\hat{\rho}$ as described in Sec.~\ref{setup}. When the system is in the $S_z=+1$ eigenspace, $\langle Z_a\rangle$ is static. $\bar{I}_a(t)$ converges to $+1$ and $\langle S_z\rangle_{\hat{\rho}}\to +1$. When the system is in the $S_z=-1$ eigenspace, $\langle Z_a\rangle$ is oscillatory. $\bar{I}_a(t)$ approaches 0 and $\langle S_z\rangle_{\hat{\rho}}\to -1$. The same detection rule applies to $S_x$.

\begin{figure}
  \includegraphics[width=8.5cm,height=8.5cm,keepaspectratio]{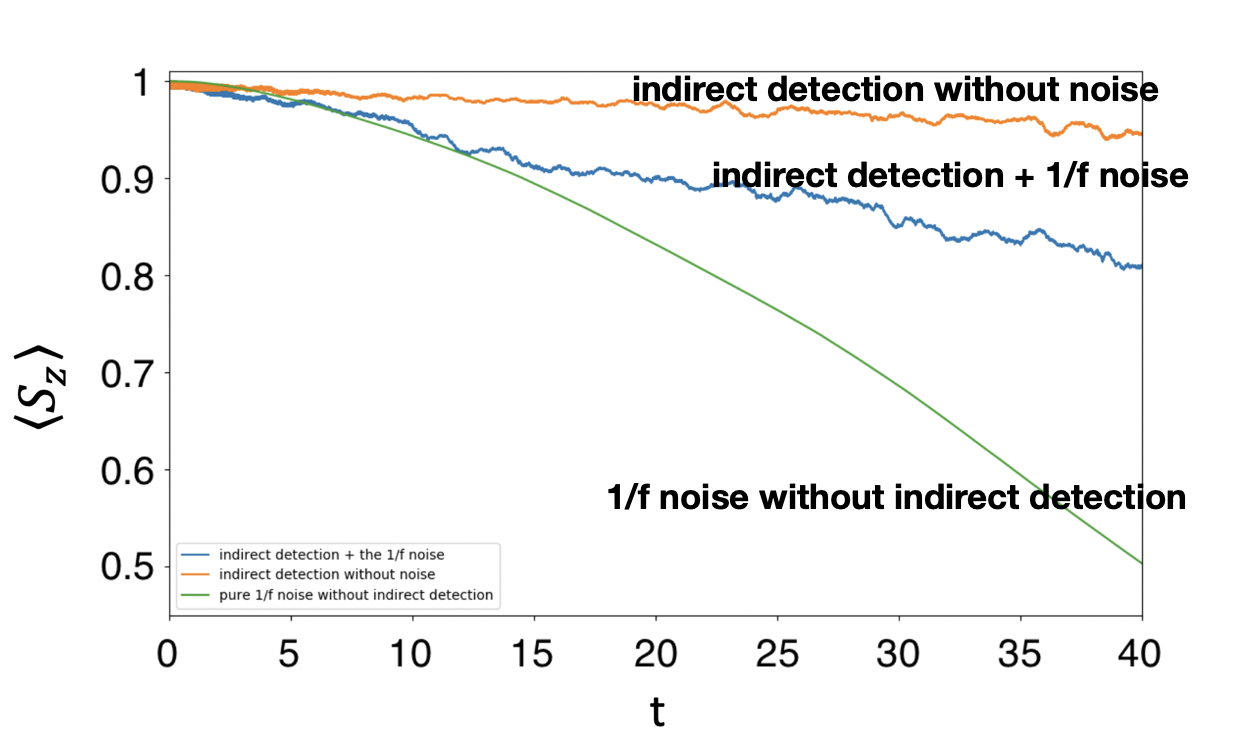}
  \caption{\label{2meas}The evolution of $\langle S_z\rangle$ under the full 8-qubit construction. The orange curve includes the indirect detection alone without any errors. The blue curve also includes 1/f Hamiltonian noise. The green curve is the 1/f Hamiltonian noise alone without any indirect measurement. They are ensemble averages over 500 trajectories.}
\end{figure}

 It is worth recalling that the constant $K\epsilon^2$ in the effective Hamiltonian is the strength $k$ of the target Hamiltonian in Eq.~(\ref{target}). The fact that $\epsilon$ needs to be small for the perturbation to work accurately implies that $K$, the strength of the base Hamiltonian, has to be large enough that $K\epsilon^2$ is large compared to the error strength (or rate). We numerically simulated an example with $\epsilon^2\sim0.001$ to demonstrate the performance of the full perturbative construction. The result for $\langle S_z\rangle$ is shown in Fig.~\ref{2meas}. ($\langle S_x\rangle$ behaves similarly.) The ensemble averages of trajectories for $\langle S_z\rangle$ are plotted for various cases.  The orange curve represents the no-error case when we apply the full construction using only 2-local Hamiltonians from Eqs.~(\ref{base}) and (\ref{perturb}) and continuous measurements of $Z_a$ and $Z_c$. When there is no error, $\langle S_z\rangle$ is expected to remain 1 throughout the detection process. This is true for the 6-qubit setup introduced in Sec.~\ref{6-qubit}. However, building the Hamiltonian perturbatively causes the stabilizers to drop slightly below 1, indicating the presence of small errors due to higher-order corrections. Nonetheless, the deviation is small as shown in Fig.~\ref{2meas}.  The blue and the green curves are the cases when the system suffers from the 1/f Hamiltonian errors defined in Eq.~(\ref{1/fnoise}), where the sum is over all physical qubits (including the monitor qubits). The blue includes continuous indirect detection while the green does not. The suppression of errors is apparent, although it is slightly less effective than the ideal 6-qubit case shown in Fig.~\ref{errorplot}(a). For most trajectories where errors are suppressed, the stabilizer values stay close to 1. For some trajectories where errors cause $\langle S_{z,x}\rangle$ to flip to $-1$, we can detect them by observing $\bar{I}_{a,c}(t)$ decaying towards 0 or $\langle S_{z,x}\rangle_{\hat{\rho}}$ flipping to $-1$. These behaviors are essentially the same as in Fig.~\ref{errsample}.

\section{Conclusion}
We have presented and analyzed a method for the continuous measurement of high-weight operators, and applied this to the problem of continuous quantum error detection by the four-qubit Bacon-Shor code. This method includes engineering an interaction Hamiltonian between the system and the continuously measured ancillary qubits. More nontrivially, the Hamiltonian can be effectively built using physically viable two-local interactions, and the measurements on the monitor qubits consist of well-studied single-qubit continuous measurements. 

One major advantage of using this type of continuous monitoring is that it can exhibit error suppression for non-Markovian noise. The traditional circuit-based model is a discrete-time scheme, which cannot generally be carried out quickly enough to produce error suppression. This continuous monitoring scheme does not replace fault-tolerance methods, but it can be incorporated into a larger code (or a larger quantum algorithm) by concatenation. We can implement continuous monitoring for the lowest level qubits and pass the error information to higher levels for error tracing and correction. More specifically, we could encode a qubit at the bottom layer of a large code as the logical qubit of an error-correcting code (or error-detecting code), where the stabilizer generators are continuously monitored by the scheme we introduce here.

In general, this detection scheme can be applied to measuring the stabilizers in any quantum code. However, as the weight of the stabilizers in a code increases, the difficulty of performing this detection scheme is also increased. This is because perturbatively constructing the Hamiltonian for the indirect detection requires applying a strong base Hamiltonian. The strength of this base Hamiltonian grows as the weight of the target term increases because these terms would appear at higher orders in the expansion. This is one of the reasons that we apply it to the four-qubit Bacon-Shor code, where the stabilizers are weight-four and their values can be obtained by measuring the two-local gauge operators. In this case, the target Hamiltonian can appear in the second order expansion, which is the minimum.  The question of how the construction scheme applies to other quantum codes remains open, but it should certainly apply to the 9-qubit and larger Bacon-Shor codes. 

Two methods are provided for retrieving the measurement outcomes. The estimator approach is computationally hard and difficult to carry out in real time but may be beneficial to theoretical analysis. By contrast, the signal time average is noisier, but more efficient to perform in real time. It is shown that errors with low rates can be detected and (in the non-Markovian case) suppressed. This is in the regime where the indirect detection is effective. For high-rate or high-strength errors that change the system too rapidly, the detection scheme becomes inapplicable. However, if the type of errors can be learned from the experiments, it may be possible to adjust the setup for better performance.

Overall, we have presented a new method for measuring high-weight operators using practical experimental resources. This is a step towards practical quantum error-correction for quantum computing.

\begin{acknowledgements}
TAB and YHC are grateful for useful conversations with Namit Anand, Justin Dressel, Daniel Lidar and Chris Sutherland. YHC acknowledges some of the simulations use the Armadillo C++ library \cite{Sanderson2016,Sanderson2018} and are performed on the High-Performance Computing Center at USC. This research was supported in part by NSF Grants QIS-1719778 and FET-1911089.

\end{acknowledgements}

\appendix
\section{Ito rule expansion}

Recall from Eqs.~(\ref{signal}), (\ref{stateevolution}) and (\ref{Aoperator}) in the paper, we have  
\begin{align}
&\mathcal{A}(dI)=e^{-iH dt -\lambda \left(\frac{dI}{dt}-Z_m\right)^2dt} \nonumber \\
&=e^{-\lambda (\frac{dI}{dt})^2 dt-\lambda dt} e^{-i H dt + 2\lambda dI Z_m} \nonumber \\
&=e^{-\lambda (\frac{dI}{dt})^2 dt-\lambda dt} \Big[ I- i H dt + 2\lambda \langle Z_m \rangle Z_m dt \nonumber \\
& \ \ \ \ \ \ \ \ \ \ \ \ \ \ \ \ \ \ + \sqrt{\lambda} Z_m dW+ \frac{1}{2}\lambda dt + \mathcal{O}(dtdW)\Big],
\end{align}
where $e^{-\lambda (\frac{dI}{dt})^2 dt-\lambda dt}$ is a constant and will be cancelled out after we normalize the state. We use Ito's rule, i.e., $dW^2=dt$, and keep terms up to $\mathcal{O}(dt)$ and drop terms of $\mathcal{O}(dtdW)$ and higher. We then have
\begin{align}
&\mathcal{A}(dI)\rho\mathcal{A}^{\dagger}(dI)  \\
&= e^{-2\lambda \left[(\frac{dI}{dt})^2+1\right] dt}\Big\{ \rho -i [H, \rho]dt+\lambda \rho dt+ \lambda Z_m \rho Z_m dt \nonumber \\
&+ 2\lambda \langle Z_m\rangle \left( Z_m \rho +\rho Z_m\right)dt  + \sqrt{\lambda} \left( Z_m \rho +\rho Z_m\right)dW \Big\} \nonumber
\end{align}
and
\begin{align}
&\Tr\left[ \mathcal{A}(dI)\rho\mathcal{A}^{\dagger}(dI)\right]= e^{-2\lambda \left[(\frac{dI}{dt})^2+1\right] dt} \Big[ 1+ 4\lambda \langle Z_m\rangle^2 dt  \nonumber \\
&\ \ \ \ \ \ \ \ \ \ \ \ \ \ \ \ \ \ \ \ \ \ \ \ \ + 2\lambda dt+ 2\sqrt{\lambda}\langle Z_m\rangle dW +\mathcal{O}(dtdW)\Big] \nonumber\\
&\implies  \frac{1}{\Tr\left[ \mathcal{A}(dI)\rho\mathcal{A}^{\dagger}(dI)\right]}=e^{2\lambda \left[(\frac{dI}{dt})^2+1\right] dt} \Big[1- 2\lambda dt   \nonumber\\
& -\cancel{4\lambda \langle Z_m\rangle^2 dt}- 2\sqrt{\lambda}\langle Z_m\rangle dW + \cancel{4\lambda \langle Z_m\rangle^2 dt} +\mathcal{O}(dtdW)\Big]. 
\end{align}
Combining together these terms, we get
\begin{align}
d\rho&=\frac{\mathcal{A}(dI)\rho\mathcal{A}^{\dagger}(dI)}{\Tr\left[ \mathcal{A}(dI)\rho\mathcal{A}^{\dagger}(dI)\right]} -\rho\nonumber \\
&=-i [H,\rho]dt +\lambda \left( Z_m \rho Z_m -\rho\right)dt  \\
& \ \ \ + \sqrt{\lambda}\left( Z_m \rho +\rho Z_m -2 \langle Z_m\rangle \rho\right)dW +\mathcal{O}(dtdW), \nonumber
\end{align}
which is Eq.~(\ref{eqn4}) in the paper.
Eq.~(\ref{eqn5}) is derived by multiplying both sides of Eq.~(\ref{eqn4}) by the operator $\mathcal{O}$ and taking the trace. Since $[H,\mathcal{O}]=[H,Z_m]=0$, Eq.~(\ref{eqn5}) can be derived as 
\begin{align}
&d\langle \mathcal{O} \rangle=\Tr \left[\mathcal{O} d\rho\right] \nonumber \\
&= -i \cancelto{0}{\Tr \left[ \mathcal{O} [H,\rho]\right]}dt+ \lambda \cancelto{0}{\Tr[\mathcal{O}Z_m \rho Z_m -\mathcal{O}\rho] }dt \nonumber \\
& \ \ \ +\sqrt{\lambda}\Tr\left[\mathcal{O} Z_m \rho +\mathcal{O} \rho Z_m -2 \langle Z_m\rangle\mathcal{O}  \rho \right]dW \nonumber \\
&=2\sqrt{\lambda}\left(\langle Z_m \mathcal{O}  \rangle -\langle Z_m\rangle \langle \mathcal{O} \rangle \right)dW.
\end{align}

\section{Bayes rule relation}
 The approximation in Eq.~(\ref{prob1}) is pure expansion based on the fact that $dt$ is infinitesimal. To the first order of $dt$, the $H$ term does not appear. The overall factor $\mathcal{N}$ is irrelevant to the ratio between $p_{\pm}$ in our argument and we do not need to expand it. Let $dI= fdt +dW/(2\sqrt{\lambda})$. We have
\begin{align}
&p_{\pm}(t+dt) \nonumber \\
&= \frac{1}{\mathcal{N}}\left\{p_{\pm}(t)\Tr \left[\rho_{\pm}(t) e^{-2\lambda\left(\frac{dI}{dt}-Z_m\right)^2dt}\right] +\mathcal{O}(dt^2)\right\} \nonumber \\
&=\frac{1}{\mathcal{N}} \left\{p_{\pm}(t)\Tr \left[\rho_{\pm}(t) e^{-2\lambda\left(f-Z_m+\frac{dW}{2\sqrt{\lambda}dt}\right)^2dt}\right] +\mathcal{O}(dt^2)\right\} \nonumber \\
&=\frac{1}{\mathcal{N}} \Bigg\{p_{\pm}(t)\Tr \Big[\rho_{\pm}(t) e^{-2\lambda\left(f-Z_m\right)^2dt -2\sqrt{\lambda}(f-Z_m)dW} \nonumber \\
& \ \ \ \ \ \ \  \ \ \ \ \ \ \ \ \ \ \ \ \ \ \ \ \ \ \ \ \  \ \ \ \ \ \ \ \  \ e^{-2\lambda\left(\frac{dW}{2\sqrt{\lambda}dt}\right)^2dt} \Big]+\mathcal{O}(dt^2)\Bigg\} \nonumber\\
&= \frac{1}{\mathcal{N}}\Bigg\{p_{\pm}(t)\Tr \Big\{\rho_{\pm}(t) \big[I-2\lambda\left(f-Z_m\right)^2dt \nonumber \\
&-2\sqrt{\lambda}(f-Z_m)dW+2\lambda\left(f-Z_m\right)^2dt\big]e^{-2\lambda\left(\frac{dW}{2\sqrt{\lambda}dt}\right)^2dt}\Big\} \nonumber \\
&\ \ \ \ \ \ \  \ \ \ \ \ \ \ \ \ \ \ \ \ \ \ \ \ \ \ \ \ \ \ \ \ \ \  \ \ \ \ \ \ \ \ \ \ \ \ \ \ \ \ \ \ \ +\mathcal{O}(dt^2)\Bigg\}\nonumber \\
&= \frac{1}{\mathcal{N}}\Bigg\{p_{\pm}(t)\left[1-2\sqrt{\lambda}(f-\langle Z_m\rangle_{\pm})dW\right]e^{-2\lambda\left(\frac{dW}{2\sqrt{\lambda}dt}\right)^2dt} \nonumber \\
& \ \ \ \ \ \ \  \ \ \ \ \ \ \ \ \ \ \ \ \ \ \ \ \ \ \ \ \ \ \ \ \ \ \  \ \ \ \ \ \ \ \ \ \ \ \ \ \ \ +\mathcal{O}(dt^2)\Bigg\},
\end{align}
where $dW^2=dt$ is used. Replacing $Z_m$ by $\langle Z_m\rangle_{\pm}$ and going backwards through the above equalities, we get
\begin{align}
&p_{\pm}(t+dt) \nonumber \\
&= \frac{1}{\mathcal{N}}\left\{p_{\pm}(t)e^{-2\lambda\left(f-\langle Z_m\rangle_{\pm}+\frac{dW}{2\sqrt{\lambda}dt}\right)^2dt}+\mathcal{O}(dt^2) \right\} \nonumber\\
&=\frac{1}{\mathcal{N}}\left\{p_{\pm}(t)e^{-2\lambda\left(\frac{dI}{dt}-\langle Z_m\rangle_{\pm}\right)^2dt}+\mathcal{O}(dt^2)\right\}, 
\end{align}
which shows the Eq.~(\ref{prob1}).

\bibliography{monitor}

\providecommand{\noopsort}[1]{}\providecommand{\singleletter}[1]{#1}%
\begin{thebibliography}{34}%
\makeatletter
\providecommand \@ifxundefined [1]{%
 \@ifx{#1\undefined}
}%
\providecommand \@ifnum [1]{%
 \ifnum #1\expandafter \@firstoftwo
 \else \expandafter \@secondoftwo
 \fi
}%
\providecommand \@ifx [1]{%
 \ifx #1\expandafter \@firstoftwo
 \else \expandafter \@secondoftwo
 \fi
}%
\providecommand \natexlab [1]{#1}%
\providecommand \enquote  [1]{``#1''}%
\providecommand \bibnamefont  [1]{#1}%
\providecommand \bibfnamefont [1]{#1}%
\providecommand \citenamefont [1]{#1}%
\providecommand \href@noop [0]{\@secondoftwo}%
\providecommand \href [0]{\begingroup \@sanitize@url \@href}%
\providecommand \@href[1]{\@@startlink{#1}\@@href}%
\providecommand \@@href[1]{\endgroup#1\@@endlink}%
\providecommand \@sanitize@url [0]{\catcode `\\12\catcode `\$12\catcode
  `\&12\catcode `\#12\catcode `\^12\catcode `\_12\catcode `\%12\relax}%
\providecommand \@@startlink[1]{}%
\providecommand \@@endlink[0]{}%
\providecommand \url  [0]{\begingroup\@sanitize@url \@url }%
\providecommand \@url [1]{\endgroup\@href {#1}{\urlprefix }}%
\providecommand \urlprefix  [0]{URL }%
\providecommand \Eprint [0]{\href }%
\providecommand \doibase [0]{https://doi.org/}%
\providecommand \selectlanguage [0]{\@gobble}%
\providecommand \bibinfo  [0]{\@secondoftwo}%
\providecommand \bibfield  [0]{\@secondoftwo}%
\providecommand \translation [1]{[#1]}%
\providecommand \BibitemOpen [0]{}%
\providecommand \bibitemStop [0]{}%
\providecommand \bibitemNoStop [0]{.\EOS\space}%
\providecommand \EOS [0]{\spacefactor3000\relax}%
\providecommand \BibitemShut  [1]{\csname bibitem#1\endcsname}%
\let\auto@bib@innerbib\@empty
\bibitem [{\citenamefont {Wiseman}\ and\ \citenamefont
  {Milburn}(2009)}]{Wiseman2009}%
  \BibitemOpen
  \bibfield  {author} {\bibinfo {author} {\bibfnamefont {H.~M.}\ \bibnamefont
  {Wiseman}}\ and\ \bibinfo {author} {\bibfnamefont {G.~J.}\ \bibnamefont
  {Milburn}},\ }\href@noop {} {\emph {\bibinfo {title} {Quantum Measurement and
  Control}}}\ (\bibinfo  {publisher} {Cambridge University Press},\ \bibinfo
  {year} {2009})\BibitemShut {NoStop}%
\bibitem [{\citenamefont {Jacobs}(2014)}]{Kurt2014}%
  \BibitemOpen
  \bibfield  {author} {\bibinfo {author} {\bibfnamefont {K.}~\bibnamefont
  {Jacobs}},\ }\href@noop {} {\emph {\bibinfo {title} {Quantum Measurement
  Theory and its Applications}}}\ (\bibinfo  {publisher} {Cambridge University
  Press},\ \bibinfo {year} {2014})\BibitemShut {NoStop}%
\bibitem [{\citenamefont {Oreshkov}\ and\ \citenamefont
  {Brun}(2005)}]{Oreshkov2005}%
  \BibitemOpen
  \bibfield  {author} {\bibinfo {author} {\bibfnamefont {O.}~\bibnamefont
  {Oreshkov}}\ and\ \bibinfo {author} {\bibfnamefont {T.~A.}\ \bibnamefont
  {Brun}},\ }\bibfield  {title} {\bibinfo {title} {Weak measurements are
  universal},\ }\href {https://doi.org/10.1103/PhysRevLett.95.110409}
  {\bibfield  {journal} {\bibinfo  {journal} {Phys. Rev. Lett.}\ }\textbf
  {\bibinfo {volume} {95}},\ \bibinfo {pages} {110409} (\bibinfo {year}
  {2005})}\BibitemShut {NoStop}%
\bibitem [{\citenamefont {Korotkov}(1999)}]{Korotkov1999}%
  \BibitemOpen
  \bibfield  {author} {\bibinfo {author} {\bibfnamefont {A.~N.}\ \bibnamefont
  {Korotkov}},\ }\bibfield  {title} {\bibinfo {title} {Continuous quantum
  measurement of a double dot},\ }\href
  {https://doi.org/10.1103/PhysRevB.60.5737} {\bibfield  {journal} {\bibinfo
  {journal} {Phys. Rev. B}\ }\textbf {\bibinfo {volume} {60}},\ \bibinfo
  {pages} {5737} (\bibinfo {year} {1999})}\BibitemShut {NoStop}%
\bibitem [{\citenamefont {Vool}\ \emph {et~al.}(2016)\citenamefont {Vool},
  \citenamefont {Shankar}, \citenamefont {Mundhada}, \citenamefont {Ofek},
  \citenamefont {Narla}, \citenamefont {Sliwa}, \citenamefont {Zalys-Geller},
  \citenamefont {Liu}, \citenamefont {Frunzio}, \citenamefont {Schoelkopf},
  \citenamefont {Girvin},\ and\ \citenamefont {Devoret}}]{Vool2016}%
  \BibitemOpen
  \bibfield  {author} {\bibinfo {author} {\bibfnamefont {U.}~\bibnamefont
  {Vool}}, \bibinfo {author} {\bibfnamefont {S.}~\bibnamefont {Shankar}},
  \bibinfo {author} {\bibfnamefont {S.~O.}\ \bibnamefont {Mundhada}}, \bibinfo
  {author} {\bibfnamefont {N.}~\bibnamefont {Ofek}}, \bibinfo {author}
  {\bibfnamefont {A.}~\bibnamefont {Narla}}, \bibinfo {author} {\bibfnamefont
  {K.}~\bibnamefont {Sliwa}}, \bibinfo {author} {\bibfnamefont
  {E.}~\bibnamefont {Zalys-Geller}}, \bibinfo {author} {\bibfnamefont
  {Y.}~\bibnamefont {Liu}}, \bibinfo {author} {\bibfnamefont {L.}~\bibnamefont
  {Frunzio}}, \bibinfo {author} {\bibfnamefont {R.~J.}\ \bibnamefont
  {Schoelkopf}}, \bibinfo {author} {\bibfnamefont {S.~M.}\ \bibnamefont
  {Girvin}},\ and\ \bibinfo {author} {\bibfnamefont {M.~H.}\ \bibnamefont
  {Devoret}},\ }\bibfield  {title} {\bibinfo {title} {Continuous quantum
  nondemolition measurement of the transverse component of a qubit},\ }\href
  {https://doi.org/10.1103/PhysRevLett.117.133601} {\bibfield  {journal}
  {\bibinfo  {journal} {Phys. Rev. Lett.}\ }\textbf {\bibinfo {volume} {117}},\
  \bibinfo {pages} {133601} (\bibinfo {year} {2016})}\BibitemShut {NoStop}%
\bibitem [{\citenamefont {Murch}\ \emph {et~al.}(2013)\citenamefont {Murch},
  \citenamefont {Weber}, \citenamefont {Macklin},\ and\ \citenamefont
  {Siddiqi}}]{Murch2013}%
  \BibitemOpen
  \bibfield  {author} {\bibinfo {author} {\bibfnamefont {K.~W.}\ \bibnamefont
  {Murch}}, \bibinfo {author} {\bibfnamefont {S.~J.}\ \bibnamefont {Weber}},
  \bibinfo {author} {\bibfnamefont {C.}~\bibnamefont {Macklin}},\ and\ \bibinfo
  {author} {\bibfnamefont {I.}~\bibnamefont {Siddiqi}},\ }\bibfield  {title}
  {\bibinfo {title} {Observing single quantum trajectories of a superconducting
  quantum bit},\ }\href {https://doi.org/10.1038/nature12539} {\bibfield
  {journal} {\bibinfo  {journal} {Nature}\ }\textbf {\bibinfo {volume} {502}},\
  \bibinfo {pages} {211} (\bibinfo {year} {2013})}\BibitemShut {NoStop}%
\bibitem [{\citenamefont {Yang}\ \emph {et~al.}(2020)\citenamefont {Yang},
  \citenamefont {Grankin}, \citenamefont {Sieberer}, \citenamefont {Vasilyev},\
  and\ \citenamefont {Zoller}}]{Zoller2020}%
  \BibitemOpen
  \bibfield  {author} {\bibinfo {author} {\bibfnamefont {D.}~\bibnamefont
  {Yang}}, \bibinfo {author} {\bibfnamefont {A.}~\bibnamefont {Grankin}},
  \bibinfo {author} {\bibfnamefont {L.~M.}\ \bibnamefont {Sieberer}}, \bibinfo
  {author} {\bibfnamefont {D.~V.}\ \bibnamefont {Vasilyev}},\ and\ \bibinfo
  {author} {\bibfnamefont {P.}~\bibnamefont {Zoller}},\ }\bibfield  {title}
  {\bibinfo {title} {Quantum non-demolition measurement of a many-body
  hamiltonian},\ }\href {https://doi.org/10.1038/s41467-020-14489-5} {\bibfield
   {journal} {\bibinfo  {journal} {Nature Communications}\ }\textbf {\bibinfo
  {volume} {11}},\ \bibinfo {pages} {775} (\bibinfo {year} {2020})}\BibitemShut
  {NoStop}%
\bibitem [{\citenamefont {Korotkov}(2001)}]{Korotkov2001}%
  \BibitemOpen
  \bibfield  {author} {\bibinfo {author} {\bibfnamefont {A.~N.}\ \bibnamefont
  {Korotkov}},\ }\bibfield  {title} {\bibinfo {title} {Selective quantum
  evolution of a qubit state due to continuous measurement},\ }\href
  {https://doi.org/10.1103/PhysRevB.63.115403} {\bibfield  {journal} {\bibinfo
  {journal} {Phys. Rev. B}\ }\textbf {\bibinfo {volume} {63}},\ \bibinfo
  {pages} {115403} (\bibinfo {year} {2001})}\BibitemShut {NoStop}%
\bibitem [{\citenamefont {Weber}\ \emph {et~al.}(2014)\citenamefont {Weber},
  \citenamefont {Chantasri}, \citenamefont {Dressel}, \citenamefont {Jordan},
  \citenamefont {Murch},\ and\ \citenamefont {Siddiqi}}]{Weber2014}%
  \BibitemOpen
  \bibfield  {author} {\bibinfo {author} {\bibfnamefont {S.~J.}\ \bibnamefont
  {Weber}}, \bibinfo {author} {\bibfnamefont {A.}~\bibnamefont {Chantasri}},
  \bibinfo {author} {\bibfnamefont {J.}~\bibnamefont {Dressel}}, \bibinfo
  {author} {\bibfnamefont {A.~N.}\ \bibnamefont {Jordan}}, \bibinfo {author}
  {\bibfnamefont {K.~W.}\ \bibnamefont {Murch}},\ and\ \bibinfo {author}
  {\bibfnamefont {I.}~\bibnamefont {Siddiqi}},\ }\bibfield  {title} {\bibinfo
  {title} {Mapping the optimal route between two quantum states},\ }\href
  {https://doi.org/10.1038/nature13559} {\bibfield  {journal} {\bibinfo
  {journal} {Nature}\ }\textbf {\bibinfo {volume} {511}},\ \bibinfo {pages}
  {570} (\bibinfo {year} {2014})}\BibitemShut {NoStop}%
\bibitem [{\citenamefont {Atalaya}\ \emph {et~al.}(2017)\citenamefont
  {Atalaya}, \citenamefont {Bahrami}, \citenamefont {Pryadko},\ and\
  \citenamefont {Korotkov}}]{Atalaya2017}%
  \BibitemOpen
  \bibfield  {author} {\bibinfo {author} {\bibfnamefont {J.}~\bibnamefont
  {Atalaya}}, \bibinfo {author} {\bibfnamefont {M.}~\bibnamefont {Bahrami}},
  \bibinfo {author} {\bibfnamefont {L.~P.}\ \bibnamefont {Pryadko}},\ and\
  \bibinfo {author} {\bibfnamefont {A.~N.}\ \bibnamefont {Korotkov}},\
  }\bibfield  {title} {\bibinfo {title} {Bacon-shor code with continuous
  measurement of noncommuting operators},\ }\href
  {https://doi.org/10.1103/PhysRevA.95.032317} {\bibfield  {journal} {\bibinfo
  {journal} {Phys. Rev. A}\ }\textbf {\bibinfo {volume} {95}},\ \bibinfo
  {pages} {032317} (\bibinfo {year} {2017})}\BibitemShut {NoStop}%
\bibitem [{\citenamefont {Atalaya}\ \emph {et~al.}(2019)\citenamefont
  {Atalaya}, \citenamefont {Korotkov},\ and\ \citenamefont
  {Whaley}}]{Atalaya2019}%
  \BibitemOpen
  \bibfield  {author} {\bibinfo {author} {\bibfnamefont {J.}~\bibnamefont
  {Atalaya}}, \bibinfo {author} {\bibfnamefont {A.~N.}\ \bibnamefont
  {Korotkov}},\ and\ \bibinfo {author} {\bibfnamefont {K.~B.}\ \bibnamefont
  {Whaley}},\ }\href@noop {} {\bibinfo {title} {Error correcting bacon-shor
  code with continuous measurement of noncommuting operators}} (\bibinfo {year}
  {2019}),\ \Eprint {https://arxiv.org/abs/1910.08272} {arXiv:1910.08272
  [quant-ph]} \BibitemShut {NoStop}%
\bibitem [{\citenamefont {Atalaya}\ \emph {et~al.}(2020)\citenamefont
  {Atalaya}, \citenamefont {Zhang}, \citenamefont {Niu}, \citenamefont
  {Babakhani}, \citenamefont {Chan}, \citenamefont {Epstein},\ and\
  \citenamefont {Whaley}}]{atalaya2020continuous}%
  \BibitemOpen
  \bibfield  {author} {\bibinfo {author} {\bibfnamefont {J.}~\bibnamefont
  {Atalaya}}, \bibinfo {author} {\bibfnamefont {S.}~\bibnamefont {Zhang}},
  \bibinfo {author} {\bibfnamefont {M.~Y.}\ \bibnamefont {Niu}}, \bibinfo
  {author} {\bibfnamefont {A.}~\bibnamefont {Babakhani}}, \bibinfo {author}
  {\bibfnamefont {H.~C.~H.}\ \bibnamefont {Chan}}, \bibinfo {author}
  {\bibfnamefont {J.}~\bibnamefont {Epstein}},\ and\ \bibinfo {author}
  {\bibfnamefont {K.~B.}\ \bibnamefont {Whaley}},\ }\href@noop {} {\bibinfo
  {title} {Continuous quantum error correction for evolution under
  time-dependent hamiltonians}} (\bibinfo {year} {2020}),\ \Eprint
  {https://arxiv.org/abs/2003.11248} {arXiv:2003.11248 [quant-ph]} \BibitemShut
  {NoStop}%
\bibitem [{\citenamefont {Paz}\ and\ \citenamefont {Zurek}(1998)}]{Paz1998}%
  \BibitemOpen
  \bibfield  {author} {\bibinfo {author} {\bibfnamefont {J.~P.}\ \bibnamefont
  {Paz}}\ and\ \bibinfo {author} {\bibfnamefont {W.~H.}\ \bibnamefont
  {Zurek}},\ }\bibfield  {title} {\bibinfo {title} {Continuous error
  correction},\ }\href {https://doi.org/https://doi.org/10.1098/rspa.1998.0165}
  {\bibfield  {journal} {\bibinfo  {journal} {Proc. R. Soc. Lond. A.}\ }\textbf
  {\bibinfo {volume} {454}},\ \bibinfo {pages} {355} (\bibinfo {year}
  {1998})}\BibitemShut {NoStop}%
\bibitem [{\citenamefont {Oreshkov}(2013)}]{OreshkovBook}%
  \BibitemOpen
  \bibfield  {author} {\bibinfo {author} {\bibfnamefont {O.}~\bibnamefont
  {Oreshkov}},\ }in\ \href@noop {} {\emph {\bibinfo {booktitle} {Quantum Error
  Correction}}},\ \bibinfo {editor} {edited by\ \bibinfo {editor}
  {\bibfnamefont {D.~A.}\ \bibnamefont {Lidar}}\ and\ \bibinfo {editor}
  {\bibfnamefont {T.~A.}\ \bibnamefont {Brun}}}\ (\bibinfo  {publisher}
  {Cambridge University Press},\ \bibinfo {year} {2013})\ Chap.~\bibinfo
  {chapter} {8}, pp.\ \bibinfo {pages} {201--228}\BibitemShut {NoStop}%
\bibitem [{\citenamefont {Hsu}\ and\ \citenamefont {Brun}(2016)}]{KC2016}%
  \BibitemOpen
  \bibfield  {author} {\bibinfo {author} {\bibfnamefont {K.-C.}\ \bibnamefont
  {Hsu}}\ and\ \bibinfo {author} {\bibfnamefont {T.~A.}\ \bibnamefont {Brun}},\
  }\bibfield  {title} {\bibinfo {title} {Method for quantum-jump
  continuous-time quantum error correction},\ }\href
  {https://doi.org/10.1103/PhysRevA.93.022321} {\bibfield  {journal} {\bibinfo
  {journal} {Phys. Rev. A}\ }\textbf {\bibinfo {volume} {93}},\ \bibinfo
  {pages} {022321} (\bibinfo {year} {2016})}\BibitemShut {NoStop}%
\bibitem [{\citenamefont {Ahn}\ \emph {et~al.}(2002)\citenamefont {Ahn},
  \citenamefont {Doherty},\ and\ \citenamefont {Landahl}}]{Ahn2002}%
  \BibitemOpen
  \bibfield  {author} {\bibinfo {author} {\bibfnamefont {C.}~\bibnamefont
  {Ahn}}, \bibinfo {author} {\bibfnamefont {A.~C.}\ \bibnamefont {Doherty}},\
  and\ \bibinfo {author} {\bibfnamefont {A.~J.}\ \bibnamefont {Landahl}},\
  }\bibfield  {title} {\bibinfo {title} {Continuous quantum error correction
  via quantum feedback control},\ }\href
  {https://doi.org/10.1103/PhysRevA.65.042301} {\bibfield  {journal} {\bibinfo
  {journal} {Phys. Rev. A}\ }\textbf {\bibinfo {volume} {65}},\ \bibinfo
  {pages} {042301} (\bibinfo {year} {2002})}\BibitemShut {NoStop}%
\bibitem [{\citenamefont {Ahn}\ \emph {et~al.}(2003)\citenamefont {Ahn},
  \citenamefont {Wiseman},\ and\ \citenamefont {Milburn}}]{Ahn2003}%
  \BibitemOpen
  \bibfield  {author} {\bibinfo {author} {\bibfnamefont {C.}~\bibnamefont
  {Ahn}}, \bibinfo {author} {\bibfnamefont {H.~M.}\ \bibnamefont {Wiseman}},\
  and\ \bibinfo {author} {\bibfnamefont {G.~J.}\ \bibnamefont {Milburn}},\
  }\bibfield  {title} {\bibinfo {title} {Quantum error correction for
  continuously detected errors},\ }\href
  {https://doi.org/10.1103/PhysRevA.67.052310} {\bibfield  {journal} {\bibinfo
  {journal} {Phys. Rev. A}\ }\textbf {\bibinfo {volume} {67}},\ \bibinfo
  {pages} {052310} (\bibinfo {year} {2003})}\BibitemShut {NoStop}%
\bibitem [{\citenamefont {Ahn}\ \emph {et~al.}(2004)\citenamefont {Ahn},
  \citenamefont {Wiseman},\ and\ \citenamefont {Jacobs}}]{Ahn2004}%
  \BibitemOpen
  \bibfield  {author} {\bibinfo {author} {\bibfnamefont {C.}~\bibnamefont
  {Ahn}}, \bibinfo {author} {\bibfnamefont {H.}~\bibnamefont {Wiseman}},\ and\
  \bibinfo {author} {\bibfnamefont {K.}~\bibnamefont {Jacobs}},\ }\bibfield
  {title} {\bibinfo {title} {Quantum error correction for continuously detected
  errors with any number of error channels per qubit},\ }\href
  {https://doi.org/10.1103/PhysRevA.70.024302} {\bibfield  {journal} {\bibinfo
  {journal} {Phys. Rev. A}\ }\textbf {\bibinfo {volume} {70}},\ \bibinfo
  {pages} {024302} (\bibinfo {year} {2004})}\BibitemShut {NoStop}%
\bibitem [{\citenamefont {van Handel}\ and\ \citenamefont
  {Mabuchi}(2005)}]{Ramon2005}%
  \BibitemOpen
  \bibfield  {author} {\bibinfo {author} {\bibfnamefont {R.}~\bibnamefont {van
  Handel}}\ and\ \bibinfo {author} {\bibfnamefont {H.}~\bibnamefont
  {Mabuchi}},\ }\href@noop {} {\bibinfo {title} {Optimal error tracking via
  quantum coding and continuous syndrome measurement}} (\bibinfo {year}
  {2005}),\ \Eprint {https://arxiv.org/abs/quant-ph/0511221}
  {arXiv:quant-ph/0511221 [quant-ph]} \BibitemShut {NoStop}%
\bibitem [{\citenamefont {Chase}\ \emph {et~al.}(2008)\citenamefont {Chase},
  \citenamefont {Landahl},\ and\ \citenamefont {Geremia}}]{Chase2008}%
  \BibitemOpen
  \bibfield  {author} {\bibinfo {author} {\bibfnamefont {B.~A.}\ \bibnamefont
  {Chase}}, \bibinfo {author} {\bibfnamefont {A.~J.}\ \bibnamefont {Landahl}},\
  and\ \bibinfo {author} {\bibfnamefont {J.}~\bibnamefont {Geremia}},\
  }\bibfield  {title} {\bibinfo {title} {Efficient feedback controllers for
  continuous-time quantum error correction},\ }\href
  {https://doi.org/10.1103/PhysRevA.77.032304} {\bibfield  {journal} {\bibinfo
  {journal} {Phys. Rev. A}\ }\textbf {\bibinfo {volume} {77}},\ \bibinfo
  {pages} {032304} (\bibinfo {year} {2008})}\BibitemShut {NoStop}%
\bibitem [{\citenamefont {Gottesman}(1997)}]{gottesman1997stabilizer}%
  \BibitemOpen
  \bibfield  {author} {\bibinfo {author} {\bibfnamefont {D.}~\bibnamefont
  {Gottesman}},\ }\href@noop {} {\bibinfo {title} {Stabilizer codes and quantum
  error correction}} (\bibinfo {year} {1997}),\ \Eprint
  {https://arxiv.org/abs/quant-ph/9705052} {arXiv:quant-ph/9705052 [quant-ph]}
  \BibitemShut {NoStop}%
\bibitem [{\citenamefont {Bravyi}\ and\ \citenamefont
  {Kitaev}(1998)}]{bravyi1998quantum}%
  \BibitemOpen
  \bibfield  {author} {\bibinfo {author} {\bibfnamefont {S.~B.}\ \bibnamefont
  {Bravyi}}\ and\ \bibinfo {author} {\bibfnamefont {A.~Y.}\ \bibnamefont
  {Kitaev}},\ }\href@noop {} {\bibinfo {title} {Quantum codes on a lattice with
  boundary}} (\bibinfo {year} {1998}),\ \Eprint
  {https://arxiv.org/abs/quant-ph/9811052} {arXiv:quant-ph/9811052 [quant-ph]}
  \BibitemShut {NoStop}%
\bibitem [{\citenamefont {Fowler}\ \emph {et~al.}(2012)\citenamefont {Fowler},
  \citenamefont {Mariantoni}, \citenamefont {Martinis},\ and\ \citenamefont
  {Cleland}}]{Fowler2012}%
  \BibitemOpen
  \bibfield  {author} {\bibinfo {author} {\bibfnamefont {A.~G.}\ \bibnamefont
  {Fowler}}, \bibinfo {author} {\bibfnamefont {M.}~\bibnamefont {Mariantoni}},
  \bibinfo {author} {\bibfnamefont {J.~M.}\ \bibnamefont {Martinis}},\ and\
  \bibinfo {author} {\bibfnamefont {A.~N.}\ \bibnamefont {Cleland}},\
  }\bibfield  {title} {\bibinfo {title} {Surface codes: Towards practical
  large-scale quantum computation},\ }\href
  {https://doi.org/10.1103/PhysRevA.86.032324} {\bibfield  {journal} {\bibinfo
  {journal} {Phys. Rev. A}\ }\textbf {\bibinfo {volume} {86}},\ \bibinfo
  {pages} {032324} (\bibinfo {year} {2012})}\BibitemShut {NoStop}%
\bibitem [{\citenamefont {Bacon}(2006)}]{Bacon2006}%
  \BibitemOpen
  \bibfield  {author} {\bibinfo {author} {\bibfnamefont {D.}~\bibnamefont
  {Bacon}},\ }\bibfield  {title} {\bibinfo {title} {Operator quantum
  error-correcting subsystems for self-correcting quantum memories},\ }\href
  {https://doi.org/10.1103/PhysRevA.73.012340} {\bibfield  {journal} {\bibinfo
  {journal} {Phys. Rev. A}\ }\textbf {\bibinfo {volume} {73}},\ \bibinfo
  {pages} {012340} (\bibinfo {year} {2006})}\BibitemShut {NoStop}%
\bibitem [{\citenamefont {Oreshkov}\ and\ \citenamefont
  {Brun}(2007)}]{Oreshkov2007}%
  \BibitemOpen
  \bibfield  {author} {\bibinfo {author} {\bibfnamefont {O.}~\bibnamefont
  {Oreshkov}}\ and\ \bibinfo {author} {\bibfnamefont {T.~A.}\ \bibnamefont
  {Brun}},\ }\bibfield  {title} {\bibinfo {title} {Continuous quantum error
  correction for non-markovian decoherence},\ }\href
  {https://doi.org/10.1103/PhysRevA.76.022318} {\bibfield  {journal} {\bibinfo
  {journal} {Phys. Rev. A}\ }\textbf {\bibinfo {volume} {76}},\ \bibinfo
  {pages} {022318} (\bibinfo {year} {2007})}\BibitemShut {NoStop}%
\bibitem [{\citenamefont {Itô}(1944)}]{ito1944}%
  \BibitemOpen
  \bibfield  {author} {\bibinfo {author} {\bibfnamefont {K.}~\bibnamefont
  {Itô}},\ }\bibfield  {title} {\bibinfo {title} {Stochastic integral},\
  }\href {https://doi.org/10.3792/pia/1195572786} {\bibfield  {journal}
  {\bibinfo  {journal} {Proc. Imp. Acad.}\ }\textbf {\bibinfo {volume} {20}},\
  \bibinfo {pages} {519} (\bibinfo {year} {1944})}\BibitemShut {NoStop}%
\bibitem [{\citenamefont {Jacobs}(2010)}]{Kurt2010}%
  \BibitemOpen
  \bibfield  {author} {\bibinfo {author} {\bibfnamefont {K.}~\bibnamefont
  {Jacobs}},\ }\href@noop {} {\emph {\bibinfo {title} {Stochastic Processes For
  Physicists}}}\ (\bibinfo  {publisher} {Cambridge University Press},\ \bibinfo
  {year} {2010})\BibitemShut {NoStop}%
\bibitem [{\citenamefont {Milotti}(2002)}]{milotti20021f}%
  \BibitemOpen
  \bibfield  {author} {\bibinfo {author} {\bibfnamefont {E.}~\bibnamefont
  {Milotti}},\ }\href@noop {} {\bibinfo {title} {1/f noise: a pedagogical
  review}} (\bibinfo {year} {2002}),\ \Eprint
  {https://arxiv.org/abs/physics/0204033} {arXiv:physics/0204033
  [physics.class-ph]} \BibitemShut {NoStop}%
\bibitem [{\citenamefont {Paz-Silva}\ \emph {et~al.}(2012)\citenamefont
  {Paz-Silva}, \citenamefont {Rezakhani}, \citenamefont {Dominy},\ and\
  \citenamefont {Lidar}}]{Paz-Silva2012}%
  \BibitemOpen
  \bibfield  {author} {\bibinfo {author} {\bibfnamefont {G.~A.}\ \bibnamefont
  {Paz-Silva}}, \bibinfo {author} {\bibfnamefont {A.~T.}\ \bibnamefont
  {Rezakhani}}, \bibinfo {author} {\bibfnamefont {J.~M.}\ \bibnamefont
  {Dominy}},\ and\ \bibinfo {author} {\bibfnamefont {D.~A.}\ \bibnamefont
  {Lidar}},\ }\bibfield  {title} {\bibinfo {title} {Zeno effect for quantum
  computation and control},\ }\href
  {https://doi.org/10.1103/PhysRevLett.108.080501} {\bibfield  {journal}
  {\bibinfo  {journal} {Phys. Rev. Lett.}\ }\textbf {\bibinfo {volume} {108}},\
  \bibinfo {pages} {080501} (\bibinfo {year} {2012})}\BibitemShut {NoStop}%
\bibitem [{\citenamefont {W\"uster}(2017)}]{Wuster2017}%
  \BibitemOpen
  \bibfield  {author} {\bibinfo {author} {\bibfnamefont {S.}~\bibnamefont
  {W\"uster}},\ }\bibfield  {title} {\bibinfo {title} {Quantum zeno suppression
  of intramolecular forces},\ }\href
  {https://doi.org/10.1103/PhysRevLett.119.013001} {\bibfield  {journal}
  {\bibinfo  {journal} {Phys. Rev. Lett.}\ }\textbf {\bibinfo {volume} {119}},\
  \bibinfo {pages} {013001} (\bibinfo {year} {2017})}\BibitemShut {NoStop}%
\bibitem [{\citenamefont {Kondo}\ \emph {et~al.}(2016)\citenamefont {Kondo},
  \citenamefont {Matsuzaki}, \citenamefont {Matsushima},\ and\ \citenamefont
  {Filgueiras}}]{Kondo_2016}%
  \BibitemOpen
  \bibfield  {author} {\bibinfo {author} {\bibfnamefont {Y.}~\bibnamefont
  {Kondo}}, \bibinfo {author} {\bibfnamefont {Y.}~\bibnamefont {Matsuzaki}},
  \bibinfo {author} {\bibfnamefont {K.}~\bibnamefont {Matsushima}},\ and\
  \bibinfo {author} {\bibfnamefont {J.~G.}\ \bibnamefont {Filgueiras}},\
  }\bibfield  {title} {\bibinfo {title} {Using the quantum zeno effect for
  suppression of decoherence},\ }\href
  {https://doi.org/10.1088/1367-2630/18/1/013033} {\bibfield  {journal}
  {\bibinfo  {journal} {New Journal of Physics}\ }\textbf {\bibinfo {volume}
  {18}},\ \bibinfo {pages} {013033} (\bibinfo {year} {2016})}\BibitemShut
  {NoStop}%
\bibitem [{\citenamefont {Jordan}\ and\ \citenamefont
  {Farhi}(2008)}]{Jordan2008}%
  \BibitemOpen
  \bibfield  {author} {\bibinfo {author} {\bibfnamefont {S.~P.}\ \bibnamefont
  {Jordan}}\ and\ \bibinfo {author} {\bibfnamefont {E.}~\bibnamefont {Farhi}},\
  }\bibfield  {title} {\bibinfo {title} {Perturbative gadgets at arbitrary
  orders},\ }\href {https://doi.org/10.1103/PhysRevA.77.062329} {\bibfield
  {journal} {\bibinfo  {journal} {Phys. Rev. A}\ }\textbf {\bibinfo {volume}
  {77}},\ \bibinfo {pages} {062329} (\bibinfo {year} {2008})}\BibitemShut
  {NoStop}%
\bibitem [{\citenamefont {Sanderson}\ and\ \citenamefont
  {Curtin}(2016)}]{Sanderson2016}%
  \BibitemOpen
  \bibfield  {author} {\bibinfo {author} {\bibfnamefont {C.}~\bibnamefont
  {Sanderson}}\ and\ \bibinfo {author} {\bibfnamefont {R.}~\bibnamefont
  {Curtin}},\ }\bibfield  {title} {\bibinfo {title} {Armadillo: a
  template-based c++ library for linear algebra},\ }\href
  {https://doi.org/10.21105/joss.00026} {\bibfield  {journal} {\bibinfo
  {journal} {Journal of Open Source Software}\ }\textbf {\bibinfo {volume}
  {1}},\ \bibinfo {pages} {26} (\bibinfo {year} {2016})}\BibitemShut {NoStop}%
\bibitem [{\citenamefont {Sanderson}\ and\ \citenamefont
  {Curtin}(2018)}]{Sanderson2018}%
  \BibitemOpen
  \bibfield  {author} {\bibinfo {author} {\bibfnamefont {C.}~\bibnamefont
  {Sanderson}}\ and\ \bibinfo {author} {\bibfnamefont {R.}~\bibnamefont
  {Curtin}},\ }\bibfield  {title} {\bibinfo {title} {A user-friendly hybrid
  sparse matrix class in c++},\ }\href@noop {} {\bibfield  {journal} {\bibinfo
  {journal} {Lecture Notes in Computer Science (LNCS)}\ }\textbf {\bibinfo
  {volume} {10931}},\ \bibinfo {pages} {422} (\bibinfo {year}
  {2018})}\BibitemShut {NoStop}%
\end{thebibliography}%

\end{document}